\documentclass[11pt,a4paper]{article}

\pdfoutput=1
\usepackage{amssymb}
\usepackage{amsmath}
\usepackage{graphicx}
\usepackage{booktabs}
\usepackage{color}
\usepackage{textcomp}
\usepackage{multirow}
\usepackage{bm}
\usepackage{caption}
\usepackage{subcaption}
\usepackage{tikz}
\usepackage{longtable}
\usepackage{colortbl}
\usepackage{rotating}
\usepackage{jheppub}
\usepackage[utf8]{inputenc}

\definecolor{jlab_red}{RGB}{192,39,45}
\definecolor{jlab_orange}{RGB}{249,102,0}
\definecolor{jlab_blue}{RGB}{47,122,121}
\definecolor{jlab_green}{RGB}{65,125,10}

\newcommand{\pmerr}[2]{\, ^{+ \, #1}_{- \, #2}}
\newcommand{\cm}{\ensuremath{\mathsf{cm}}}
\newcommand{\nn}{\mathfrak{n}}

\newcommand{\singleop}{\ensuremath{{\bar{\psi} \mathbf{\Gamma} \psi}}}
\newcommand{\Ndf}{\ensuremath{N_\mathrm{dof}}}
\newenvironment{smallbmatrix}{\left[\begin{smallmatrix}}{\end{smallmatrix}\right]}

\hypersetup{%
pdftitle = {D K I=0, D Kbar I=0,1 scattering and the Ds0*(2317) from lattice QCD},
pdfauthor = {Hadron Spectrum Collaboration},
colorlinks = {true},
filecolor = {black},
linkcolor = {jlab_blue},
menucolor = {black},
citecolor = {jlab_green},
urlcolor = {jlab_green},
}{}

\begin{document}

\title{$DK$ $I=0,$ $D\bar{K}$ $I=0,1$ scattering and the $D_{s0}^\ast(2317)$ from lattice QCD}

\author{Gavin~K.~C.~Cheung,}
\author[a]{Christopher~E.~Thomas,} \emailAdd{c.e.thomas@damtp.cam.ac.uk}
\author[a]{David~J.~Wilson,} \emailAdd{d.j.wilson@damtp.cam.ac.uk}
\author[b]{Graham~Moir,}
\author[c]{Michael~Peardon,}
\author[c]{Sin\'{e}ad~M.~Ryan}
\author{\\(for the Hadron Spectrum Collaboration)}

\affiliation[a]{DAMTP, University of Cambridge, Centre for Mathematical Sciences, Wilberforce Road, \mbox{Cambridge}, CB3 0WA, UK}
\affiliation[b]{Hurstpierpoint College, College Lane, Hurstpierpoint, West Sussex, BN6 9JS, UK}
\affiliation[c]{School of Mathematics and Hamilton Mathematics Institute, Trinity College, Dublin 2, Ireland}

\abstract{Elastic scattering amplitudes for $I=0$ $DK$ and $I=0,1$ $D\bar{K}$ are computed in $S$, $P$ and $D$ partial waves using lattice QCD with light-quark masses corresponding to $m_\pi = 239$~MeV and $m_\pi = 391$~MeV. The $S$-waves contain interesting features including a near-threshold $J^P=0^+$ bound state in $I=0$ $DK$, corresponding to the $D_{s0}^\ast(2317)$, with an effect that is clearly visible above threshold, and suggestions of a $0^+$ virtual bound state in $I=0$ $D\bar{K}$. The $S$-wave $I=1$ $D\bar{K}$ amplitude is found to be weakly repulsive. The computed finite-volume spectra also contain a deeply-bound $D^\ast$ vector resonance, but negligibly small $P$-wave $DK$ interactions are observed in the energy region considered; the $P$ and $D$-wave $D\bar{K}$ amplitudes are also small. There is some evidence of $1^+$ and $2^+$ resonances in $I=0$ $DK$ at higher energies.}

\arxivnumber{2008.06432}

\maketitle

\section{Introduction}
\label{sec:introduction}

Many questions in hadron spectroscopy remain unanswered following recent
precision data from experiments investigating the charm quark.
These studies revealed a number of surprising features. While the best known 
are the so-called ``$XYZ$s'', some of which 
appear as charmonium-like states close to thresholds for two-meson decay,
open-charm mesons also show unexpected properties. Ultimately, forming 
a complete understanding of mesons requires calculation of the scattering 
amplitudes involving meson states. Since the dynamics of hadrons is governed by 
quantum chromodynamics (QCD) which is strongly coupled at hadronic energy 
scales, a non-perturbative method is needed. A lattice regularisation of the 
QCD path integral provides just such a technique, amenable to 
large-scale numerical investigation. The aim of lattice spectroscopy studies 
is to connect experimentally-observed hadron resonances directly with the QCD 
Lagrangian in order to learn more about the confined constituents of these 
states. With a more complete understanding of scattering involving mesons made 
of charm quarks as a goal, the work described in this paper investigates the 
charm-anti-strange and charm-strange sectors. 

The $D_{s0}^\ast(2317)$ charm-anti-strange scalar meson, first observed by the
BaBar collaboration~\cite{Aubert:2003fg}, has attracted significant attention. 
The mass is much lower than predicted by the quark model~\cite{Godfrey:1985xj}, where it 
is described as a spin-triplet $P$-wave orbital excitation, and is surprisingly close
to the corresponding charm-light scalar state, $D_0^\ast(2400)$. The 
$D_{s0}^\ast(2317)$ appears as a very narrow resonance below the $DK$ 
threshold and decays almost entirely to $D_s^+ \pi^0$, breaking isospin 
symmetry. This contrasts strongly with the $D_0^\ast(2400)$, found 
above the corresponding $D\pi$ threshold and seen as a broad resonance close to
predictions from the quark model. A number of explanations for the failure
of the quark model to predict the $D_{s0}^\ast(2317)$ meson mass reliably 
have been postulated, the most common modelling it as a $DK$ molecule, a tetraquark
or a conventional meson which has its properties modified by coupling to $DK$
(see for example Ref.~\cite{Chen:2016spr} for a recent review). 
An axial-vector state, $D_{s1}(2460)$, was also discovered~\cite{Besson:2003jp} 
and similarly does not fit easily into expectations based on a quark model.
While this article was being finalised, LHCb announced the observation of a structure
in the exotic-flavour $D^- K^+$ ($\bar{c}\bar{s}du$) channel at an energy
$\sim 2.9$ GeV~\cite{LHCbCharmStrange:2020,Aaij:2020hon,Aaij:2020ypa}.

We have recently performed a lattice QCD calculation of the coupled-channel $D\pi$, $D \eta$, $D_s \bar{K}$ isospin-1/2 scattering amplitudes \cite{Moir:2016srx} in an attempt to understand the $D_0^\ast(2400)$. The scattering amplitudes were analytically continued in the complex plane and a scalar bound state was found just below the $D\pi$ threshold. This calculation was performed with unphysically-heavy light quarks corresponding to $m_\pi = 391$ MeV and calculations closer to 
the physical point are required to make a more definite comparison with the experimentally-observed state.

This work focuses on the charm-anti-strange and charm-strange sectors,
calculating the isoscalar ($I=0$) $DK$ elastic scattering amplitudes in $S$, $P$ and
$D$ partial waves, along with isoscalar and isovector ($I=0$ and $1$) $D\bar K$
scattering. Computations are performed on two sets of Monte Carlo ensembles
of gauge configurations, 
sampled with light-quark masses corresponding to pion masses of approximately $239$ and $391$ MeV. A large range of
finite-volume spectra is extracted for various irreducible representations of
the group of spatial rotations, including for systems with overall non-zero momentum
with respect to the lattice. From the energy levels, the
infinite-volume scattering amplitudes are constrained using the L\"uscher
method and then analytically continued in the complex plane where the pole singularities
correspond to bound states and resonances. A scalar bound
state is found in $I=0$ $DK$ which we identify with the $D_{s0}^\ast(2317)$. In addition, a
deeply bound state and a resonance are found in the $P$ and $D$ wave scattering
amplitudes respectively. For both the $S$ and $P$-wave bound states, comparisons 
between the two sets of ensembles show little light-quark mass dependence.  In
addition, we find suggestions of a virtual bound state in $S$-wave $D\bar{K}$.
Some preliminary results from these calculations have appeared in
Refs.~\cite{Thomas:2016euu,Cheung:2019fsx}.

The rest of this paper is organised as follows: Sections \ref{sec:calculation} and \ref{sec:lattice} give the calculation details and lattice parameters. Results for the finite-volume spectra are presented in Section \ref{sec:spectrum} and the scattering amplitudes, derived using L\"uscher's framework, are given in Section \ref{sec:scattering}. In Section \ref{sec:interpretation} we analytically continue the amplitudes to the complex energy plane and find the location of pole singularities, interpret the results and compare with previous lattice calculations which have studied $DK$/$D\bar{K}$ scattering~\cite{Liu:2012zya,Mohler:2013rwa,Lang:2014yfa,Bali:2017pdv,Alexandrou:2019tmk}. A summary and outlook are presented in Section \ref{sec:conclusion}.

\section{Calculation details}
\label{sec:calculation}

Lattice QCD calculations are performed in a finite volume leading to the quantisation of momentum and a discrete spectrum. For a cubic spatial volume with periodic boundary conditions such as that used in this work, momentum is quantised as $\vec{P} = \frac{2\pi}{L} (n_x,n_y,n_z)$, where $L$ is the spatial extent and $n_i$ are integers; we will use a shorthand notation $[n_xn_yn_z]$.
Furthermore, the finite volume and lattice discretisation break the continuous rotational symmetry of an infinite-volume continuum. This means that angular momentum, $J$, is not a good quantum number and states must instead be labelled by the irreducible representations (\emph{irreps}) of the remaining symmetry group. For a cubic lattice and spatial volume, the relevant group is the octahedral group with parity, $O_h$, for mesons at rest~\cite{Johnson:1982yq} and the smaller little group, $\text{LG}(\vec{P})$, for mesons at non-zero momentum $\vec{P}$~\cite{Moore:2005dw}.

We will follow our well-established procedure to determine the finite-volume spectrum in each channel. A matrix of correlation functions is computed,
\begin{equation}
C_{ij}(t) = \langle 0 | \mathcal{O}_i^{\vphantom{\dagger}} (t) \mathcal{O}_j^\dagger(0) | 0 \rangle \,
\end{equation}
for a basis of interpolating operators, $\{\mathcal{O}_i\}$, with the appropriate quantum numbers. The spectrum is then extracted using a variational method where a generalised eigenvalue problem $C_{ij}(t) v^{\nn}_j = \lambda^\nn(t,t_0) C_{ij}(t_0) v^\nn_j$ is solved for some appropriate choice of $t_0$ \cite{Michael:1985ne,Luscher:1990ck}. The $\nn$'th eigenvalue $\lambda^\nn$, referred to as a \emph{principal correlator}, is related to the energy $E_\nn$ of the $\nn$'th energy eigenstate $|\nn\rangle$. In our implementation, detailed in Refs.~\cite{Dudek:2007wv,Dudek:2010wm}, the energies are obtained by fitting the principal correlators to the form $\lambda^\nn(t,t_0) = (1-A_\nn) e^{-E_\nn(t-t_0)} + A_\nn e^{-E'_\nn(t-t_0)}$; the fit parameters are $E_\nn, E_\nn'$ and $A_\nn$, and the second exponential is used to account for possible contamination from excited states. The eigenvectors, $v^\nn_j$, are related to the matrix elements, $Z_i^{\nn} = \langle \nn | \mathcal{O}_i^\dagger | 0 \rangle$, and can be used to construct the variationally-optimal combination of $\{\mathcal{O}_i\}$ to interpolate $|\nn\rangle$, $\Omega_\nn^\dagger = \sum_i v^\nn_i \mathcal{O}^\dagger_i$.

The basis of operators used must have an appropriate variety of structures to robustly extract the finite-volume states of interest (see e.g.~Refs.~\cite{Dudek:2012xn,Wilson:2015dqa,Cheung:2017tnt}). In this work, we use \emph{single-meson operators}, $\mathcal{O}^{\Lambda \mu}_{\mathbb{M}}(\vec{P})$, with a structure resembling a single meson, and \emph{meson-meson operators}, $\mathcal{O}^{\Lambda \mu}_{\mathbb{M}_1\mathbb{M}_2}(\vec{P})$, with a structure resembling two mesons, projected onto definite momentum, $\vec{P}$, and transforming in an irrep, $\Lambda$, and row, $\mu$, of the relevant symmetry group.
Single-meson operators~\cite{Dudek:2010wm,Thomas:2011rh} are constructed from fermion bilinears, $\sum_{\vec{x}} e^{i \vec{P} \cdot \vec{x}}\bar{\psi} \Gamma \overleftrightarrow{D} \dots \psi$, with a definite $J$ that are then subduced to the lattice irrep.
Meson-meson operators~\cite{Dudek:2012gj,Dudek:2012xn} are constructed from the product of two single-meson operators, $\sum_{\vec{p}_1, \vec{p}_2} \mathcal{C}(\vec{p}_1, \vec{p}_2) \Omega_{\mathbb{M}_1}^\dagger(\vec{p}_1) \Omega^\dagger_{\mathbb{M}_2}(\vec{p}_2)$ where $\mathcal{C}$ is a generalised Clebsch-Gordan coefficient and and $\Omega_{\mathbb{M}_i}^\dagger(\vec{p}_i)$ is a variationally-optimised operator for interpolating meson $\mathbb{M}_i$.
Previous calculations have suggested that local tetraquark-like operators have little effect on the spectra~\cite{Cheung:2017tnt,Padmanath:2015era}.
Because we only consider an energy region well below any relevant three-meson thresholds, we do not include operators constructed with a structure resembling three or more mesons~\cite{Cheung:2017tnt,Woss:2019hse}. The specific bases of operators we use for each channel are presented in Appendix~\ref{app:operators}.

\begin{figure}
\begin{center}
\includegraphics[width=0.99\textwidth]{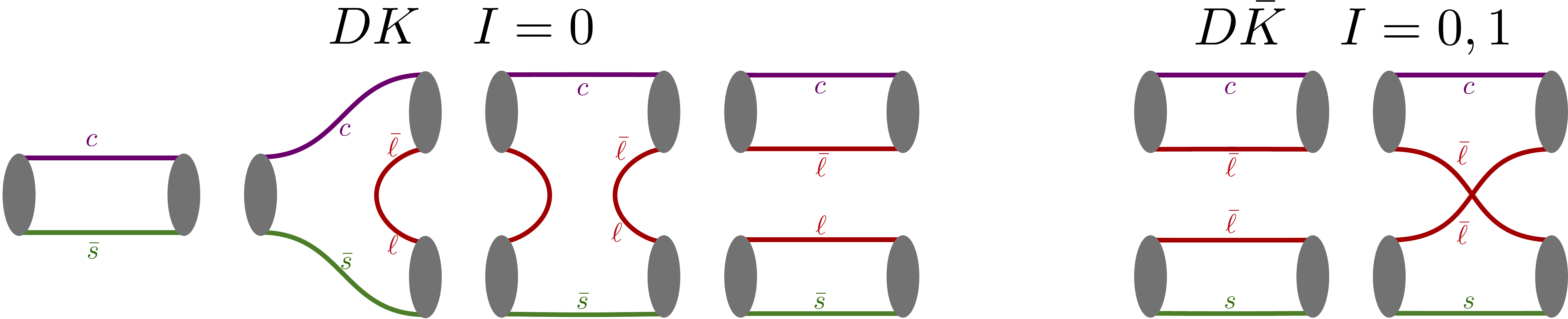}
\caption{A schematic representation of the types of Wick contraction topologies that contribute to isospin-0 $DK$ and isospin-0 and 1 $D\bar{K}$ correlation functions involving $D_s$ (single-meson) and $DK/\bar{K}$ (meson-meson) operators. The grey ellipses represent fermion-bilinear operators with various structures or sums of such operators.
For each irrep we compute sums of several thousand Wick contractions with these sorts of topologies.}
\label{fig:wx}
\end{center}
\end{figure}

To calculate the matrix of two-point correlation functions, we use the distillation framework~\cite{Peardon:2009gh}. In this approach, the quark fields appearing in interpolating operators are smeared with a distillation operator, $\Box(t) = \sum^{N_{\text{vecs}}}_{i=1} \xi_i(t) \xi_i^\dagger(t)$, where $\xi_i$ are the lowest $N_{\text{vecs}}$ eigenvectors of the discretised gauge-covariant Laplacian. This procedure factorises the computation of correlation functions and allows us to consider operators with various structures where each operator is projected onto a definite momentum. In addition, distillation enables contributions to Wick contractions where quark fields annihilate to be computed efficiently. A schematic representation of the diagrams contributing to isospin-0 $DK$ and isospin-0 and 1 $D\bar{K}$ is presented in Fig.~\ref{fig:wx}.

\section{Lattice details}
\label{sec:lattice}

\begin{table}
\centering
\begin{tabular}{c|c|ccc}
 Ensemble & $(L/a_s)^3 \times (T/a_t)$ & $N_{\text{cfgs}}$ & $ N_{\text{tsrcs}}$ & $N_{\text{vecs}}$ \\
\hline
$m_\pi = 239 $ MeV & $32^3 \times 256$ & 484 & 1 ($DK$, $D\bar{K}$, disp) & 256 \\
\hline
\multirow{3}{*}{$m_\pi = 391$ MeV}  & $16^3 \times 128$ & 478 & 8 ($DK$, $D\bar{K}$, disp) & 64\\
                                          & $20^3 \times 128$ & 603 & 3 ($DK$, $D\bar{K}$), 4 (disp) & 128 \\
                                          & $24^3 \times 128$ & 553 & 1 ($DK$), 1-3 ($D\bar{K}$), 4 (disp) & 162 
\end{tabular}
\caption{Summary of the lattice ensembles used in this work. The volume is given by $(L/a_s)^3 \times (T/a_t)$ where $L$ and $T$ are the spatial and temporal extents of the lattice. $N_{\text{cfgs}}$ is the number of gauge field configurations used, $ N_{\text{tsrcs}}$ is the number of time-sources used per configuration, and $N_{\text{vecs}}$ is the number of distillation vectors in the distillation framework. The number of time sources used to compute the momentum dependence of the $D_{(s)}^{(\ast)}$ energies for the dispersion relation fits, Fig.~\ref{fig:840disp}, are indicated by ``disp''. }
\label{table:ensembles}
\end{table}

We perform calculations on anisotropic lattices where the temporal lattice spacing, $a_t$, is smaller than the spatial lattice spacing, $a_s \approx 0.12$ fm, such that $\xi = a_s/a_t \approx 3.5$ -- the finer resolution in time enables a better extraction of finite-volume energies from two-point correlation functions. The discretised QCD action consists of a tree-level Symanzik-improved gauge action and a Wilson-clover fermion action with $N_f = 2+1$ flavours of dynamical quarks (two degenerate light up and down quarks and a heavier strange quark)~\cite{Edwards:2008ja,Lin:2008pr}. In this work we use two sets of ensembles: one volume where the light quark mass parameter is tuned such that $m_\pi = 239$~MeV and three volumes with $m_\pi = 391$~MeV; in both cases the strange quark is tuned to approximate the physical strange quark mass. The quenched charm quark is described by the same action with the mass parameter tuned to reproduce the physical $\eta_c$ meson mass~\cite{Liu:2012ze,Cheung:2016bym}. Details of the lattice ensembles used are summarised in Table~\ref{table:ensembles}.

When quoting results in physical units, we set the scale using the mass of the $\Omega$ baryon to determine $a_t^{-1} = m_\Omega^{\textrm{phys}}/a_t m_\Omega$. This gives $a_t^{-1} = 6079$~MeV for the $m_\pi = 239$~MeV ensemble~\cite{Wilson:2019wfr} and $a_t^{-1} = 5667$~MeV for $m_\pi = 391$~MeV~\cite{Edwards:2011jj}.

\begin{table}
\begin{center}
\begin{minipage}[c]{0.45\textwidth}
\begin{tabular}{c|c|c}
            & \multicolumn{2}{c}{$a_t m$} \\
\hline
$m_\pi$      &  239 MeV      & 391 MeV \\
\hline
$\pi$       & 0.03928(18)~\cite{Wilson:2015dqa}   & 0.06906(13)~\cite{Dudek:2012gj} \\
$K$         & 0.08344(7) ~\cite{Wilson:2015dqa}   & 0.09698(9)~\cite{Wilson:2014cna} \\
$\eta$      & 0.09299(56)~\cite{Wilson:2015dqa}   & 0.10364(19)~\cite{Dudek:2016cru} \\
$D$         & 0.30923(11)~\cite{Cheung:2016bym}   & 0.33303(31) \\
$D_s$       & 0.32356(12)                         & 0.34441(29) \\
$D^\ast$    & 0.33058(24)                         & 0.35494(46) \\
$D_s^\ast$  & 0.34448(15)                         & 0.36587(35) \\
\end{tabular} 
\end{minipage}  
\hspace{0.05\textwidth}
\begin{minipage}[c]{0.45\textwidth}
\begin{tabular}{c|c|c}
& \multicolumn{2}{c}{$a_t E_\mathrm{threshold}$} \\
\hline
$m_\pi$             & $239$ MeV   & $391$ MeV \\
\hline
$DK$                & 0.39267(13) & 0.4300(3) \\
$D_s \eta$          & 0.4166(6)   & 0.4481(3) \\
$D^\ast K$          & 0.4140(3)   & 0.4519(5) \\
$D_s^\ast \eta$     & 0.4375(6)   & 0.4695(4) \\
$D_s \pi \pi$       & 0.4021(3)   & 0.4825(3) \\
$D_s^\ast \pi \pi$  & 0.4230(3)   & 0.5040(4) \\
\end{tabular} 
\end{minipage}
\end{center}
\caption{Left: relevant stable meson masses for the $m_\pi = 239$~MeV and $m_\pi = 391$~MeV ensembles from dispersion fits. Right: relevant kinematic thresholds.}
\label{table:mesons}
\end{table}

The masses of stable mesons used when extracting scattering amplitudes, along with masses of other relevant stable mesons and kinematic thresholds, are summarised in Table~\ref{table:mesons}.
The anisotropy, $\xi$, can be determined by fitting the momentum dependence of a stable hadron's energy to the relativistic dispersion relation~\cite{Dudek:2012gj}. On the $m_\pi = 239$ MeV ensemble, $\xi_\pi = 3.453(6)$, $\xi_K = 3.462(4)$~\cite{Wilson:2015dqa} and $\xi_D = 3.443(7)$~\cite{Cheung:2016bym} from the $\pi$, $K$ and $D$ meson respectively, which are broadly consistent within statistical uncertainties. In the calculations of scattering amplitudes we will take $\xi = \xi_\pi = 3.453(6)$, but when estimating contributions to the systematic uncertainties we use values spanning the range of these anisotropies, $\xi = 3.436$ and $3.466$.  
For the $D$ meson mass on this ensemble we use the result of the fit to the dispersion relation in Ref.~\cite{Cheung:2016bym} and analogous fits for the $D_s$ and $D_{(s)}^\ast$ mesons.

\begin{figure}[tb]
\includegraphics[width=1.0\textwidth]{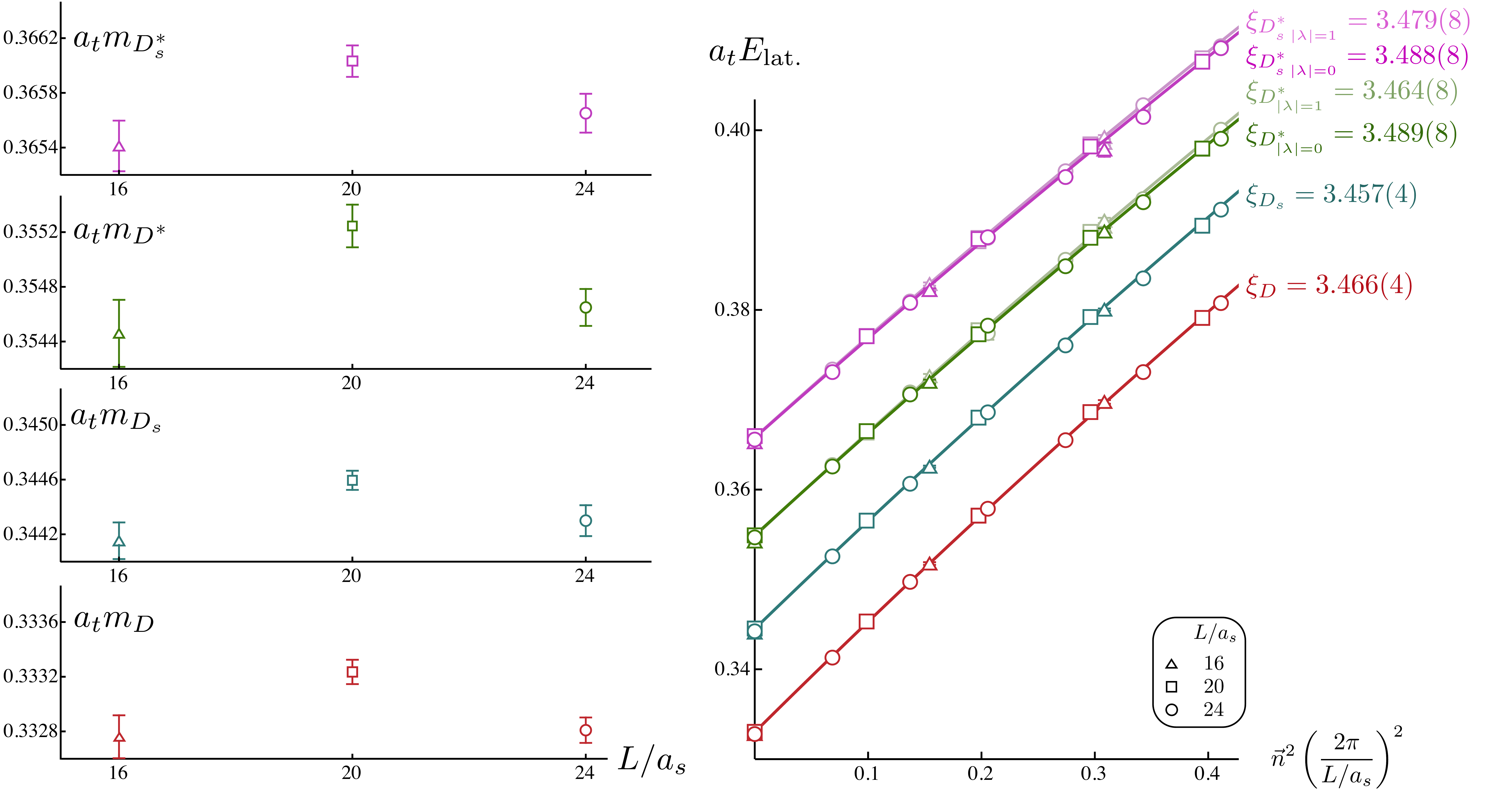}
\caption{Charmed meson masses and dispersion relation fits on the ensembles with $m_\pi~=~391$~MeV. Left panel shows meson masses extracted on each volume from a fit to the dispersion relation on that volume. Right panel shows the meson energies from all three volumes as a function of momentum and results of fits to the dispersion relation on the largest volume -- these all give a reasonable fit to the data.}
\label{fig:840disp}
\end{figure}

On the $m_\pi = 391$~MeV ensembles, $\xi_\pi = 3.444(6)$~\cite{Dudek:2012gj} and $\xi_K = 3.449(4)$~\cite{Wilson:2014cna}.
For the $D$ meson we update the original dispersion fit presented in Ref.~\cite{Liu:2012ze} with a fit that includes all the propagators now available and the $20^3$ volume -- Fig.~\ref{fig:840disp} summarises the results. The right panel shows a fit on the largest volume which gives $\xi_D = 3.466(4)$. We will use $\xi = \xi_\pi = 3.444(6)$, but when estimating some of the systematic uncertainties we use values spanning the range of the $\pi$, $K$ and $D$ anisotropies, $\xi = 3.438$ and $3.470$.
As the central values for the charm meson masses (see Table~\ref{table:mesons}) we take the average from fits to the dispersion relations on the $20^3$ and $24^3$ volumes separately,\footnote{We do not include fits on the $16^3$ volume because this is a relatively small spatial volume and so exponentially-suppressed finite-volume effects may be more significant.}
and for the uncertainty we quote a value which spans the central values and uncertainties obtained from those fits.\footnote{We quote a symmetric uncertainty which is the larger of the $+$ and $-$ uncertainties.}

In the left panel of Fig.~\ref{fig:840disp} it can be seen that on each volume the mass of the $D$ meson is extracted with high statistical precision and there are small but statistically significant differences between the different volumes. To allow for this systematic effect in our analyses, we add an additional systematic uncertainty to each finite-volume energy level on the $m_\pi = 391$~MeV ensembles via,
\vspace{-0.2cm}
$$a_t \delta E_\mathrm{lat.}\to a_t \delta E_\mathrm{lat.+syst.} = a_t\left( \delta E_\mathrm{lat.}^2 + \delta E_{\mathrm{syst.}}^2 \right)^\frac{1}{2} \, ,$$
\vspace{-0.6cm}

\noindent
where we estimate $\delta E_{\mathrm{syst.}} = 0.00021$ to span an envelope of the central values of the $D$ meson masses on the $20^3$ and $24^3$ volumes.


\section{Finite-volume spectra}
\label{sec:spectrum}

In this section we first present the finite-volume spectra obtained in the isospin-0 $DK$ sector on the $m_\pi = 239$~MeV and $m_\pi = 391$~MeV ensembles, and then show the analogous isospin-0 and isospin-1 $D\bar{K}$ results. The $J^P$ and pseudoscalar-pseudoscalar partial waves, $\ell$, which contribute to the different irreps are presented in Table \ref{table:subductions}. These spectra are analysed to determine scattering amplitudes in Section~\ref{sec:scattering}. 

\begin{table}[t!]
\begin{center}
\begin{tabular}{cc|c|l|l}
           &&&&\\[-1.7ex]
      \multirow{2}{*}{$\vec{P}$} & \multirow{2}{*}{LG$(\vec{P})  $ }  & \;  \multirow{2}{*}{$\Lambda$} \;& $\,\,J^P (\vec{P}=\vec{0})$    & \; \multirow{2}{*}{$\ell^N$} \\  
      &               &       & $\left|\lambda\right|^{({\tilde{\eta}})}(\vec{P}\neq\vec{0})$ & \\[0.5ex]
      \hline \hline
           &&&&\\[-1.5ex]
      \multirow{7}{*}{$\left[0,0,0\right]$}&     \multirow{7}{*}{$\textrm{O}_h^\textrm{D}$ ($\textrm{O}_h$)} & $A_1^+$            
      & $0^+,\, 4^+$          &\; $0^1,\, 4^1$\\
      && $T_1^-$    & $1^-,\, 3^-,\, \mathit{(4^-)}$ &\; $1^1,\, 3^1$\\
      && $E^+$      & $2^+,\, 4^+$          &\; $2^1,\, 4^1$\\
      && $T_2^+$    & $2^+,\, 4^+,\,  \mathit{(3^+)}\,$  &\; $2^1,\, 4^1$\\
      && $T_1^+$    & $4^+, \, \mathit{(1^+,3^+)}$ &\; $4^1$ \\
      && $T_2^-$    & $3^-,\, \mathit{(2^-,4^-)}$ &\; $3^1$ \\
      && $A_2^-$    & $3^-$               &\; $3^1$\\[0.5ex]
      \hline
      \hline 
      &&&&\\[-1.5ex]
      
      \multirow{5}{*}{$\left[0,0,n\right]$} & \multirow{5}{*}{Dic$_4$ ($\textrm{C}_{4\textrm{v}}$)}
      & $A_1$      & $0^+,\, 4$        &\; $0^1,\, 1^1,\, 2^1,\, 3^1,\, 4^2$ \\
      && $E_2$      & $1,\, 3$          &\; $1^1,\, 2^1,\, 3^2,\, 4^2$\\
      && $B_1$      & $2$               &\; $2^1,\, 3^1,\, 4^1$ \\
      && $B_2$      & $2$               &\; $2^1,\, 3^1,\, 4^1$ \\ 
      && $A_2$      & $4,\, \mathit{(0^-)}$      &\; $4^1$ \\[0.5ex]    
      \hline
      &&&&\\[-1.5ex]
      
      \multirow{4}{*}{$\left[0,n,n\right]$} & \multirow{4}{*}{Dic$_2$ ($\textrm{C}_{2\textrm{v}}$)} 
      & $A_1$     & $0^+,\, 2,\, 4$   &\; $0^1,\, 1^1,\, 2^2,\, 3^2,\, 4^3$ \\
      && $B_1$     & $1,\, 3$          &\; $1^1,\, 2^1,\, 3^2,\, 4^2$ \\
      && $B_2$     & $1,\, 3$          &\; $1^1,\, 2^1,\, 3^2,\, 4^2$ \\
      && $A_2$     & $2,\, 4, \, \mathit{(0^-)}$ &\; $2^1,\, 3^1,\, 4^2$ \\[0.5ex]      
      \hline
      &&&&\\[-1.5ex]
   
      \multirow{3}{*}{$\left[n,n,n\right]$} &       \multirow{3}{*}{Dic$_3$ ($\textrm{C}_{3\textrm{v}}$)}
      & $A_1$     & $0^+,\, 3$        &\;  $0^1,\, 1^1,\, 2^1,\, 3^2,\,  4^2$\\
      && $E_2$     & $1,\, 2,\, 4$     &\;  $1^1,\, 2^2,\, 3^2,\,  4^3$ \\
      && $A_2$     & $3, \, \mathit{(0^-)}$      &\;  $3^1,\, 4^1$\\[0.5ex]
      \hline
\end{tabular}
\end{center}
\caption{The pattern of subductions of pseudoscalar-pseudoscalar partial waves, $\ell \leq 4$, into lattice irreps, $\Lambda$, when the pseudoscalars have unequal mass, e.g.\ $DK$ or $D\bar{K}$ (from Table III of Ref.~\cite{Wilson:2014cna}).
Here $N$ is the number of embeddings of this $\ell$ in the irrep and $n$ is a non-zero integer. LG$(\vec{P})$ is the double-cover little group and the corresponding single-cover little group relevant for only integer spin is given in parentheses.  Also shown are the various $J \leq 4$ or $|\lambda| \leq 4$ that appear in each of the relevant irreps. The $J^P$ values and $|\lambda|^{\tilde{\eta}}=0^-$ in italics are in the ``unnatural parity'' [$P = (-1)^{J+1}$] series and do not contribute to pseudoscalar-pseudoscalar scattering.}
\label{table:subductions}
\end{table}

\subsection{$DK$ $I=0$ with $m_\pi = 239$ MeV}
\label{sec:spectrum:DK_860}

\begin{figure}
\includegraphics[width=1.0\textwidth]{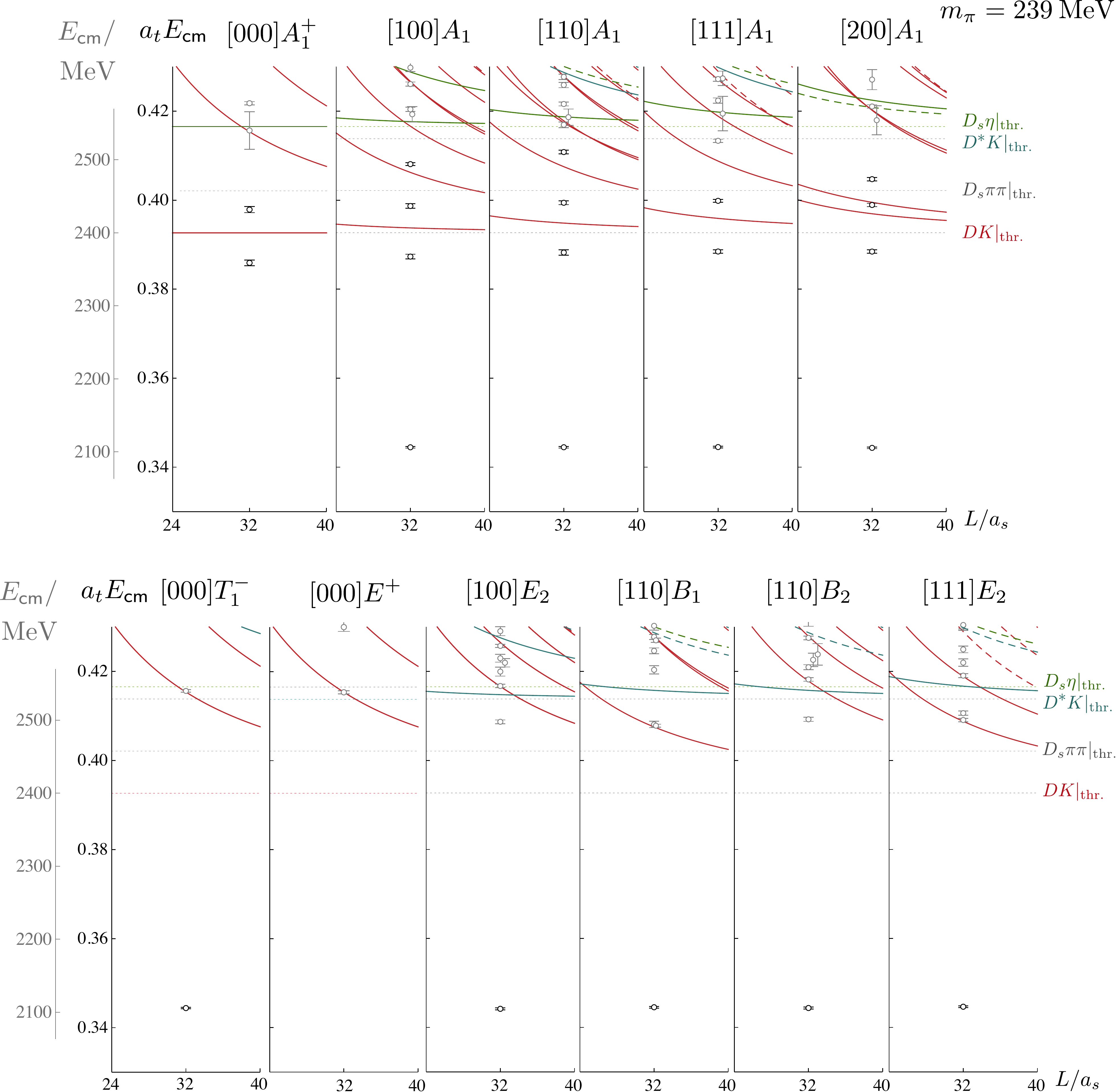}
\caption{$DK$ $I=0$ finite-volume spectra on the $m_\pi = 239$ MeV ensemble labelled by $[\vec{P}] \Lambda^{(P)}$. Points show the extracted finite-volume energy levels transformed to the centre-of-momentum frame; black points are used in the scattering analysis whilst grey points are not. Dotted lines show relevant kinematic thresholds. Solid lines/curves show non-interacting energies with dashed curves indicating non-interacting energies where the corresponding operator was not included in the basis. }
\label{fig:860spectrum:DK}
\end{figure}

The extracted $DK$ $I=0$ finite-volume spectra on the $m_\pi = 239$ MeV ensemble, computed using the operators listed in Table~\ref{tab:ops:860:DK} in Appendix~\ref{app:operators}, are shown in Fig.~\ref{fig:860spectrum:DK} labelled by $[\vec{P}] \Lambda^{(P)}$, where parity, $P$, is a good quantum number for $\vec{P} = \vec{0}$. Along with the computed energies, kinematic thresholds and non-interacting energies are also shown. In the scattering analyses we use the energy levels below $D^\ast K$ threshold in the $A_1^+$ and $A_1$ irreps, and levels below $D_s \pi \pi$ threshold in other irreps\footnote{three pseudoscalars cannot appear in $J^P=0^+$} -- these are shown as black points in the figure. Some representative examples of the underlying principal correlator fits are given in Appendix~\ref{app:prin_corrs}.

We will analyse these spectra to determine scattering amplitudes in Section~\ref{sec:scattering}, but here we make some qualitative observations. The top row of Fig.~\ref{fig:860spectrum:DK} shows irreps where the lowest contributing partial wave is $\ell=0$. The presence of an `extra' energy level in the energy region $a_t E_{\text{cm}} \approx 0.38 - 0.41$ compared to the number expected in the absence of meson-meson interactions, as well as shifts away from the non-interacting energy levels, suggests there are significant $DK$ interactions. This feature could arise from a $J^P = 0^+$ bound state or resonance in $S$-wave $DK$ scattering.

An additional `extra' energy level is seen at $a_t E_{\text{cm}} \approx 0.34$ far below $DK$ threshold in $A_1$ irreps with $\vec{P} \neq \vec{0}$ -- these contain $\ell=1$ contributions, suggesting a deeply bound $1^-$ state in $P$-wave. Further evidence for this can be seen in the irreps in the bottom row of Fig.~\ref{fig:860spectrum:DK} which have $\ell=1$ as the lowest contributing partial wave (all except for $[000]E^+$ where $\ell=2$ is the lowest partial wave), where a similar energy level is found.

It is interesting to note the presence of another `extra' level below $D^\ast K$ threshold in the $[100]E_2$, $[110]B_{1,2}$ and $[111]E_2$ irreps at $a_t E_{\text{cm}} \approx 0.41$. Such an energy level could arise from a bound state or resonance in $D^\ast K$ scattering with $J^P = 1^+$. We only consider elastic $DK$ scattering in this study and an extended coupled-channel analysis would be needed to draw further conclusions on this -- we return to this point in Section \ref{sec:scattering:DK:1p}.

The $[000]E^+$ irrep shown in the bottom row of Fig.~\ref{fig:860spectrum:DK} has $\ell=2$ as the lowest contributing partial wave. The lowest level shows no significant shift away from the associated non-interacting energy, suggesting no significant interactions in this energy region. However, at higher energies we observe `extra' levels which have dominant overlap with $c\bar{s}$ fermion-bilinear constructions subduced from $J^P=2^+$ continuum operators.

\subsection{$DK$ $I=0$ with $m_\pi = 391$ MeV}
\label{sec:spectrum:DK_840}

\begin{figure}[p]
\centering
\includegraphics[width=0.99\textwidth]{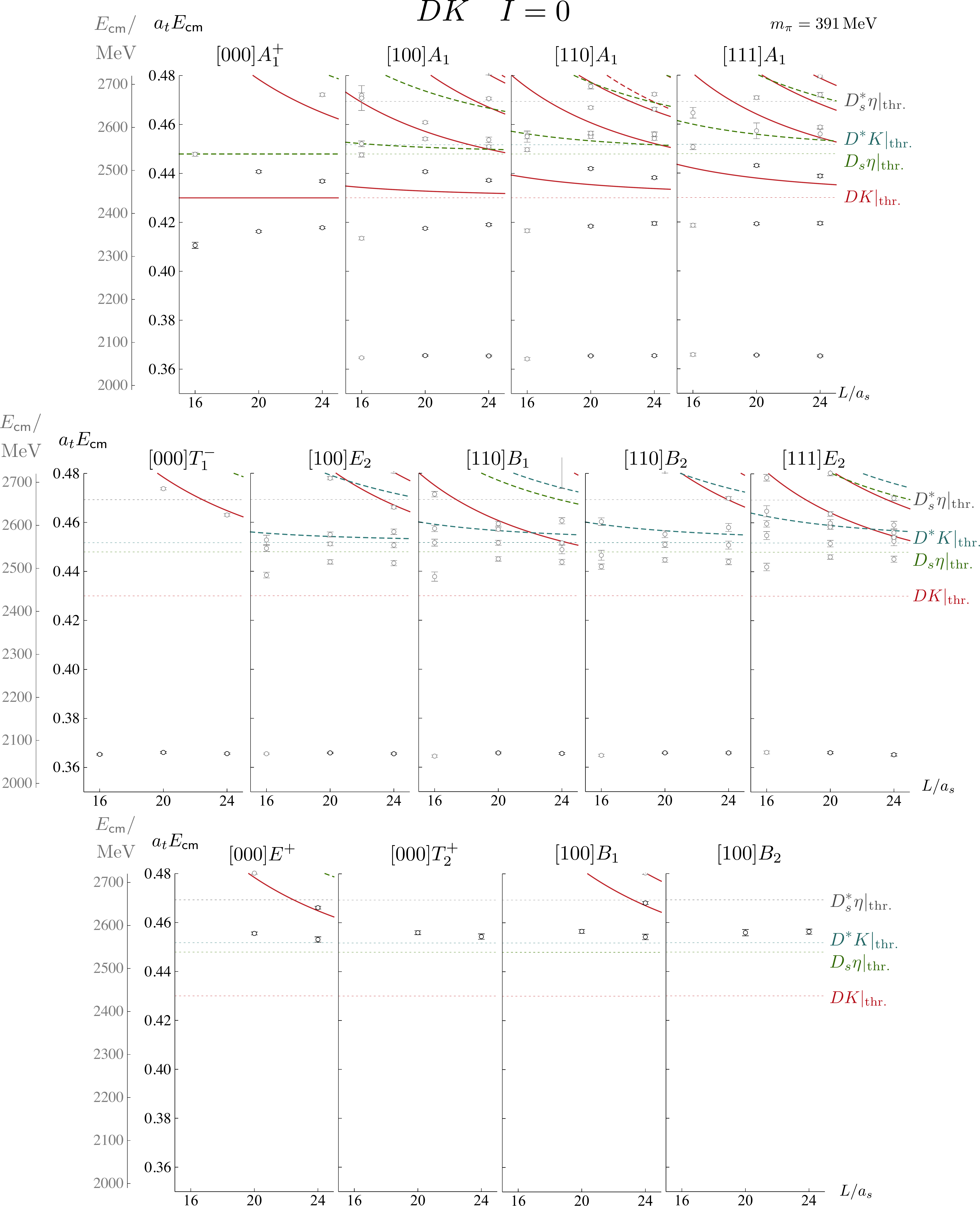}
\caption{As Figure~\ref{fig:860spectrum:DK} but for $I=0$ $DK$ on the $m_\pi = 391$~MeV ensembles.}
\label{fig:840spectrum:DK}
\end{figure}

Finite-volume spectra for $DK$ $I=0$ with $m_\pi = 391$ MeV are presented in Fig.~\ref{fig:840spectrum:DK}, computed using the operators listed in Table~\ref{tab:ops:840:DK}. The energy levels we use in the scattering analyses are shown as black points -- these are the levels well below $D_s \eta$ threshold in the irreps where the lowest contribution is $\ell=0$ or $1$ $DK$, and the low-lying levels in other irreps; on the $16^3$ volume we only use the $[000]A_1^+$ and $[000]T_1^+$ irreps. Some representative examples of the underlying principal correlator fits are again given in Appendix~\ref{app:prin_corrs}.

The spectra show similar qualitative features to the $m_\pi = 239$ MeV spectra in Fig.~\ref{fig:860spectrum:DK}. An `extra' level is observed in the energy region $a_t E_{\text{cm}} \approx 0.41 - 0.45$ in irreps which have contributions from $\ell=0$. A level is found around $a_t E_{\text{cm}} \approx 0.36$ far below $DK$ threshold in irreps which contain contributions from $\ell=1$, suggesting a $1^-$ deeply bound state.

The bottom row of Fig.~\ref{fig:840spectrum:DK} shows irreps which have $\ell=2$ as the lowest contributing partial wave. The presence of an `extra' energy level in the energy region $a_t E_{\text{cm}} \approx 0.45 - 0.46$ and shifts of levels away from non-interacting energies suggests that there is a non-trivial interaction in $D$-wave $DK$.

\subsection{$D\bar{K}$ $I=0,1$ with $m_\pi = 239$ MeV}
\label{sec:spectrum:DKbar_860}

\begin{figure}[p]
\centering
\includegraphics[width=0.99\textwidth]{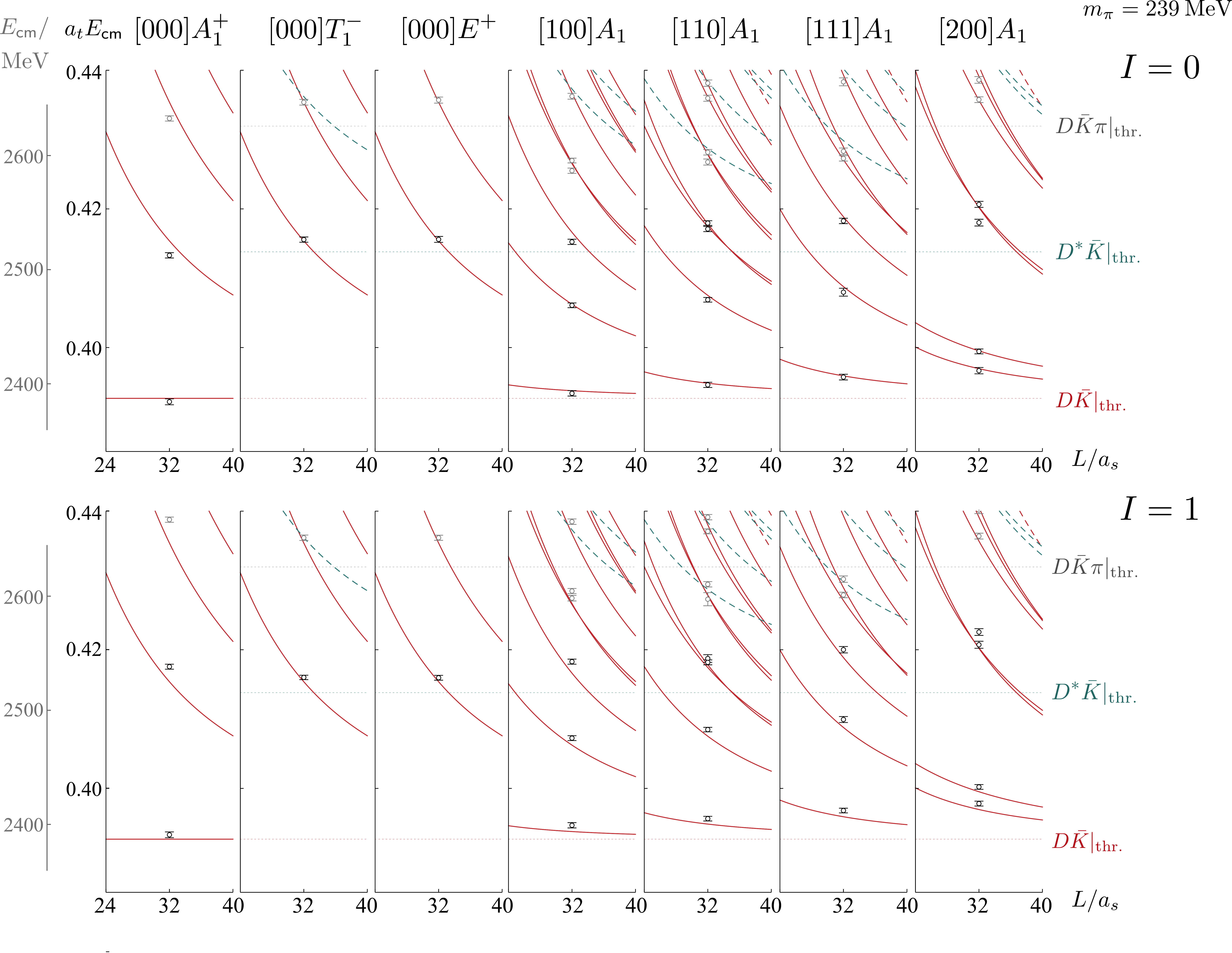}
\caption{As Figure~\ref{fig:860spectrum:DK} but for $I=0$ and $I=1$ $D\bar{K}$ on the $m_\pi = 239$ MeV ensemble.}
\label{fig:860spectrum:DKbar}
\end{figure}

Finite-volume spectra for $D\bar{K}$ with $I=0$ and $I=1$ on the $m_\pi = 239$ MeV ensemble, computed using the operators listed in Table~\ref{tab:ops:860:DKbar}, are shown in Fig.~\ref{fig:860spectrum:DKbar}. These exotic-flavour combinations cannot be interpolated by single-meson (fermion-bilinear) operators, quark-line annihilations are not possible and, as shown in Fig.~\ref{fig:wx}, each Wick diagram contains exactly four quark lines that propagate from source to sink. In the scattering analyses we use the energy levels shown in black in Fig.~\ref{fig:860spectrum:DKbar} -- these levels are well below $D\bar{K}\pi$ threshold and the lowest non-interacting $D^\ast \bar{K}$ energy level. Some representative examples of the underlying principal correlator fits are given in Appendix~\ref{app:prin_corrs}.

The qualitative pattern of energy levels is very different to the case of $I=0$ $DK$. For both isospins, we see no sign of any `extra' levels and the energies show only small shifts away from the non-interacting energies, suggesting weak interactions. However, in the $I=0$ channel there are negative shifts in irreps where $\ell=0$ contributes, suggesting an attractive $S$-wave interaction, with very small shifts in other irreps. In the $I=1$ channel there are small positive shifts in the irreps with $\ell=0$ suggesting weak repulsion.

\subsection{$D\bar{K}$ $I=0,1$ with $m_\pi = 391$ MeV}
\label{sec:spectrum:DKbar_840}

\begin{figure}[p]
\centering
\includegraphics[width=0.99\textwidth]{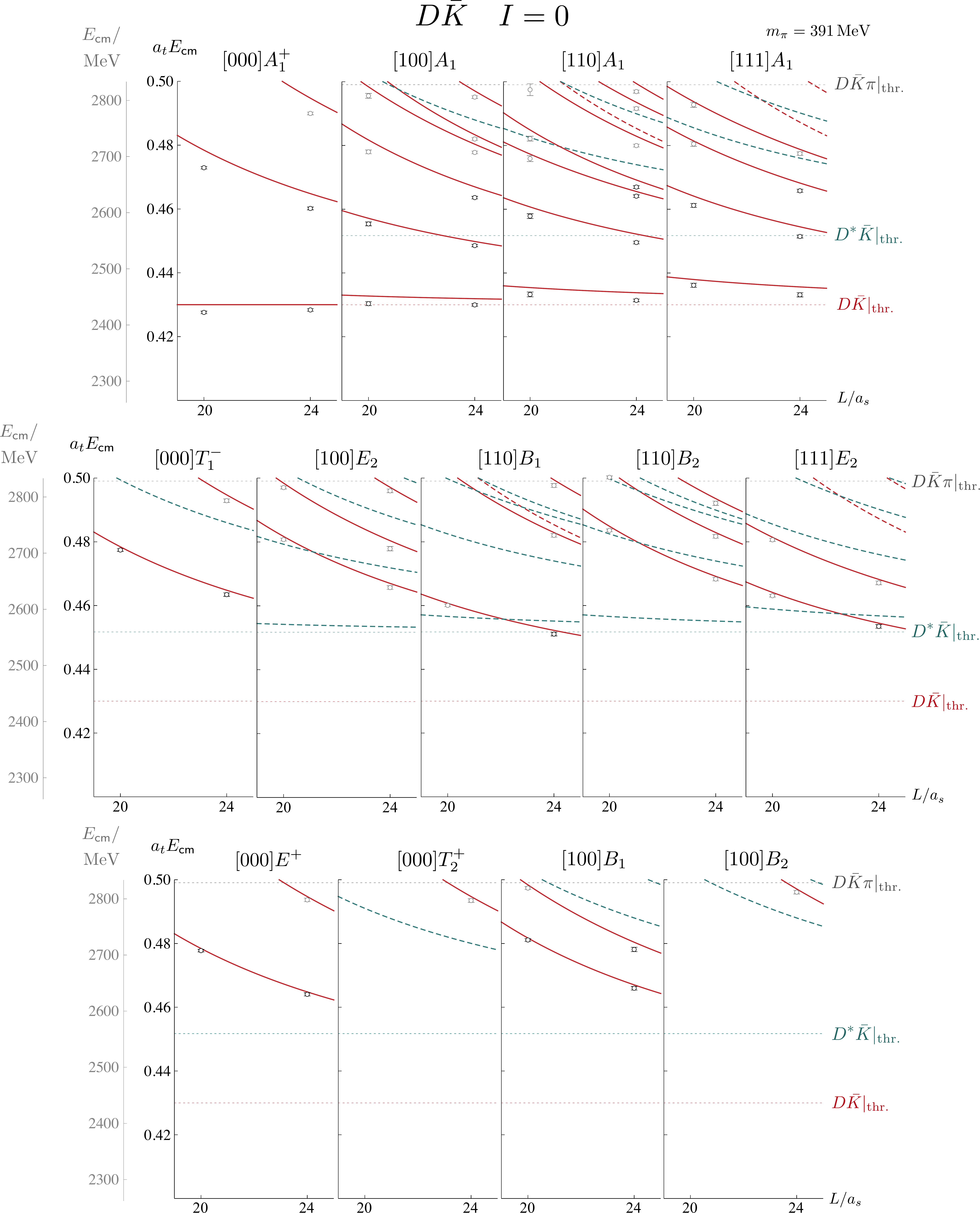}
\caption{As Figure~\ref{fig:860spectrum:DK} but for $I=0$ $D\bar{K}$ on the $m_\pi = 391$ MeV ensembles.}
\label{fig:840spectrum:DKbarI0}
\end{figure}

\begin{figure}[p]
\centering
\includegraphics[width=0.99\textwidth]{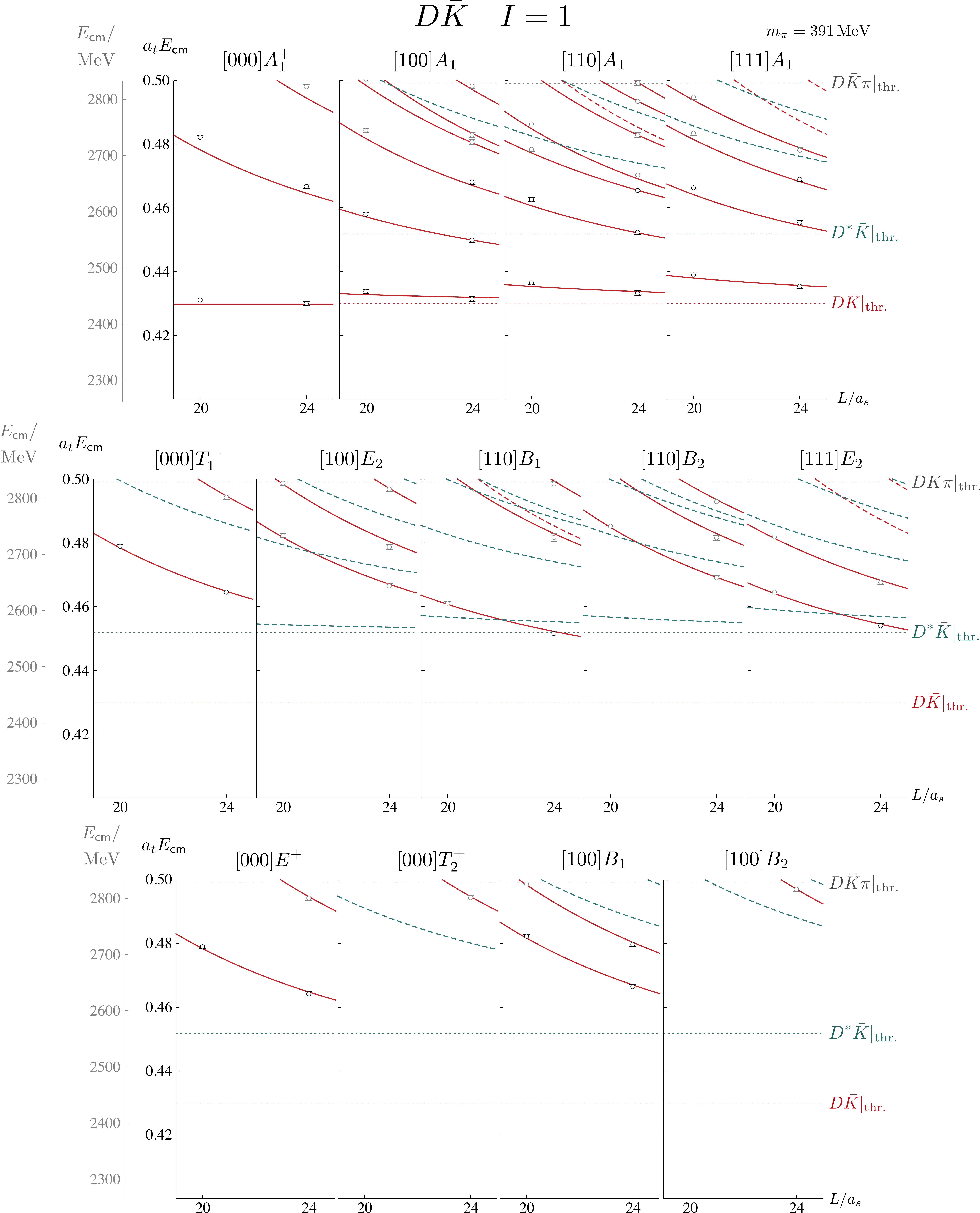}
\caption{As Figure~\ref{fig:860spectrum:DK} but for $I=1$ $D\bar{K}$ on the $m_\pi = 391$ MeV ensembles.}
\label{fig:840spectrum:DKbarI1}
\end{figure}

Finite-volume spectra for $D\bar{K}$ with $I=0$ and $I=1$ on the $m_\pi = 391$ MeV ensemble, computed using the operators listed in Table~\ref{tab:ops:840:DKbar}, are shown in Figs.~\ref{fig:840spectrum:DKbarI0} and \ref{fig:840spectrum:DKbarI1} respectively. These spectra show the same qualitative features as those on the $m_\pi = 239$ MeV ensemble in Fig.~\ref{fig:860spectrum:DKbar}. In the scattering analyses we use the energy levels shown in black in the figures -- these are levels below the lowest non-interacting $D^\ast \bar{K}$ energy level. Some representative examples of the underlying principal correlator fits are again given in Appendix~\ref{app:prin_corrs}.

\clearpage

\section{Scattering amplitudes}
\label{sec:scattering}

Infinite-volume scattering amplitudes are related to finite-volume energy levels through a quantisation condition derived by L\"{u}scher and subsequently extended by many others.  For the case of elastic hadron-hadron scattering\footnote{Ref.~\cite{Hansen:2019nir} reviews recent progress in developing a quantisation condition for three-hadron systems.}, the quantisation condition~\cite{Luscher:1986pf,Luscher:1990ux,Luscher:1991cf,Rummukainen:1995vs,Kim:2005gf,Christ:2005gi,Fu:2011xz,Leskovec:2012gb,Briceno:2014oea} for lattice irrep $\Lambda$ with overall momentum $\vec{P}$ can be written as,
\begin{equation}
\det \left[ \delta_{\ell \ell'} \delta_{n n'} + i \, \rho(E_\cm) \, t^{(\ell)}(E_\text{cm}) \left( \delta_{\ell \ell'} \delta_{n n'} + i \, \mathcal{M}^{\vec{P},\Lambda}_{\ell n; \ell' n'}(E_\cm,L) \right) \right] = 0 \, ,
\label{eqn:det}
\end{equation}
where the determinant is over $\ell$, the partial waves which subduce into $\Lambda$, and $n$, indexing the embeddings of $\ell$ in $\Lambda$; the pattern of subductions is shown in Table~\ref{table:subductions}. Here $\rho(E_\cm) = 2k/E_\cm$ is the phase space, $E_\cm$ and $k$ are the energy and momentum in the centre-of-momentum ($\cm$) frame, and $t^{(\ell)}$ is the infinite-volume $t$-matrix in partial wave $\ell$, which for elastic scattering can be written in terms of a single energy-dependent scattering phase-shift, $\delta_\ell(E_\text{cm})$, as $t^{(\ell)} = \tfrac{1}{\rho} e^{i \delta_\ell } \sin \delta_\ell$. The volume-dependent matrix of known functions of energy, $\mathcal{M}^{\vec{P},\Lambda}_{\ell n; \ell' n'}$, is not diagonal in partial waves, a manifestation of the reduced symmetry of a finite cubic volume. Ref.~\cite{Briceno:2017max} reviews this formalism and its application to the determination of elastic and coupled-channel scattering amplitudes.

In principle, an infinite set of partial waves contributes to the finite-volume spectrum in each irrep and the determinant in Eq.~\ref{eqn:det} is over an infinite-dimensional space. However, the kinematic suppression of higher partial waves close to threshold (in the absence of any dynamical enhancement) enables us to consider only the lower partial waves which contribute to each irrep. If, in a particular energy region, only one partial wave is relevant for a given irrep, Eq.~\ref{eqn:det} gives a one-to-one relation between each energy level, $E_\cm$, and the scattering $t$-matrix (or phase shift) at that energy.

When more than one partial wave is relevant there is no longer a one-to-one mapping between energy levels and phase shifts. We will follow the the approach of Ref.~\cite{Dudek:2012gj} where the energy dependence of the $t$-matrix is parametrised in terms of a small number of parameters and the best fit to the extracted finite-volume spectra is found by varying the parameters, minimising a $\chi^2$ function (Eq.~(9) in Ref.~\cite{Dudek:2012xn}). In order to ensure that results are not biased by a particular choice of parametrisation, we consider a variety of different forms.

One parametrisation of the $t$-matrix for elastic scattering is the effective range expansion,
an expansion about threshold, given by,
\begin{equation}
k^{2\ell+1} \cot \delta_\ell = \frac{1}{a_\ell} + \frac{1}{2} r_\ell k^2 + \mathcal{O}(k^4) \, ,
\label{eqn:effrange}
\end{equation}
where the parameters $a_\ell$ and $r_\ell$ are known as the scattering length and the effective range.
The scattering-length parametrisation corresponds to setting $r_\ell = 0$.
Another form, commonly used to parametrise a narrow resonance, is the relativistic Breit-Wigner,
\begin{equation}
t^{(\ell)}(s) = \frac{1}{\rho(s)} \frac{\sqrt{s} \Gamma_\ell(s)}{m_R^2 - s - i \sqrt{s} \Gamma_\ell(s)} \,
\label{eqn:BW}
\end{equation}
where $s = E^2_{\text{cm}}$, the width, $\displaystyle \Gamma_\ell(s) = \frac{g_R^2}{6 \pi}\frac{k^{2\ell+1}}{s \, m_R^{2(\ell-1)}}$, ensures the correct threshold behaviour, and the parameters are the Breit-Wigner mass, $m_R$, and the coupling constant, $g_R$.

A more general approach is to write the $t$-matrix in terms of a real symmetric $K$-matrix,
\begin{equation}
t^{-1}(s) = \frac{1}{(2k)^\ell} K^{-1}(s) \frac{1}{(2k)^\ell} + I(s) \, ,
\end{equation}
where for brevity we have dropped the partial-wave label $\ell$. Unitarity of the $S$-matrix is guaranteed if $\text{Im}[I(s)] = -\rho(s)$ above threshold and zero below. There is some remaining freedom in choosing $I$ and one of the simplest choices is $I(s) = -i \rho(s)$. Another choice is the Chew-Mandelstam prescription~\cite{Chew:1960iv} that defines $\text{Re}[I(s)]$ in terms of a dispersive integral and has a better analytic structure and a smoother transition across threshold; our implementation is described in Ref.~\cite{Wilson:2014cna}. This is a particularly convenient class of parametrisations for inelastic scattering, where $K$ is a matrix, but can equally be used to give a variety of forms for elastic scattering where $K$ is a real function.

We parametrise $K$ in terms of a pole and a polynomial,
\begin{equation}
K(s) = \left( \sum^{N_g}_{n=0} g^{(n)} s^n \right)^2\frac{1}{m^2 -s} + \sum^{N_\gamma}_{n=0} \gamma^{(n)} s^n \, ,
\end{equation} 
where $g^{(n)}$, $\gamma^{(n)}$ and $m$ are real parameters, or $K^{-1}$ in terms of a polynomial,
\begin{equation}
[K(s)]^{-1} = \sum^{N_c}_{n=0} c^{(n)} s^n \, ,
\end{equation}
where $c^{(n)}$ are real parameters. When we use the Chew-Mandelstam prescription, we will subtract at the $K$-matrix pole at $s=m^2$ when present or at threshold when there is no $K$-matrix pole term.

We now determine scattering amplitudes on the $m_\pi = 239$ and $m_\pi = 391$ MeV ensembles using the extracted finite-volume energy levels described above. Our discussion begins with $I=0$ $DK$ scattering before moving on to the flavour-exotic $I=0$ and $1$ $D\bar{K}$-channels. In Section~\ref{sec:interpretation} we then analytically continue these amplitudes into the complex energy plane to find the location of pole singularities and interpret the results.

\subsection{$DK$ scattering in $I=0$}
\label{sec:scattering:DK}

Starting with the $m_\pi = 239$~MeV ensemble, we analyse the 22 energy levels shown in black in Fig.~\ref{fig:860spectrum:DK} -- these are below $D^\ast K$ threshold in the $A_1^+$ and $A_1$ irreps where the lowest contribution is from $\ell=0$, and below $D_s \pi \pi$ threshold in other irreps where the lowest contribution is from $\ell=1$. For now we neglect partial waves with $\ell \geq 2$, but we return to this point in Section~\ref{sec:scattering:DK:Dwave}. A good fit to the spectra is obtained using a $K$-matrix parametrisation with a pole in $\ell=0$ and a pole plus a constant term in $\ell=1$, with Chew-Mandelstam phase space,
\begin{center}
\begin{tabular}{rll}   
$m_0 \;\; =$ & $(0.38868 \pm 0.00029 \, ^{+ \, 0.00004}_{- \, 0.00003} ) \cdot a_t^{-1}$ & 
\multirow{5}{*}{ $\begin{bmatrix}
1  & 0.61 &  0.62 & -0.17 & -0.50 \\
   & 1    &  0.40 & -0.15 & -0.48 \\
   &      &  1    & -0.40 & -0.59 \\
   &      &       &  1    &  0.79 \\
   &      &       &       &  1    \\
\end{bmatrix}$ } \\
$g^{(0)}_0 \;\;       =$ & $(0.629 \pm 0.053 \, ^{+ \, 0.009}_{- \, 0.008}) \cdot a_t^{-1} $ & \\
$m_1 \;\;            =$ & $(0.34432 \pm 0.00014 \pm 0.00002) \cdot a_t^{-1} $ & \\
$g^{(0)}_1 \;\;       =$ & $2.94 \pm 0.37 \, ^{+ \, 0.03}_{- \, 0.21}$ & \\
$\gamma^{(0)}_1  \;\; =$ & $(119 \pm 69 \, ^{+ \, 10}_{- \, 23}) \cdot a_t^2 $ & \\
&\multicolumn{2}{l}{ $\chi^2 / \Ndf = \frac{15.1}{22 - 5} = 0.89,$}  \\
\end{tabular}
\end{center}
\vspace{-1.27cm}
\begin{equation}\label{eqn:860DK:ref}\end{equation}\\
where the subscript on the parameters labels the partial wave, $\ell$, the first uncertainty is statistical, and the second is an envelope over the uncertainties from varying the $D$ and $K$ meson masses and the anisotropy within their uncertainties.\footnote{To be precise, we vary in turn each of $m_D$, $m_K$ and $\xi$ by $\pm \sigma$ using the uncertainties given in Section~\ref{sec:lattice}. The second uncertainty quoted on each parameter $x$ is $+ \text{max}_i  \left[ (\bar{x}_i + \sigma_{x_i}) - (\bar{x}_0 + \sigma_{x_0}) \right]$ and $- \text{min}_i  \left[ (\bar{x}_i - \sigma_{x_i}) - (\bar{x}_0 - \sigma_{x_0}) \right]$, where $i$ indexes the variations and $0$ corresponds to using the mean values for the meson masses and anisotropy.}  The matrix on the right gives the correlations between the parameters.

Similarly, on the $m_\pi = 391$~MeV ensemble we analyse the 34 energy levels in irreps where the lowest contribution is from $\ell=0$ or $1$, the black points in the top two rows of Fig.~\ref{fig:840spectrum:DK}. A reasonable fit to the data is again obtained using a $K$-matrix parametrisation with a pole in $\ell=0$ and a pole plus a constant term in $\ell=1$, with Chew-Mandelstam phase space,
\begin{center}
\begin{tabular}{rll}   
$m_0 \;\; =$ & $(0.41981 \pm 0.00029 \pmerr{0.00001}{0.00002}) \cdot a_t^{-1}$ & 
\multirow{5}{*}{ $\begin{bmatrix}
1  & 0.47 &  0.71 & -0.35 & -0.52 \\
   & 1    &  0.33 & -0.19 & -0.14 \\
   &      &  1    & -0.65 & -0.71 \\
   &      &       &  1    &  0.87 \\
   &      &       &       &  1    \\
\end{bmatrix}$ } \\
$g^{(0)}_0        \;\; =$ & $(0.536 \pm 0.019 \pmerr{0.022}{0.006}) \cdot a_t^{-1}$ & \\
$m_1      \;\; =$ & $(0.36564 \pm 0.00019 \pmerr{0.00002}{0.00001})  \cdot a_t^{-1}$   & \\
$g^{(0)}_1      \;\; =$ & $0.72 \pm 0.82 \pmerr{0.03}{0.07}$  & \\
$\gamma^{(0)}_1 \;\; =$ & $(58 \pm 25 \pmerr{8}{7}) \cdot a_t^2 $  & \\
&\multicolumn{2}{l}{ $\chi^2 / \Ndf = \frac{45.4}{34 - 5} = 1.56.$}  \\
\end{tabular}
\end{center}
\vspace{-1cm}
\begin{equation}\label{eqn:840DK:ref}\end{equation}\\

\begin{figure}[tb]
\centering
\includegraphics[width=0.99\textwidth]{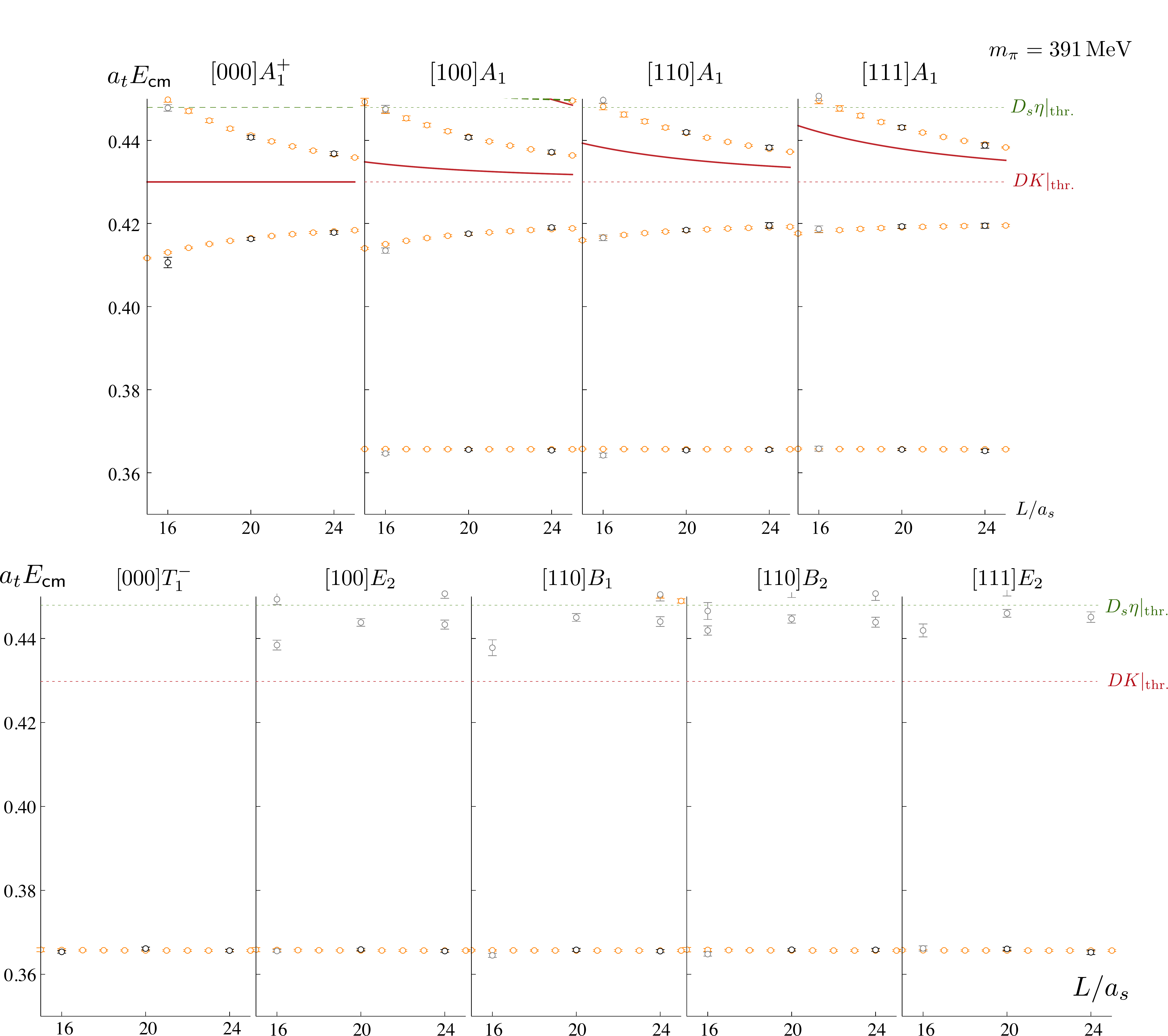} 
\caption{$DK$ $I=0$ finite-volume spectra on the $m_\pi = 391$~MeV ensembles, as in the top two rows of Fig.~\ref{fig:840spectrum:DK}, with the addition of orange points which show the energy levels from the parametrisation in Eq.~(\ref{eqn:840DK:ref}). }
\label{fig:840DK:paramspectrum}
\end{figure}

In Fig.~\ref{fig:840DK:paramspectrum} we show a comparison between the computed finite-volume energy levels and the energy levels which follow from Eq.~(\ref{eqn:840DK:ref}) using the quantisation condition, Eq.~(\ref{eqn:det}).  This demonstrates the good description of the data provided by the parametrisation, reflecting the reasonable $\chi^2 / \Ndf$ of the fit. It also shows the good description of computed energy levels below $D_s \eta$ threshold which were not used in the fit (grey points), with the exception of those above $DK$ threshold in the lower row which are expected to be $J^P = 1^+$ (and/or higher $J$) and so not described by $S$ and $P$-wave $DK$ amplitudes.

\hfill

\begin{figure}[tb]
\centering
\includegraphics[width=0.99\textwidth]{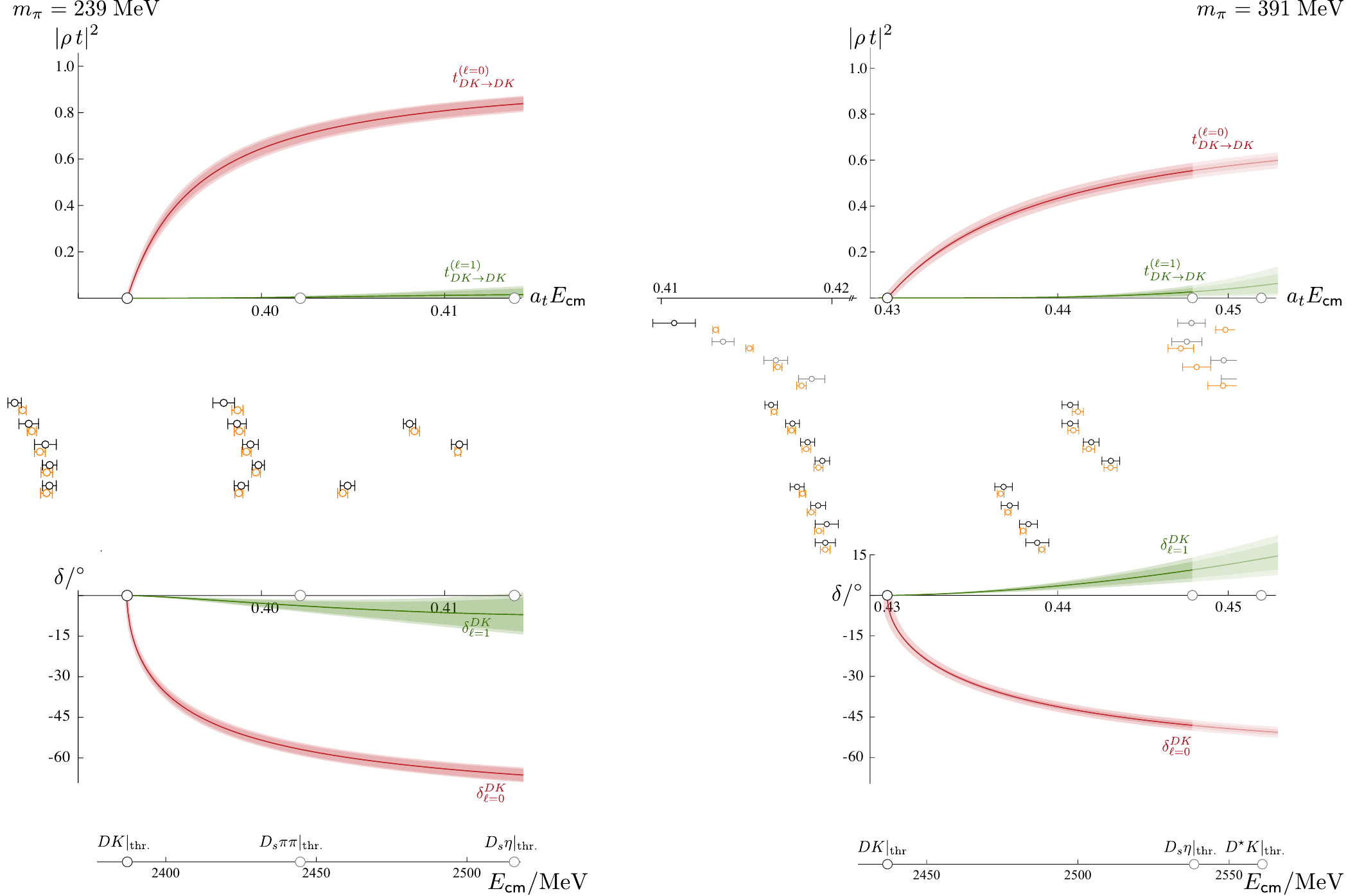} 
\caption{$S$ and $P$-wave squared-amplitudes (upper) and scattering phase shifts (lower) for $I=0$ $DK$ on the $m_\pi = 239$~MeV (left) and $m_\pi = 391$~MeV (right) ensembles. The darker inner bands are the reference parametrisations given in Eqs.~(\ref{eqn:860DK:ref}) and (\ref{eqn:840DK:ref}) with statistical uncertainties, while the lighter outer bands reflect the uncertainty from varying the $D$ and $K$ meson masses and the anisotropy as described in the text.  The black points between the plots show the computed finite-volume energy levels from Section~\ref{sec:spectrum} used to constrain the amplitudes, grey points are finite-volume energy levels which were not used, and orange points show the energy levels following from the reference parametrisations.}
\label{fig:DK:ref_amps}
\end{figure}

The amplitudes for $m_\pi = 239$~MeV and $m_\pi = 391$~MeV presented in Eqs.~(\ref{eqn:860DK:ref}) and (\ref{eqn:840DK:ref}), which we refer to as \emph{reference parametrisations}, are shown in Fig.~\ref{fig:DK:ref_amps}. The figure also shows the good agreement between the computed finite-volume energy levels and the energy levels from these parametrisations.
It can be seen that, while there is a significant amplitude for scattering in $S$ wave, the $P$-wave amplitude is close to zero for physical scattering energies. There is an energy level very far below threshold in irreps where the lowest contribution is from $P$-wave scattering with an energy that shows very little dependence on the volume or irrep, suggestive of a $J^P=1^-$ bound state. Because this bound state is far below threshold, it does not appear to influence the physical $DK$ scattering region strongly.

\hfill

A convenient alternative parametrisation for elastic scattering is the effective range expansion, Eq.~(\ref{eqn:effrange}). Using such a parametrisation in $S$-wave and a $K$-matrix with a pole plus a constant term in $P$-wave (as in the reference parameterisations) gives a good description of the finite-volume spectra for $m_\pi = 239$~MeV,
\begin{center}
\begin{tabular}{rll}   
$a_0 \;\; =$ & $(-43.6 \pm 2.0 \pm 0.5) \cdot a_t$ &
\multirow{5}{*}{ $\begin{bmatrix}
 & 1  & -0.77 &  -0.61 & 0.18 &  0.54 \\
 &    &  1    &  0.41 & -0.16 & -0.49 \\
 &    &       &  1    & -0.40 & -0.59 \\
 &    &       &       &  1    &  0.79 \\
 &    &       &       &       &  1    \\
\end{bmatrix}$ } \\
$r_0 \;\;            =$ & $(-0.79 \pm 1.04 \pmerr{0.14}{0.20}) \cdot a_t$   & \\
$m_1 \;\;            =$ & $(0.34432 \pm 0.00014 \pm 0.00002) \cdot a_t^{-1} $ & \\
$g^{(0)}_1 \;\;       =$ & $2.94 \pm 0.34 \pmerr{0.05}{0.16}$ & \\
$\gamma^{(0)}_1  \;\; =$ & $(119 \pm 69 \pmerr{10}{23}) \cdot a_t^2 $ & \\
&\multicolumn{2}{l}{ $\chi^2 / \Ndf = \frac{15.1}{22 - 5} = 0.89.$}  \\
\end{tabular}
\end{center}
\begin{equation}\label{eqn:860DK:eff_range}\end{equation}\\
The same parameterisation gives a reasonable description of the $m_\pi = 391$~MeV spectra,
\begin{center}
\begin{tabular}{rll}   
$a_0 \;\; =$ & $(-24.01 \pm 0.53 \pmerr{0.48}{0.52}) \cdot a_t$ & 
\multirow{5}{*}{ $\begin{bmatrix}
 & 1  & -0.82 & -0.64 &  0.33 &  0.42 \\
 &    &  1    &  0.35 & -0.19 & -0.15 \\
 &    &       &  1    & -0.65 & -0.71 \\
 &    &       &       &  1    &  0.87 \\
 &    &       &       &       &  1    \\
\end{bmatrix}$ } \\
$r_0           \;\; =$ & $(-4.26 \pm 0.62 \pmerr{0.64}{0.28}) \cdot a_t$   & \\
$m_1           \;\; =$ & $(0.36564 \pm 0.00019 \pmerr{0.00002}{0.00001}) \cdot a_t^{-1}$   & \\
$g^{(0)}_1      \;\; =$ & $0.72 \pm 0.82 \pmerr{0.03}{0.07}$  & \\
$\gamma^{(0)}_1 \;\; =$ & $(58 \pm 25 \pmerr{8}{7}) \cdot a_t^2 $  & \\
&\multicolumn{2}{l}{ $\chi^2 / \Ndf = \frac{45.6}{34 - 5} = 1.57.$}  \\
\end{tabular}
\end{center}
\vspace{-1cm}
\begin{equation}\label{eqn:840DK:eff_range}\end{equation}\\

\subsubsection{Varying the parametrisations}

\begin{figure}[tb]
\centering
\includegraphics[width=0.99\textwidth]{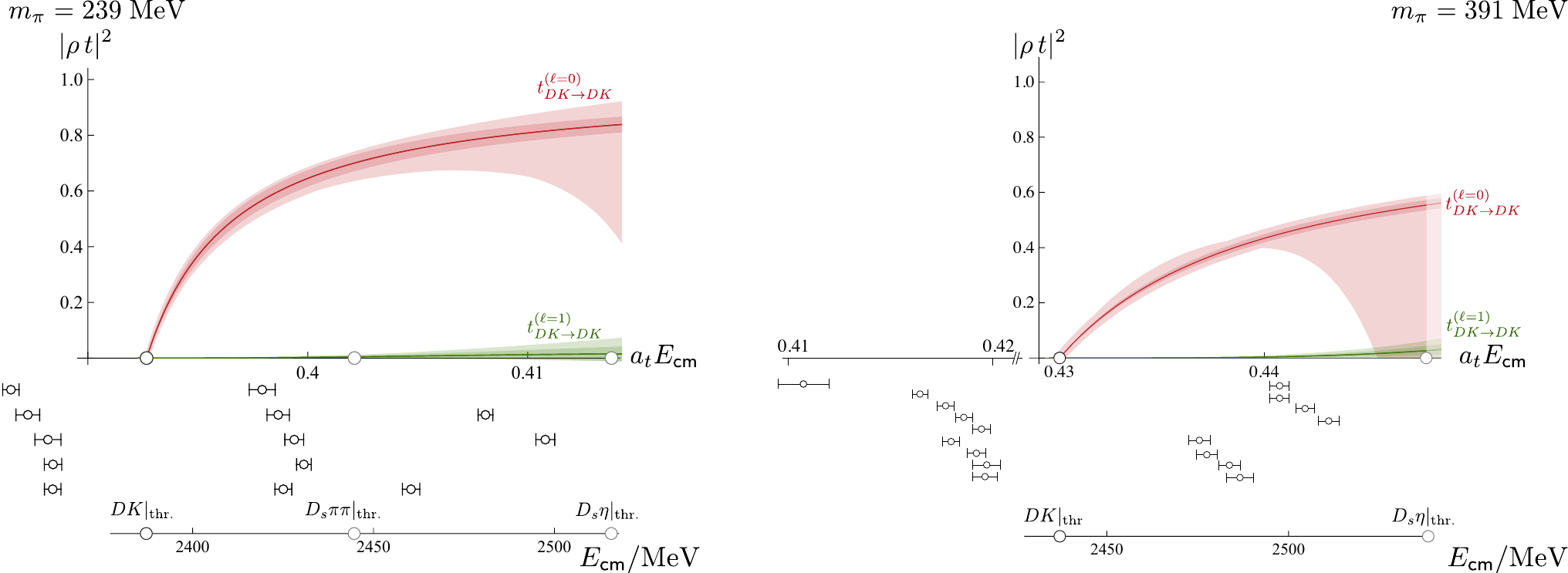} 
\caption{As in the upper plots in Fig.~\ref{fig:DK:ref_amps}, but the lighter outer bands now also include an envelope over the acceptable parameterisations in Table~\ref{tab:DK:variations} including their statistical uncertainties.
The black points at the bottom show the computed finite-volume energy levels from Section~\ref{sec:spectrum} used to constrain the amplitudes.}
\label{fig:DK:varying_amps}
\end{figure}

The parametrisations used in Eq.~(\ref{eqn:det}) were varied to investigate how much dependence there is on the exact form chosen. The $K$-matrix type, in particular, can be used with a wide variety of different functional forms -- we considered higher order polynomials, forms with and without explicit poles, and forms with and without Chew-Mandelstam phase-space. The amplitudes used and associated $\chi^2$'s are summarised in Table~\ref{tab:DK:variations} -- those parametrisations with a $\chi^2/\Ndf$ in italics are not considered an acceptable description of the data for the reason given in the table. We reject parametrisations which have a $\chi^2/\Ndf > 1.6$, an additional level near the top of the fitted region which is not supported by the lattice QCD spectra, poles on the physical sheet located away from the real axis, or an additional finite-volume energy level below threshold. 
The scattering amplitudes from acceptable parametrisations are shown in Fig.~\ref{fig:DK:varying_amps}. It can be seen that there is relatively little dependence on the particular form used in the energy region where we are constraining the amplitudes, but, as might be expected, there is more variation where we do not have constraints.

\begin{figure}[tb]
\centering
\includegraphics[width=0.99\textwidth]{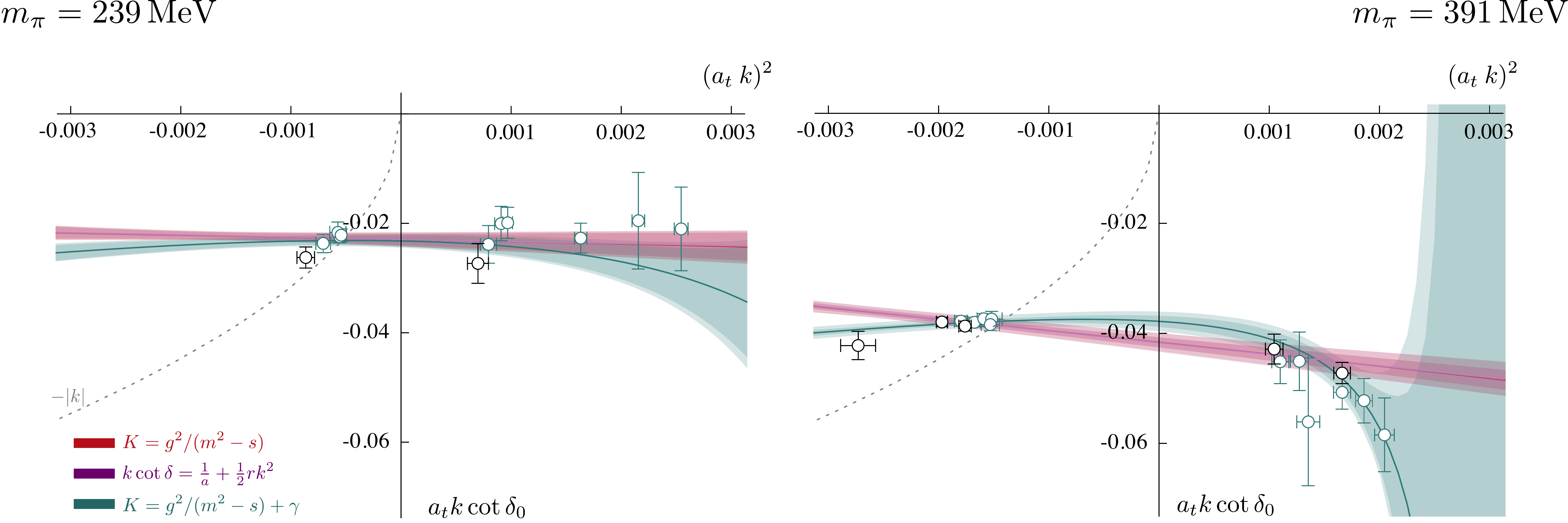}
\caption{$DK$ $I=0$ $S$-wave scattering amplitudes, plotted as $a_t k \cot \delta_{\ell=0}$, using the reference (aa), effective range (x) and pole plus constant $K$-matrix (a) parametrisations for $m_\pi = 239$~MeV (left) and $m_\pi = 391$~MeV (right). Black points are determined from energy levels in the $[000]A_1^+$ irrep assuming $\ell > 4$ amplitudes are negligible. Blue-green points are from the non-zero momentum $A_1$ irreps with amplitudes for higher partial waves fixed to zero.}
\label{fig:DK:kcotdelta}
\end{figure}

In Fig.~\ref{fig:DK:kcotdelta} we present the $S$-wave scattering amplitudes as $a_t k \cot \delta_{\ell=0}$ for the reference amplitudes, the effective range parameterisations described above and a $K$-matrix pole plus constant form (parametrisations (aa), (x) and (a) in Table~\ref{tab:DK:variations}). This again shows that the amplitudes are relatively insensitive to the parametrisation chosen, though there is slightly more freedom in the shape of the amplitude for $m_\pi = 391$~MeV around threshold, between the two clusters of constraints. Also plotted is $-|k|$ for $k^2 < 0$ -- the intersection of this curve with $a_t k \cot \delta_{\ell=0}$ indicates the presence of a pole singularity in the scattering amplitude below $DK$ threshold on the real axis of the physical sheet. We examine the pole singularities of the amplitudes in more detail and discuss their interpretation in Section~\ref{sec:interpretation}, but first we investigate a couple of partial waves which we have so far neglected.

\begin{table}[tb]
\vspace{-1cm}
\begin{center}
\begin{tabular}{lcccc}
Parametrisation & $N_\mathrm{pars}^{(\ell=0)}$ & $N_\mathrm{pars}^{(\ell=1)}$ & ${m_\pi = 239\:\mathrm{MeV}}$ & ${m_\pi = 391\:\mathrm{MeV}}$ \\
 & & & $\chi^2/\Ndf$ & $\chi^2/\Ndf$ \\
 & & & {\small 22 levels} & {\small 34 levels} \\
\hline
\multicolumn{4}{l}{K-matrix with Chew-Mandelstam $I(s)$, $K_1=\frac{g_1^2}{m_1^2-s}+\gamma_1$,}\\
(aa)\quad $K=\frac{g^2}{m^2-s}$                             & 2 & 3 & \bf{0.89} & \bf{1.56} \\
(a) \quad $K=\frac{g^2}{m^2-s}+\gamma^{(0)}$                 & 3 & 3 & 0.81 & 1.11 \\
(b) \quad $K=\frac{g^2}{m^2-s}+\gamma^{(1)} s$               & 3 & 3 & 0.83 & 1.17 \\
(c) \quad $K=\frac{(g^{(0)}+g^{(1)}s)^2}{m^2-s}$              & 3 & 3 & {\it0.81$\;^\ddagger$} & {\it1.08$\;^\ddagger$} \\
(d) \quad $K^{-1}=c_0+c_1\hat{s}$                            & 2 & 3 & 0.89 & 1.56 \\
(e) \quad $K=(1+d_1\hat{s})/(c_0+c_1\hat{s})$                & 3 & 3 & 0.81 & 1.11 \\
[0.3ex]
\hline
\multicolumn{4}{l}{K-matrix with $I(s)=-i\rho(s)$ \& $K_1=\frac{g_1^2}{m_1^2-s}+\gamma_1$}\\
(f) \quad $K=\frac{g^2}{m^2-s}$                              & 2 & 3 & 0.95 & 1.58 \\
(g) \quad $K=\frac{g^2}{m^2-s}+\gamma^{(0)}$                 & 3 & 3 & 0.91 & {\it1.36$\;^\dagger$} \\ 
(h) \quad $K=\frac{g^2}{m^2-s}+\gamma^{(1)} s$               & 3 & 3 & {\it0.89$\;^\dagger$} & {\it1.45$\;^\dagger$} \\
(i) \quad $K=\frac{(g^{(0)}+g^{(1)}s)^2}{m^2-s}$             & 3 & 3 & {\it0.88$\;^\ddagger$} & {\it1.32$\;^\ddagger$} \\
(j) \quad $K^{-1}=c_0+c_1\hat{s}$                            & 3 & 3 & 0.94 & 1.58 \\
(k) \quad $K=(1+d_0\hat{s})/(c_0+c_1\hat{s})$                & 3 & 3 & {\it0.90$\;^\ddagger$} & {\it1.36$\;^\dagger$} \\
[0.3ex]
\hline
\multicolumn{4}{l}{K-matrix with Chew-Mandelstam $I(s)$ \& $K_1=\frac{g_1^2}{m_1^2-s}$}\\
(l) \quad $K=\frac{g^2}{m^2-s}$                              & 2 & 2 & 0.93 & {\it2.03$\;^\diamond$} \\
(m) \quad $K=\frac{g^2}{m^2-s}+\gamma^{(0)}$                 & 3 & 2 & 0.84 & {\it1.09$\;^\dagger$} \\
(n) \quad $K=\frac{g^2}{m^2-s}+\gamma^{(1)} s$               & 3 & 2 & 0.85 & 1.52 \\
(o) \quad $K=\frac{(g^{(0)}+g^{(1)}s)^2}{m^2-s}$              & 3 & 2 & {\it0.83$\;^\ddagger$} & {\it1.06$\;^\ddagger$} \\
(p) \quad $K^{-1}=c_0+c_1\hat{s}$                            & 3 & 2 & 0.93 & {\it2.03$\;^\diamond$} \\
(q) \quad $K=(1+d_0\hat{s})/(c_0+c_1\hat{s})$                & 3 & 2 & 0.83 & {\it1.09$\;^\dagger$} \\
[0.3ex]
\hline
\multicolumn{4}{l}{Effective range expansion in $\ell=0$ \& $\ell=1$, $k^{2\ell+1} \cot\delta_\ell = $} \\
(r) \quad $\frac{1}{a_0}$, \quad $\frac{1}{a_1}$                      & 1 & 1 & {\it3.37$\;^\diamond$} & {\it3.72$\;^\diamond$} \\
(s) \quad $\frac{1}{a_0}+\frac{1}{2}r_0 k^2$, \quad $\frac{1}{a_1}$  & 2 & 1 & {\it2.36$\;^\diamond$} & {\it1.93$\;^\diamond$} \\
(t) \quad $\frac{1}{a_0}$, \quad $\frac{1}{a_1}+\frac{1}{2}r_1 k^2$  & 1 & 2 & 1.50 & {\it3.69$\;^\diamond$} \\
(u) \quad $\frac{1}{a_0}+\frac{1}{2}r_0 k^2$, \quad $\frac{1}{a_1}+\frac{1}{2}r_1 k^2$           & 2 & 2 & 1.33 & {\it1.68$\;^\diamond$} \\
(v) \quad $\frac{1}{a_0}+\frac{1}{2}r_0 k^2 + P_2 k^4$, \quad $\frac{1}{a_1}+\frac{1}{2}r_1 k^2$ & 3 & 2 & {\it0.92$\;^{\ddagger}$} & {\it1.18$\;^{\ddagger}$} \\
(w) \quad $\frac{1}{a_0}+\frac{1}{2}r_0 k^2$, \quad $\frac{1}{a_1}+\frac{1}{2}r_1 k^2 + P_2 k^4$ & 2 & 3 & 1.40 & {\it1.61$\;^\diamond$} \\
[0.3ex]
\hline
\multicolumn{5}{l}{Effective range expansion in $\ell=0$ \& $K_1=\frac{g_1^2}{m_1^2-s}+\gamma_1$ with Chew-Mandelstam $I(s)$}\\
(x) \quad $k\cot\delta_0 = \frac{1}{a}+\frac{1}{2}r k^2$               & 2 & 3 & {\textbf{0.89}} & \textbf{1.57} \\ 
(y) \quad $k\cot\delta_0 = \frac{1}{a}+\frac{1}{2}r k^2 + P_2 k^4$     & 3 & 3 & 0.81 & {\it1.13$\;^{\ddagger}$} \\ 
[0.3ex]
\hline
\multicolumn{5}{l}{Breit-Wigner in each partial waves}\\
(z) \quad \quad $t=\frac{1}{\rho}\frac{m\Gamma}{m^2-s-im\Gamma}$ & 2 & 2 & 1.28 & {\it2.09$\;^\diamond$} \\
\hline
\end{tabular}
{
\flushleft
\small
$\diamond$ -- rejected due to high $\chi^2/\Ndf > 1.6$ \\
$\dagger$ -- rejected due to an additional level near top of the fit region that is not supported by the data \\
$\ddagger$ -- rejected due to physical-sheet complex poles or additional finite-volume level below threshold \\
}
\caption{A selection of the $S$ and $P$-wave parametrisations used for elastic $I=0$ $DK$ scattering. $N^{(\ell)}_\mathrm{pars}$ indicates the number of free parameters for partial wave $\ell$.
The $\chi^2/\Ndf$ in bold indicate the reference and effective range fits described in the text.
Parametrisations with $\chi^2/\Ndf$ in italics are not considered an acceptable description of the data for the reason given in the table.}
\label{tab:DK:variations}
\end{center}
\end{table}

\clearpage

\subsubsection{$D$-wave scattering}
\label{sec:scattering:DK:Dwave}

Considering irreps where the lowest contribution is from $\ell=2$, on the ensemble with the lightest pion mass ($m_\pi = 239$~MeV) there are no levels in the energy range where we can rigorously extract energies. For example, the lowest energy in the $[000]E^+$ irrep is above $D_s\pi\pi$ threshold as shown in Fig.~\ref{fig:860spectrum:DK}. Nevertheless, the fact that this level is consistent with the non-interacting energy suggests that there is no significant $D$-wave $DK$ interaction below $D_s\eta$ threshold. Neglecting potential coupling to inelastic channels, that level corresponds to a scattering phase shift, $\delta_2 = 0.51(33)^\circ$. The results of Ref.~\cite{Cheung:2016bym} suggest that the lightest $2^+$ tensor resonance is at $a_t E \sim 0.43$, above the energy region we are considering.

\begin{figure}[tb]
\centering
\includegraphics[width=0.8\textwidth]{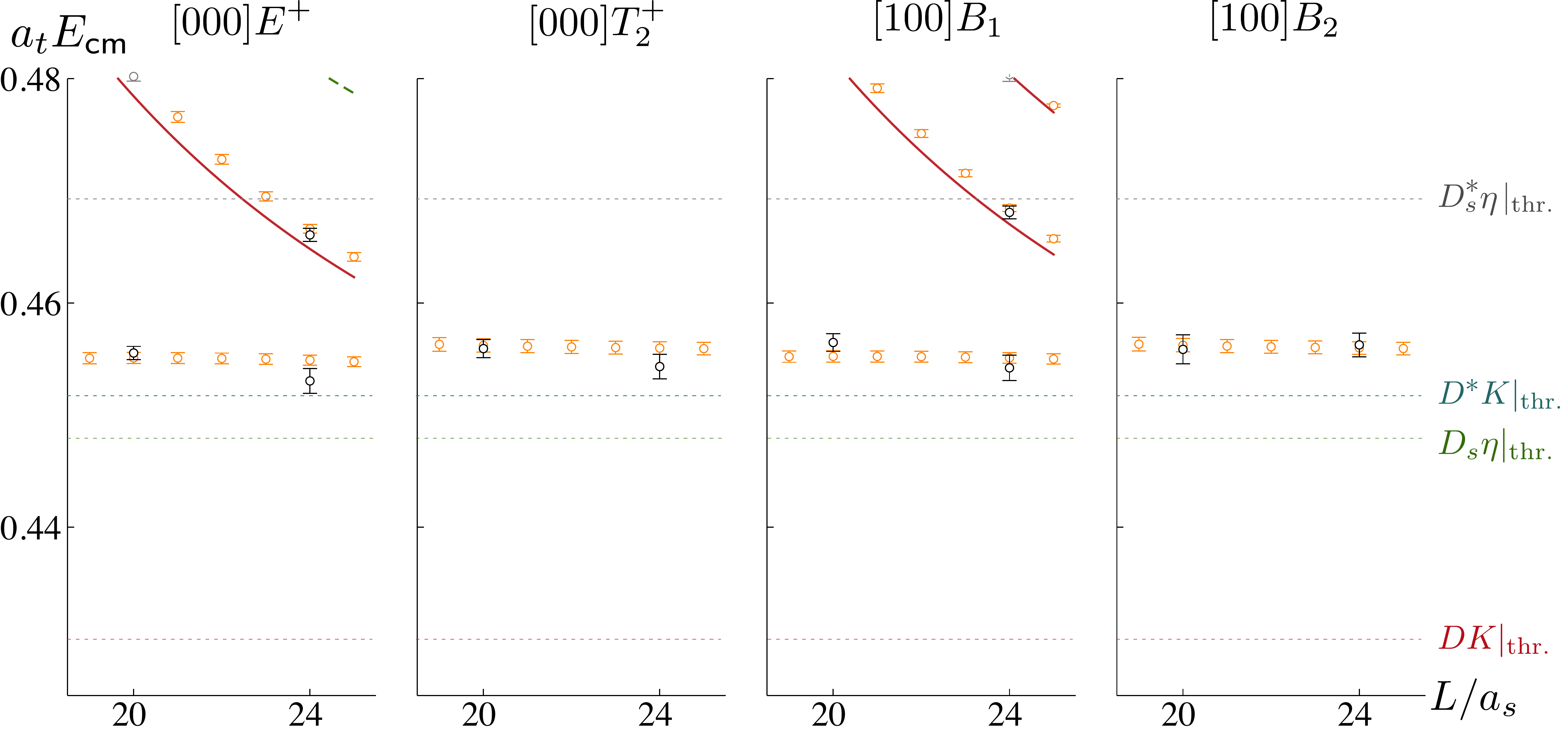}
\caption{$DK$ $I=0$ finite-volume spectra on the $m_\pi = 391$~MeV ensembles, as in the bottom row of Fig.~\ref{fig:840spectrum:DK} where black points indicate the computed finite-volume energy levels, with the addition of orange points which show the energy levels from the parametrisation in Eq.~(\ref{eqn:840DK:Dwave:ref}). }
\label{fig:DK:840:Dwave:paramspectrum}
\end{figure}

\begin{figure}[tb]
\centering
\includegraphics[width=0.8\textwidth]{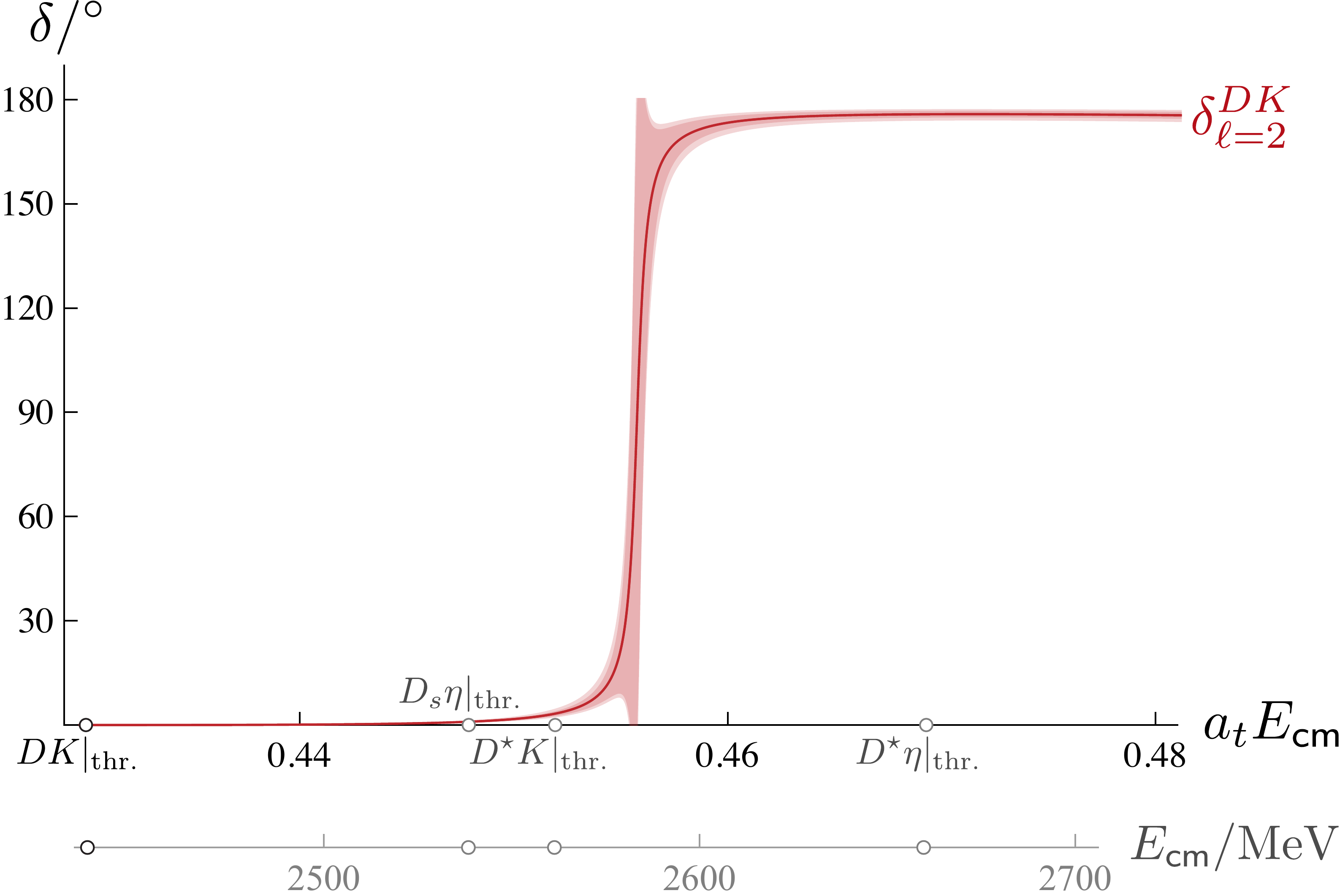}
\caption{$D$-wave scattering phase shift for $I=0$ $DK$ on the $m_\pi = 391$ MeV ensembles using the parametrisation in Eq.~(\ref{eqn:840DK:Dwave:ref}). The darker inner band indicates the statistical uncertainties, while the lighter outer band reflects the uncertainty from varying the $D$ and $K$ meson masses and the anisotropy as described in the text.}
\label{fig:DK:840:Dwave:phase}
\end{figure}

The corresponding spectra for $m_\pi = 391$~MeV are shown in the bottom row of Fig.~\ref{fig:840spectrum:DK}. Although there are no energy levels below the $D_s\eta$ and $D^\ast K$ thresholds, there are a number which are well below the lowest non-interacting energies associated with these inelastic channels.  The `extra' level below the first non-interacting $DK$ energy, roughly consistent across volumes and irreps, is suggestive of a reasonably narrow resonance.
A good description of the 10 levels shown in black in the bottom row of Fig.~\ref{fig:840spectrum:DK} is obtained using a relativistic Breit-Wigner parametrisation, Eq.~(\ref{eqn:BW}),
\begin{center}
\begin{tabular}{rll}
$m_R \;\; =$ & $(0.45578 \pm 0.00052) \cdot a_t^{-1}$ & 
\multirow{2}{*}{ $\begin{bmatrix}
 1     &  0.35 \\
       &     1 \\
\end{bmatrix}$ } \\
$g_R \;\;   =$ & $ 21.9 \pm 2.4 \, ^{+2.0}_{-1.3}$   & \\
&\multicolumn{2}{l}{ $\chi^2/ N_\mathrm{dof} = \frac{9.14}{10 - 2} = 1.14,$}  \\
\end{tabular}
\end{center}
\vspace{-1cm}
\begin{equation}\label{eqn:840DK:Dwave:ref}\end{equation}\\
where, as usual, the first uncertainty is statistical and the second uncertainty on $g_R$ is from varying the $D$ and $K$ meson masses and the anisotropy (the uncertainty on $m_R$ from varying these is negligible). In Fig.~\ref{fig:DK:840:Dwave:paramspectrum} we show a comparison between the computed finite-volume energy levels and the energy levels which follow from this parametrisation. It can be seen that there is good agreement between the two sets of energy levels, reflecting the reasonable $\chi^2 / \Ndf$ of the fit. Fig.~\ref{fig:DK:840:Dwave:phase} presents the amplitude following from this parametrisation. The phase shift is observed to rise steeply from near 0$^\circ$ towards 180$^\circ$, indicative of a narrow elastic resonance, and we examine the pole singularity in Section~\ref{sec:interpretation}.

These results show that the $D$-wave amplitude is small in the energy region where we determined $S$ and $P$-wave amplitudes, justifying our neglect of $\ell \geq 2$ there.

\subsubsection{Positive-parity $J=1$ partial waves}
\label{sec:scattering:DK:1p}

Because of the reduced symmetry of the finite-volume lattice, $J^P = 1^+$ can appear in $[100]E_2$, $[110]B_{1,2}$ and $[111]E_2$ irreps as well as $1^-$, with the lowest $1^+$ two-meson channel being $D^\ast K$.
On the $m_\pi = 239$ MeV ensemble, the results of Ref.~\cite{Cheung:2016bym} suggest that two $1^+$ axial-vector resonances should be expected at $a_t E \sim 0.41$ -- $0.42$. We included a few $D^\ast K$-like operators in the operator bases for these irreps, as shown in Table \ref{tab:ops:860:DK}. The finite-volume spectra in Fig.~\ref{fig:860spectrum:DK} have several levels in the vicinity of $D^\ast K$ threshold which have significant overlap with both $D^\ast K$ and fermion-bilinear operators, where the latter are subduced from $1^+$. One of these appears to be an `extra' level at $a_t E \sim 0.41$, below $D^\ast K$ threshold, with the possibility of another a bit higher in energy.

Considering the first excited levels in $[100]E_2$ and $[110]B_2$ which have dominant overlap with $D^\ast K$ operators and fermion-bilinear operators subduced from $J^P=1^+$, we can make a rough estimate of the $J^P=1^+$ amplitude.\footnote{Similar levels also occur in $[110]B_1$ and $[111]E_2$ but there $D K$ levels also occur nearby.} The $[100]E_2$ level is found at $a_tE_\mathsf{cm}=0.4087(4)$ corresponding to $a_t k \cot\delta=-0.0233(14)$, and the $[110]B_2$ level is found at $a_tE_\mathsf{cm}=0.4088(5)$ corresponding to $a_t k \cot\delta=-0.0242(14)$. These values imply large negative scattering lengths ($a_{1^+}\approx-40 a_t$), usually indicative of a near-threshold pole. Introducing a $D^\ast K$ $J^P=1^+$ amplitude with such a scattering length has negligible impact on the extraction of the $DK$ $1^-$ amplitude and thus on our determination of the $DK$ $0^+$ amplitude.

A similar pattern is observed on the ensembles with $m_\pi = 391$~MeV when comparing with spectra obtained using only fermion-bilinear operators~\cite{Moir:2013ub}. From this and the expected masses of the lightest axial-vector resonances, we expect the $D^\ast K$ $1^+$ amplitude to again have a negligible impact on our extraction of the $DK$ $1^-$ and $0^+$ amplitudes. However, no $D^\ast K$ operators were used here and so we cannot estimate the size of this $D^\ast K$ $1^+$ amplitude.


\subsection{$D\bar{K}$ scattering in $I=0$}
\label{sec:scattering:DKbarI0}

Compared to the $DK$ scattering amplitudes discussed above, the spectra in Figs.~\ref{fig:860spectrum:DKbar}, \ref{fig:840spectrum:DKbarI0} and \ref{fig:840spectrum:DKbarI1} suggest that the interactions in the exotic-flavour $D\bar{K}$ channels are relatively weak, with a slight attraction and repulsion in, respectively, $I=0$ and $I=1$.
When meson-meson interactions are weak, as they are in these channels, we find that the correlations between different finite-volume energy levels on the same volume can be considerable. To assess a possible impact from imprecisely estimating these correlations, we perform some ``covariance-adjusted'' fits where the eigenvalues, $\lambda$, of the data correlation matrix for which $\lambda < r \lambda_\textrm{max}$ are replaced by $r \lambda_\textrm{max}$, where $\lambda_\textrm{max}$ is the largest eigenvalue; our default fit corresponds to $r=0$.

To start our analysis of isospin-0 $D\bar{K}$ on the $m_\pi = 239$~MeV ensemble, we consider only the lowest 6 energy levels near threshold from irreps where the leading contribution is $S$-wave scattering.\footnote{One from each of $[000]A_1^+$, $[100]A_1$, $[110]A_1$ and $[111]A_1$, and two from $[200]A_1$.} Fitting to an $S$-wave scattering length gives $a_0/a_t = 4.9 \pm 3.8$ with $\chi^2/\Ndf = 5.0/5 = 1.0$.

Expanding the energy range, we fit the 18 levels which are well below the lowest $D^\ast \bar{K}$ non-interacting energy; these are the black points in Fig.~\ref{fig:860spectrum:DKbar}. A good description of the data is obtained with an effective range parametrisation in $S$-wave, and a scattering length in $P$-wave and $D$-wave,
\begin{center}
\begin{tabular}{rll}   
$a_0/a_t \;\; =$ & $9.4 \pm 1.8 \, ^{+ \, 0.8}_{- \, 0.5}$ & 
\multirow{4}{*}{ $\begin{bmatrix}
    & 1   & 0.92  & 0.59  & 0.45  \\
    &     & 1     & 0.30  & 0.24  \\
    &     &       & 1     & 0.73  \\
    &     &       &       & 1
\end{bmatrix}$ } \\
$r_0/a_t   \;\;   =$ & $32 \pm 11 \, ^{+ \, 4}_{- \, 2}$      \\
$a_1/a_t^3 \;\;   =$ & $-75 \pm 90 \pm 42$      \\
$a_2/a_t^5 \;\;   =$ & $-15200 \pm 8100 \pm 5500$      \\
&\multicolumn{2}{l}{ $\chi^2 / \Ndf = \frac{11.4}{18 - 4} = 0.81$,}  \\
\end{tabular}
\end{center}
\vspace{-0.8cm}
\begin{equation}\label{eqn:860DKbarI0:ref}\end{equation}
where, as usual, the first uncertainty is statistical. The second uncertainty is an envelope over the uncertainties from varying the $D$ and $\bar{K}$ meson masses and the anisotropy, as well as the uncertainty from a covariance-adjusted fit with $r=0.002$. Details of covariance-adjusted fits with $r=0.001$, $0.002$, $0.005$ and $0.01$ are given in Table~\ref{tab:860DKbarI0:limiteigs} in Appendix~\ref{app:DKbarfits} -- these show that the central values of the fit parameters do not change significantly as $r$ is increased, while the uncertainties increase slightly. For comparison, that table also shows a fit where correlations between energy levels have been ignored.

\begin{figure}[tb]
\centering
\includegraphics[width=0.99\textwidth]{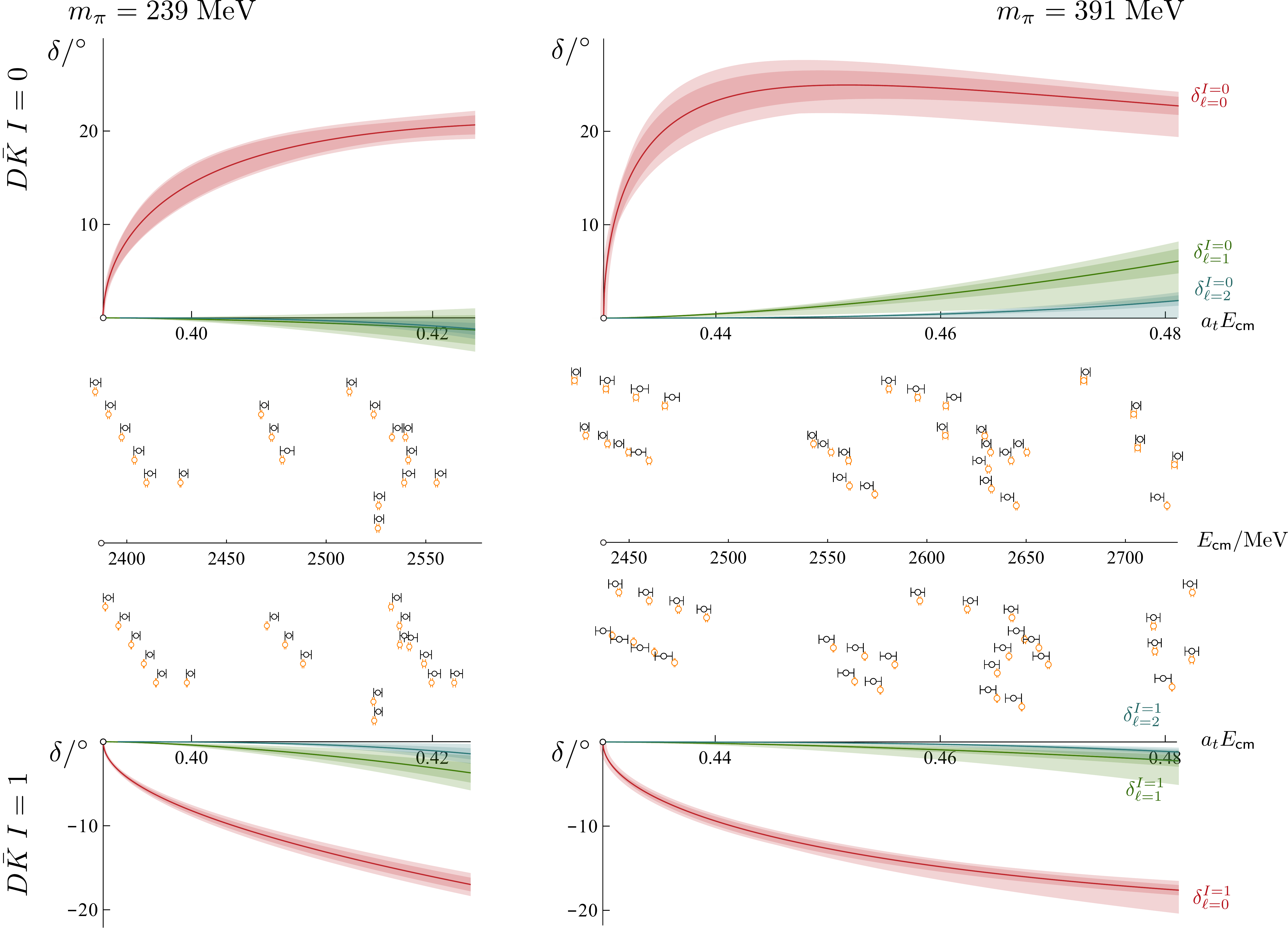} 
\caption{$S$, $P$ and $D$-wave scattering phase shifts for $D\bar{K}$ in $I=0$ (upper) and $I=1$ (lower) on the $m_\pi = 239$~MeV (left) and $m_\pi = 391$~MeV (right) ensembles.  The darker inner bands are the reference parametrisations given in Eqs.~(\ref{eqn:860DKbarI0:ref}), (\ref{eqn:840DKbarI0:ref}), (\ref{eqn:860DKbarI1:ref}) and (\ref{eqn:840DKbarI1:ref}) with statistical uncertainties, while the lighter outer bands give an envelope over the uncertainties from varying the $D$ and $\bar{K}$ meson masses and the anisotropy within their uncertainties, as well as a covariance-adjusted fit as described in the text with its statistical uncertainty. The black points in the middle show the computed finite-volume energy levels from Section~\ref{sec:spectrum} used to constrain the amplitudes, while the orange points show the energy levels following from the reference parametrisations.}
\label{fig:DKbar:ref_amps}
\end{figure}

The phase shifts resulting from this parametrisation are presented in Fig.~\ref{fig:DKbar:ref_amps} (top left plot). The figure also shows the good agreement between the computed finite-volume energy levels and the energy levels which follow from this parametrisation, reflecting the reasonable $\chi^2 / \Ndf$ of the fit.

\hfill

On the $m_\pi = 391$~MeV ensembles, fitting only the lowest energy level in each irrep where the leading contribution is from $S$-wave scattering (8 levels in total) to an $S$-wave scattering length gives $a_0/a_t = 16.5 \pm 3.1$ with $\chi^2/\Ndf = 8.9/7 = 1.3$.

Opening up the energy range, we fit the 29 levels shown as black points in Fig.~\ref{fig:840spectrum:DKbarI0}. A good description is obtained with an effective range parametrisation in $S$-wave, and a scattering length in $P$-wave and $D$-wave,
\begin{center}
\begin{tabular}{rll}   
$a_{0}/a_t \;\; =$ & $15.8 \pm 2.2 \, ^{+ \, 1.9}_{- \, 1.5}$ & 
\multirow{4}{*}{ $\begin{bmatrix}
    & 1   &  0.41  &  0.73 &  0.75   \\
    &     &  1     & -0.11 & -0.10  \\
    &     &        &  1    &  0.98   \\
    &     &        &       &  1 
\end{bmatrix}$ } \\
$r_0/a_t   \;\;   =$ & $36.5 \pm 1.8 \, ^{+ \, 7.2}_{- \, 1.5}$      \\
$a_1/a_t^3 \;\;   =$ & $128 \pm 26 \, ^{+ \, 19}_{- \, 63}$      \\
$a_2/a_t^5 \;\;   =$ & $4500 \pm 1300 \, ^{+ \, 800}_{- \, 3100}$      \\
&\multicolumn{2}{l}{ $\chi^2 / \Ndf = \frac{19.3}{29 - 4} = 0.77$.}  \\
\end{tabular}
\end{center}
\vspace{-0.8cm}
\begin{equation}\label{eqn:840DKbarI0:ref}\end{equation}
The phase shifts resulting from this parametrisation are shown in Fig.~\ref{fig:DKbar:ref_amps} (top right plot), along with a comparison between the computed energy levels and those which follow from the parametrisation. Covariance-adjusted fits are presented in Table~\ref{tab:840DKbarI0:limiteigs} in Appendix~\ref{app:DKbarfits} and, again, show that the results are insensitive to the value of $r$.

\hfill

\begin{figure}[tb]
\centering
\includegraphics[width=0.99\textwidth]{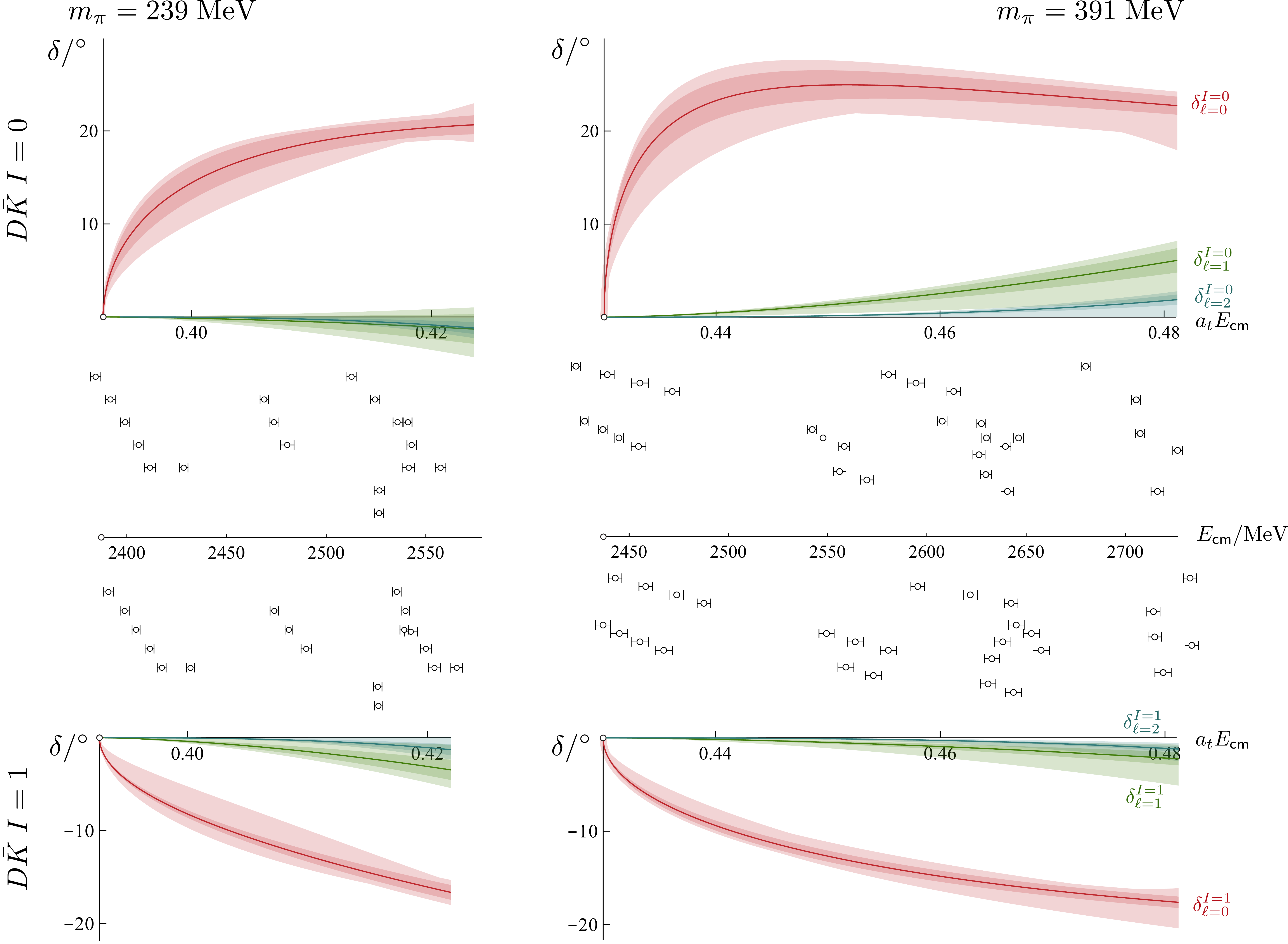} 
\caption{As Fig.~\ref{fig:DKbar:ref_amps}, but the lighter outer bands now also include an envelope over the acceptable parameterisations in Table~\ref{tab:DKbar:variations} including their statistical uncertainties.
The black points in the middle show the computed finite-volume energy levels from Section~\ref{sec:spectrum} used to constrain the amplitudes.}
\label{fig:DKbar:varying_amps}
\end{figure}

To avoid the results being biased by a particular choice, we varied the $S$-wave parametrisation, making use of higher orders in the effective range expansion and using various $K$-matrix amplitudes with simple or Chew-Mandelstam phase-space. For the $P$ and $D$-wave amplitudes we use either a scattering length or a constant $K$-matrix.
Results are summarised in Table~\ref{tab:DKbar:variations} -- those parametrisations with a $\chi^2/\Ndf$ in italics are not considered an acceptable description of the data for the reason given in the table.  The scattering amplitudes from acceptable parametrisations are shown in Fig.~\ref{fig:DKbar:varying_amps} where it can be seen that there is relatively little difference between the form of the amplitudes in the energy region where we are constraining the parametrisations.

\begin{figure}[tb]
\centering
\includegraphics[width=0.99\textwidth]{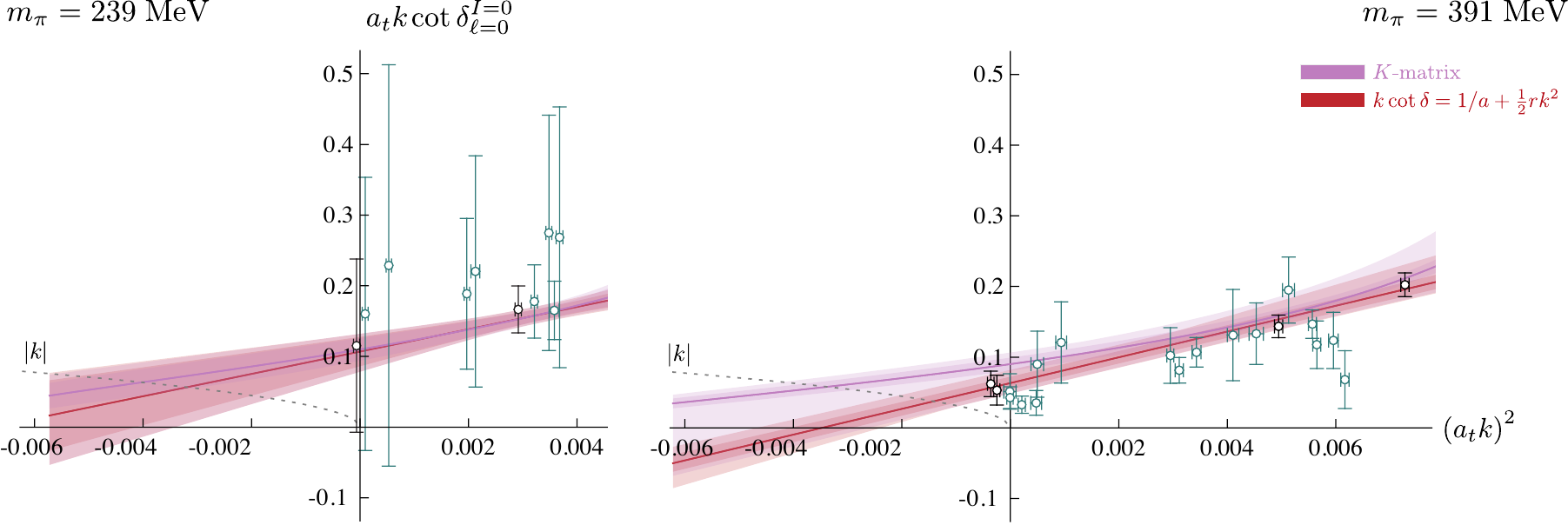} 
\caption{$S$-wave scattering amplitudes, plotted as $a_t k \cot \delta_{\ell=0}$, for $D\bar{K}$ in $I=0$ on the $m_\pi = 239$~MeV (left) and $m_\pi = 391$~MeV (right) ensembles. The bands show $K$-matrix parameterisation (e) in Table~\ref{tab:DKbar:variations} and the reference effective range parameterisation, Eqs.~(\ref{eqn:860DKbarI0:ref}) and (\ref{eqn:840DKbarI0:ref}), with their statistical uncertainties.
Black points are determined from energy levels in the $[000]A_1^+$ irrep assuming $\ell > 4$ amplitudes are negligible. Blue-green points are from the non-zero momentum $A_1$ irreps allowing for the $\ell = 1$ amplitude which has been fixed from other irreps. Some points on the left plot with very large uncertainties have been removed.}
\label{fig:DKbar:kcotdelta}
\end{figure}

As an alternative presentation of the $S$-wave scattering amplitudes, Fig.~\ref{fig:DKbar:kcotdelta} shows $a_t k \cot \delta_{\ell=0}$ in the region around $D\bar{K}$ threshold. Also shown is $|k|$ for $k^2 < 0$ -- the intersection of this curve with $a_t k \cot \delta_{\ell=0}$ indicates the presence of a pole singularity below threshold on the real axis of the \emph{unphysical sheet}, and we examine this in Section~\ref{sec:interpretation}.

\begin{table}
\begin{center}
\begin{tabular}{clcccccc}
 &           &           &          & \multicolumn{2}{c}{$m_\pi = 239$ MeV}  & \multicolumn{2}{c}{$m_\pi = 391$ MeV} \\
\multicolumn{2}{l}{\multirow{2}{*}{Parametrisation details}} & $\quad$ & $\quad$ & $\chi^2/\Ndf$ & $\chi^2/\Ndf$ & $\chi^2/\Ndf$ & $\chi^2/\Ndf$\\
 & & & & $I=0$ & $I=1$ & $I=0$ & $I=1$ \\
\hline
\hline
\multicolumn{6}{l}{$P$-wave and $D$-wave scattering lengths,}\\
\multicolumn{6}{l}{$S$-wave effective range expansion:}\\
(a) & $\frac{1}{a}$                                  &   &   & 1.11 & \textbf{1.05} & \it{3.00$\;^\diamond$} & \it{1.74$\;^\diamond$} \\
(b) & $\frac{1}{a}+\frac{1}{2}rk^2$                  &   &   & \textbf{0.81} & 1.12 & \textbf{0.77} & \textbf{1.04} \\
(c) & $\frac{1}{a}+\frac{1}{2}rk^2+P_2 k^4$          &   &   & \it{0.86$\;^\ddagger$} & \it{1.05$\;^\ddagger$} & \it{0.75$\;^\ddagger$} & \it{0.88$\;^\dagger$} \\
\hline
\multicolumn{6}{l}{$K$-matrix with Chew-Mandelstam $I(s)$,}\\
\multicolumn{6}{l}{$P$-wave and $D$-wave with a constant $K$, $S$-wave with: }\\
(d) & $K=\gamma^{(0)}$                               &   &   & 0.94 & 1.05 & \it{2.48$\;^\diamond$} & \it{1.60$\;^\diamond$} \\
(e) & $K=\gamma^{(0)}+\gamma^{(1)}s$                  &   &   & 0.81 & 1.11 & 1.16 & 0.99 \\
(f) & $K=\gamma^{(0)}+\gamma^{(1)}s+\gamma^{(2)}s^2$  &   &   & \it{0.86$\;^\ddagger$} & \it{1.03$\;^\ddagger$} & \it{0.88$\;^\ddagger$} & 0.89 \\
(g) & $K^{-1}=c^{(0)}+c^{(1)}s$                       &   &   & 0.82 & 1.12 & 0.78 & 1.04 \\
\hline
\multicolumn{6}{l}{$K$-matrix with simple phase space $I(s)=-i\rho(s)$,}\\
\multicolumn{6}{l}{$P$-wave and $D$-wave with a constant $K$, $S$-wave with: }\\
(h) & $K=\gamma^{(0)}$                                &   &   & 1.03 & 1.06 & \it{2.72$\;^\diamond$} & 1.30 \\
(i) & $K=\gamma^{(0)}+\gamma^{(1)}s$                  &   &   & 0.80 & 1.11 & 1.28 & 1.02 \\
(j) & $K=\gamma^{(0)}+\gamma^{(1)}s+\gamma^{(2)}s^2 $  &   &   & \it{0.86$\;^\dagger$} & \it{1.02$\;^\ddagger$} & \it{0.94$\;^\ddagger$} & 0.89 \\
(k) & $K^{-1}=c^{(0)}+c^{(1)}s$                        &   &   & 0.82 & 1.12 & 0.79 & 1.04 \\
\hline
\end{tabular}
{
\flushleft
\small
$\diamond$ -- rejected due to high $\chi^2/\Ndf > 1.5$ \\
$\dagger$ -- rejected due to high parameter correlations with parameter(s) consistent with zero \\
$\ddagger$ -- rejected due to physical-sheet complex poles or additional finite-volume level below threshold \\
}
\caption{A selection of the parametrisations used for elastic $D\bar{K}$ scattering with $I=0$ and $1$.
The $\chi^2/\Ndf$ in bold indicate the reference fits which are described in the text.
Parametrisations with $\chi^2/\Ndf$ shown in italics are not considered an acceptable description of the data for the reason given in the table.}
\label{tab:DKbar:variations}
\end{center}
\end{table}

\subsection{$D\bar{K}$ scattering in $I=1$}
\label{sec:scattering:DKbarI1}

Moving to isospin-1 $D\bar{K}$ on the $m_\pi = 239$~MeV ensemble, fitting only the 6 lowest energy levels from irreps where the leading contribution is $S$-wave scattering to an $S$-wave scattering length gives $a_0/a_t = a = -4.4 \pm 2.1$ with $\chi^2/\Ndf = 8.0/5 = 1.6$. Expanding the energy range, we fit the 18 energy levels which are shown as black points in Fig.~\ref{fig:860spectrum:DKbar}. A good description of the data is obtained with a scattering length parametrisation in $S$-wave, $P$-wave and $D$-wave,
\begin{center}
\begin{tabular}{rll}   
$a_0/a_t \;\; =$ & $-4.58 \pm 0.23 \,  ^{+ 0.15}_{- 0.16}$ & 
\multirow{3}{*}{ $\begin{bmatrix}
    & 1   & 0.81 & 0.68  \\
    &     & 1    & 0.73  \\
    &     &      & 1
\end{bmatrix}$ } \\
$a_1/a_t^3 \;\;   =$ & $-218 \pm 68 \, ^{+ 53}_{- 54}$      \\
$a_2/a_t^5 \;\;   =$ & $-19000 \pm 8500 \, ^{+ 6700}_{- 6900}$      \\
&\multicolumn{2}{l}{ $\chi^2 / \Ndf = \frac{15.7}{18 - 3} = 1.05$.}  \\
\end{tabular}
\end{center}
\vspace{-0.8cm}
\begin{equation}\label{eqn:860DKbarI1:ref}\end{equation}
The phase shifts resulting from this parametrisation are given in Fig.~\ref{fig:DKbar:ref_amps} (bottom left plot), along with a comparison between the computed energy levels and those which follow from the parametrisation.
Covariance-adjusted fits are presented in Table~\ref{tab:860DKbarI1:limiteigs} in Appendix~\ref{app:DKbarfits} and show that, again, the central values do not change significantly when $r$ is increased, whereas uncertainties increase slightly.

\hfill

On the $m_\pi = 391$~MeV ensembles, if only the lowest energy level in each irrep where the leading contribution is from $S$-wave scattering (8 levels in total) is fitted to an $S$-wave scattering length, we obtain, $a_0/a_t = -2.9 \pm 1.0$ with $\chi^2/\Ndf = 6.1/7 = 0.87$.
Using the full energy range, we fit the 28 energy levels shown as black points in Fig.~\ref{fig:840spectrum:DKbarI1}. A good description is obtained with an effective range parametrisation in $S$-wave, and a scattering length in $P$-wave and $D$-wave,
\begin{center}
\begin{tabular}{rll}   
$a_0/a_t \;\; =$ & $-4.43 \pm 0.30 \, ^{+ 0.12}_{- 0.10}$ & 
\multirow{4}{*}{ $\begin{bmatrix}
    & 1   &  0.80  & 0.63  & 0.61  \\
    &     &  1     & 0.18  & 0.15  \\
    &     &        & 1     & 0.95  \\
    &     &        &       & 1 
\end{bmatrix}$ } \\
$r_0/a_t   \;\;   =$ & $-16.0 \pm 3.2 \, ^{+ 8.6}_{- 2.4}$      \\
$a_1/a_t^3 \;\;   =$ & $-47 \pm 14 \, ^{+ 10}_{- 46}$      \\
$a_2/a_t^5 \;\;   =$ & $-2800 \pm 680 \, ^{+ 490}_{- 2400}$      \\
&\multicolumn{2}{l}{ $\chi^2 / \Ndf = \frac{25.0}{28 - 4} = 1.04$.}  \\
\end{tabular}
\end{center}
\vspace{-0.8cm}
\begin{equation}\label{eqn:840DKbarI1:ref}\end{equation}
Fig.~\ref{fig:DKbar:ref_amps} (bottom right plot) shows the phase shifts resulting from this parametrisation along with a comparison between the computed energy levels and those which follow from the parametrisation.
Covariance-adjusted fits are presented in Table~\ref{tab:840DKbarI1:limiteigs} in Appendix~\ref{app:DKbarfits} and, again, results show little dependence on $r$ with uncertainties increasing slightly as $r$ is increased.

\hfill

In a similar way to $I=0$, we varied the $S$-wave parametrisation, using a scattering length or constant $K$-matrix for the $P$ and $D$-wave amplitudes. The results are summarised in Table~\ref{tab:DKbar:variations} and the acceptable scattering amplitudes are shown in Fig.~\ref{fig:DKbar:varying_amps}. It can be seen that there is relatively little variation between the amplitudes from different parametrisations in the energy region where we have data constraining them. In the following section we examine the pole singularities in the $DK$ and $D\bar{K}$ amplitudes, interpret the results and compare with previous work in the literature.



\section{Poles, interpretation and comparisons}
\label{sec:interpretation}

Once the scattering $t$-matrix has been determined, by analytically continuing into the complex $s = E^2_\cm$ plane, the location of pole singularities can be found and hence the bound-state and resonance content determined. For elastic scattering there is a branch point at threshold, with the usual approach taking a branch cut from threshold along the positive real axis, leading to two Riemann sheets labelled by the sign of the imaginary part of the momentum $k$. Bound states appear as poles lying on the real axis below threshold on the \emph{physical sheet} ($\text{Im}[k] > 0$). Poles that correspond to a resonance occur in complex-conjugate pairs on the \emph{unphysical sheet} where $\text{Im}[k] < 0$.
In the proximity of a pole, the elastic $t$-matrix has the form,
\begin{equation}
\label{eqn:tmatrixpole}
t \sim \frac{c^2}{s_{\text{pole}}-s},
\end{equation}
where $c$ gives the coupling of the pole to the scattering channel.

\subsection{$S$ and $P$-wave $I=0$ $DK$}
\label{sec:interpretation:DK:SPwave}

Analytically continuing the $S$ and $P$-wave isospin-0 $DK$ amplitudes, we find that, for both values of $m_\pi$, all the acceptable parametrisations have a pole on the real axis of the physical sheet below $DK$ threshold in the $S$-wave scattering amplitude, corresponding to a bound state. In $P$-wave there is a pole on the real axis of the physical sheet very far below threshold, corresponding to a deeply bound state. There are no other poles in the region where we are constraining the amplitudes.

\begin{figure}
\centering
\includegraphics[width=0.99\textwidth]{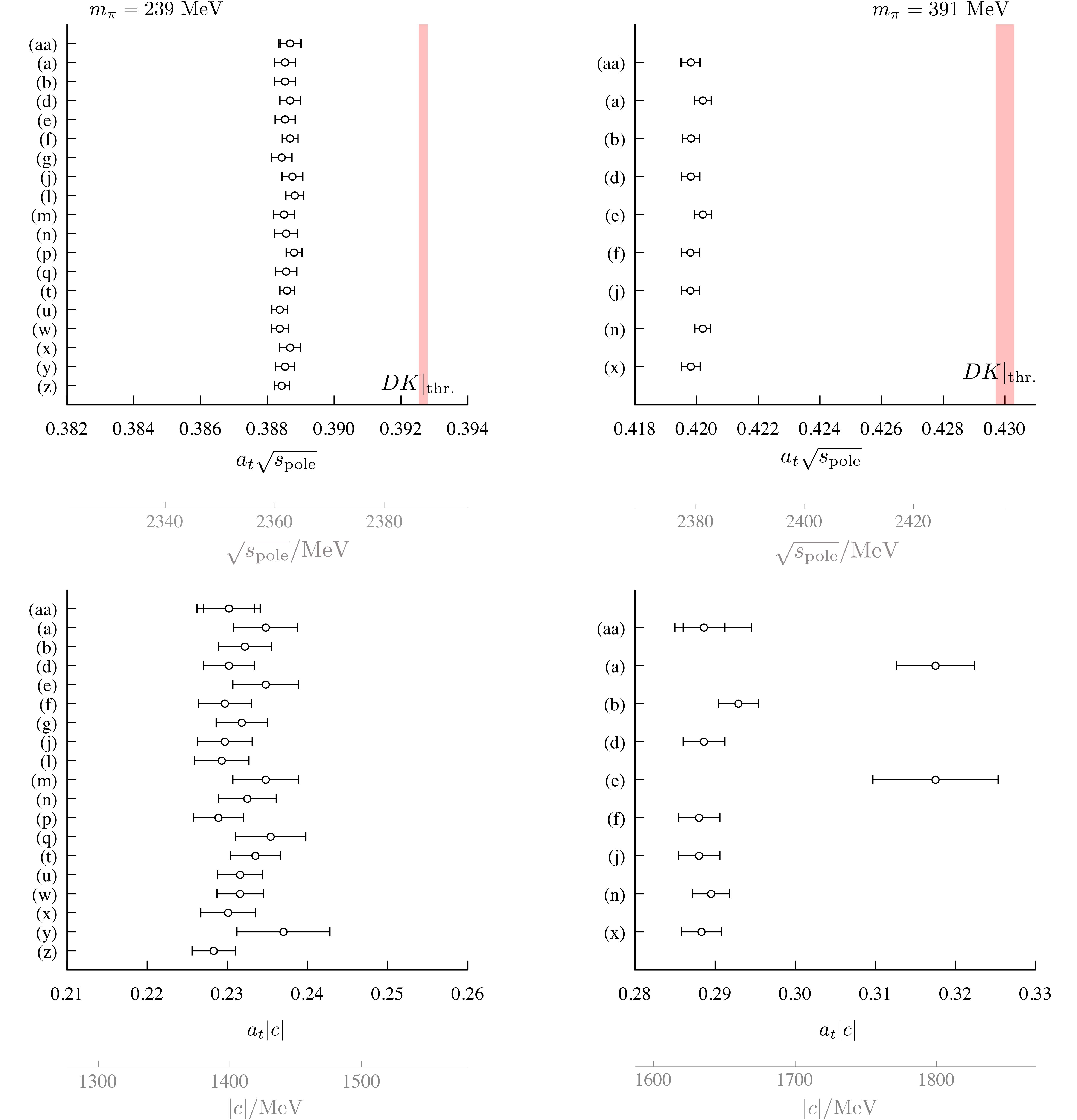}
\caption{\emph{Upper}: The location of the pole singularity on the real axis of the physical sheet in the $I=0$ $DK$ $S$-wave scattering amplitudes, for all the acceptable parametrisations in Table~\ref{tab:DK:variations}, on the $m_\pi = 239$~MeV (left) and $m_\pi = 391$~MeV (right) ensembles. The error bars give the statistical uncertainty, and the outer error bars on the reference amplitudes include also the uncertainty from varying the $D$ and $K$ meson masses and the anisotropy as described in the text.
The red band shows the location of $DK$ threshold.
\emph{Lower}: The corresponding coupling, $|c|$.}
\label{fig:DK:pole}
\end{figure}

The location of the $S$-wave pole and its coupling $c$, defined in Eq.~(\ref{eqn:tmatrixpole}), are presented in Fig.~\ref{fig:DK:pole}. The pole position is very insensitive to the choice of parametrisations, while the coupling for $m_\pi = 391$~MeV is slightly more sensitive, reflecting the smaller constraint on the amplitude around threshold as seen in Fig.~\ref{fig:DK:kcotdelta}. As our final results for the pole position and coupling, we quote an envelope that spans the values from different parametrisations, including their statistical uncertainties,\footnote{To be precise, if $A = \text{max}_i (\bar{x}_i + \sigma_{x_i})$ and $B = \text{min}_i (\bar{x}_i - \sigma_{x_i})$, where $i$ indexes the parametrisations, as the central value and uncertainty we quote $(A + B)/2$ and $(A - B)/2$ respectively.} 
\begin{center}
\begin{tabular}{rll}
                                  & $m_\pi = 239$~MeV & $m_\pi = 391$~MeV \\
$a_t \sqrt{s_\text{pole}} \;\; =$ & 0.3886(5) & 0.4200(5) \\
$a_t |c| \;\; =$                  & 0.234(9)  & 0.305(20) \\
\\
$\sqrt{s_\text{pole}} \;\; =$     & 2362(3) MeV   & 2380(3) MeV \\
$|c| \;\; =$                      & 1420(50) MeV  & 1730(110) MeV. \\
\end{tabular}
\end{center}

This scalar bound state ($J^P = 0^+$) has a large coupling to $DK$ and, at both $m_\pi$, its influence is strongly felt above threshold as seen in Fig.~\ref{fig:DK:varying_amps}. The binding energy decreases with decreasing light-quark mass: $\Delta E = m_D + m_K - \sqrt{s_\text{pole}} = 57(3)$~MeV for $m_\pi = 391$~MeV and $\Delta E = 25(3)$~MeV for $m_\pi= 239$~MeV. It appears to correspond to the experimentally observed $D_{s0}^{\ast}(2317)$ which appears below $DK$ threshold with $\Delta E \approx 45$~MeV~\cite{Tanabashi:2018oca} -- this is somewhat larger than our value on the ensemble with light-quark mass closest to the physical value. We note that there are unquantified systematic uncertainties in these calculations, such as discretisation effects~\cite{Liu:2012ze,Moir:2013ub,Cheung:2016bym}, which may contribute to this difference. In addition, experimentally the $D_{s0}^\ast(2317)$ is unstable and decays through the isospin-breaking $D_s\pi$ channel, whereas isospin symmetry is exact in our calculation.

We observe that the finite-volume eigenstates in the region of $DK$ threshold overlap significantly with both $c\bar{s}$ and $DK$-like operators; both structures appear to be essential in extracting the spectrum. While we do not use explicit compact-tetraquark operators in this calculation, extended four-quark components are present through the meson-meson operators.
Given the proximity to threshold, one may question whether this state is composed primarily of a short-distance $c\bar{s}$ component or whether a longer-distance $DK$ component is dominant. One method to assess this from the scattering amplitude itself is the Weinberg compositeness condition~\cite{Weinberg:1965zz}, valid for weakly-bound states when the scattering hadrons interact in a relative $S$-wave.
Assuming that the binding is weak enough so that corrections can be ignored,
\begin{equation}
a_0 = - 2 \, \frac{1-Z}{2-Z} \, \frac{1}{\sqrt{2 \, \mu \, \Delta E}} \, , \qquad
r_0 = - \, \frac{Z}{1-Z} \, \frac{1}{\sqrt{2 \, \mu \, \Delta E}} \, ,
\label{eqn:Z}
\end{equation}
where $0 \le Z \le 1$, with $Z \approx 0$ indicating that a composite or molecular component dominates and $Z \approx 1$ indicating an elementary configuration, $\mu$ is the $DK$ reduced mass, $\Delta E$ is the binding energy, and $a_0$ and $r_0$ use the same conventions as in Eq.~(\ref{eqn:effrange}).  Combining the two expressions to give $Z$ in terms of $a_0$ and $r_0$,
\begin{equation}
Z = 1 - \sqrt{\frac{a}{a+2r}} \, .
\label{eqn:Zar}
\end{equation}

For $m_\pi = 239$~MeV, $r_0$ is consistent with zero and supposing that the bound state is purely composite, $Z = 0$, gives $a_0 / a_t \approx -41$ to $-46$, consistent with Eq.~(\ref{eqn:860DK:eff_range}).  Alternatively, using the results in Eq.~(\ref{eqn:860DK:eff_range}) and inverting the expressions for $a_0$ and $r_0$ in Eq.~(\ref{eqn:Z}) gives $Z \lesssim 0.11$ and $Z \lesssim 0.04$ respectively, whereas Eq.~(\ref{eqn:Zar}) gives $Z \lesssim 0.04$.
Performing the same analysis for $m_\pi = 391$~MeV, assuming $Z=0$ (even though $r_0$ is not consistent with zero) gives $a_0 / a_t \approx -25$ to $-26$, roughly consistent with Eq.~(\ref{eqn:840DK:eff_range}). Instead using the results for $a_0$ and $r_0$ in Eq.~(\ref{eqn:840DK:eff_range}) gives $Z \approx 0.13(6)$ and $Z \approx 0.14(3)$ respectively, whereas using Eq.~(\ref{eqn:Zar}) gives $Z \approx 0.14(3)$.\footnote{None of the rough estimates for $Z$ given here take into account correlations between $a_0$, $r_0$ and $\Delta E$.}
This analysis suggests that a molecular $DK$ component dominates for both $m_\pi$.

\hfill

Elastic $S$-wave $DK$ scattering has been studied by several other groups~\cite{Mohler:2013rwa,Lang:2014yfa,Bali:2017pdv,Alexandrou:2019tmk} with a range of light-quark masses and lattice methodologies (see also re-analyses of lattice results in \cite{Torres:2014vna,Guo:2018tjx}), and an earlier calculation~\cite{Liu:2012zya} computed the scattering length indirectly by using a chiral unitary approach to relate it to channels where there are no annihilation contributions. Broadly speaking, these calculation all find a bound-state pole and are consistent with a dominant $DK$ configuration. However, it is not straightforward to compare them because some have only dynamical light quarks with quenched strange quarks ($N_f = 2$) rather than dynamical light and strange quarks ($N_f=2+1$), they use light-quark masses corresponding to different $m_\pi$, and some use very small volumes where corrections to the L\"uscher quantisation condition may be significant. Ref.~\cite{Alexandrou:2019tmk} is the most directly comparable to our work with $N_f=2+1$, a reasonable volume ($m_\pi L \approx 4.4$) and $m_\pi = 296$~MeV. They find a pole $\sim 51$~MeV below $DK$ threshold, consistent with our result for $m_\pi=391$~MeV, but somewhat larger than that for $m_\pi= 239$~MeV.

\hfill

For the $P$-wave deeply bound state, using a similar procedure we find a pole at,
\begin{center}
\begin{tabular}{rll}
                                   & $m_\pi = 239$~MeV & $m_\pi = 391$~MeV \\
$a_t \sqrt{s_\text{pole}} \;\; =$  & 0.34432(22) & 0.36579(43) \\
$\sqrt{s_\text{pole}} \;\; =$      & 2093(1) MeV & 2073(2) MeV \\
\end{tabular}
\end{center}
This vector bound state ($J^P = 1^-$) corresponds to the $D_{s}^{\ast}$ and does not appear to significantly influence $DK$ scattering in the physical scattering region. Its mass is consistent with the results obtained using only $\bar{q}q$ operators, see Section~\ref{sec:lattice} and Refs.~\cite{Liu:2012ze,Cheung:2016bym}, as expected for a state so far below threshold.

\subsection{$D$-wave $I=0$ $DK$}

The $D$-wave isospin-0 $DK$ amplitude on the $m_\pi = 391$~MeV ensemble, Eq.~(\ref{eqn:840DK:Dwave:ref}), has a complex-conjugate pair of pole singularities on the unphysical sheet. They have,
\begin{eqnarray*}
a_t \sqrt{s_\text{pole}} &=& \bigl( 0.4558 \pm 0.0005 \bigr) \pm \tfrac{i}{2} \bigl( 0.0006 \, ^{+ \, 0.0003}_{- \, 0.0002} \bigr) \\
a_t |c| &=& 0.033 \, ^{+ \, 0.007}_{- \, 0.006} \\  
\text{arg}[c] &=& \pm \left(0.0098 \, ^{+ \, 0.004}_{- \, 0.003} \right) \pi \\
\\
\sqrt{s_\text{pole}} &=& \bigl(2583 \pm 3\bigr) \pm \tfrac{i}{2} \bigl(3.4 \pmerr{1.7}{1.1} \bigr)\text{MeV} \\
|c| &=& 190 \pmerr{40}{30} ~ \text{MeV} \, .
\end{eqnarray*}
Any analogous pole on the $m_\pi = 239$~MeV ensemble would be expected to be above the energy region where we can constrain the scattering amplitudes -- there appears to be an extra level in the relevant irreps, suggesting a reasonably narrow resonance.

This tensor resonance ($J^P = 2^+$) corresponds to the experimentally-observed $D^\ast_{s2}(2573)$. We find that its Breit-Wigner coupling, $g_R$, is similar to the $D^\ast_2$ resonance found in $D\pi$ scattering at the same pion mass in Ref.~\cite{Moir:2016srx}.

\subsection{$I=0,1$ $D\bar{K}$}

In $I=0$ $D\bar{K}$ scattering the $P$ and $D$-wave amplitudes are very small, but a sharper rise in the $S$-wave phase shift indicates that a virtual bound-state pole may be present, a pole on the real axis below threshold on the \emph{unphysical sheet}, corresponding to an attractive interaction but one that is not strong enough to form a bound state. The $S$-wave $I=1$ channel appears to be weakly repulsive with $P$ and $D$-wave $I=1$ amplitudes roughly consistent with zero.

Analytically continuing the scattering amplitudes into the complex $s = E^2_\cm$ plane, we find that, for both $m_\pi$, most of the acceptable parametrisations do have a virtual bound-state pole below $D\bar{K}$ threshold in the $S$-wave $I=0$ scattering amplitude. There are no other poles in $I=0$ or $I=1$ $D\bar{K}$ in the region where the amplitudes are well-constrained.

\begin{figure}
\centering
\includegraphics[width=0.99\textwidth]{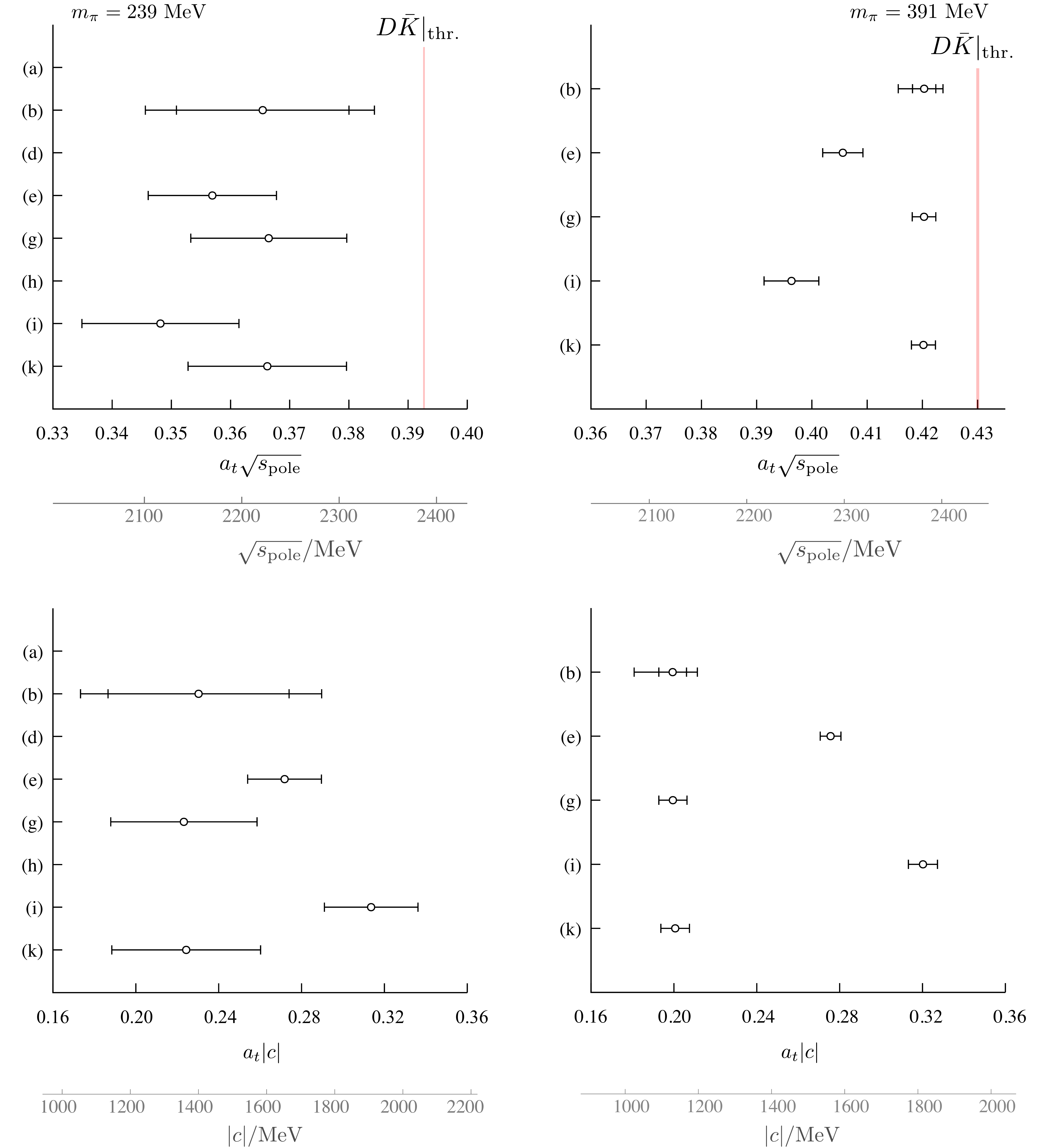}
\caption{\emph{Upper}: The location of the pole singularity on the real axis of the unphysical sheet in the $I=0$ $D\bar{K}$ $S$-wave scattering amplitudes, for acceptable parametrisations in Table~\ref{tab:DKbar:variations}, on the $m_\pi = 239$~MeV (left) and $m_\pi = 391$~MeV (right) ensembles. The error bars give the statistical uncertainty, and the outer error bars on the reference amplitudes include also the uncertainty from varying the $D$ and $K$ meson masses and the anisotropy, as well as a covariance-adjusted fit as described in the text. The red band shows the location of $D\bar{K}$ threshold.
\emph{Lower}: The corresponding coupling, $|c|$.}
\label{fig:DKbarI0:pole}
\end{figure}

The location and coupling $c$, defined in Eq.~(\ref{eqn:tmatrixpole}), of this exotic-flavour $J^P=0^+$ pole are presented in Fig.~\ref{fig:DKbarI0:pole}. The location is roughly consistent across parametrisations within statistical uncertainties, but there is somewhat more variation in the coupling. However, no pole is found for $m_\pi = 239$~MeV on parametrisations (a), (d) and (h) which give a reasonable description of the finite-volume energy levels. We can therefore only say that there is a suggestion of a pole in the $I=0$ $D\bar{K}$ amplitudes, not that a pole is absolutely required. As our results for the position and coupling of this suggested pole, we quote an envelope that spans the values from different parametrisations where a pole is found, including their statistical uncertainties,
\begin{center}
\begin{tabular}{rll}
                                  & $m_\pi = 239$~MeV & $m_\pi = 391$~MeV \\
$a_t \sqrt{s_\text{pole}} \;\; =$  & 0.357(23) & 0.407(16) \\
$a_t |c| \;\; =$                  & 0.26(7)   & 0.26(7). \\  
\\
$\sqrt{s_\text{pole}} \;\; =$      & 2170(140) MeV  & 2310(90) MeV \\
$|c| \;\; =$                      & 1600(400) MeV  & 1500(400) MeV. \\  
\end{tabular}
\end{center}

In some of the parameterisations used, a physical-sheet pole appears in the $I=0$ $D\bar{K}$ amplitudes at an energy much lower than the unphysical-sheet pole, well below where they are constrained by the finite-volume energy levels. The presence of this pole can be inferred from Fig.~\ref{fig:DKbar:kcotdelta} where an intersection of the effective range curve, $1/a+\tfrac{1}{2}rk^2$, and $-|k|$ is inevitable at large negative $k^2$. Investigating further, we find that the residue of this pole has the ``wrong'' sign (see e.g.~\cite{Sitenko:ScatteringTheory,Taylor:ScatteringTheory,Iritani:2017rlk}) and so signals that the amplitude is unphysical in this energy region -- it has been extrapolated too far in energy from where it is being constrained. The $K$-matrix amplitude shown in Fig.~\ref{fig:DKbar:kcotdelta} does not suffer this problem.

\hfill

In Ref.~\cite{Liu:2012zya}, $S$-wave $D\bar{K}$ $I=0,1$ scattering lengths were computed for a range of $m_\pi$ using a partially-quenched lattice formulation. They found attraction in $I=0$ and repulsion in $I=1$, consistent with our results, but there was no suggestion of a virtual bound state in $I=0$. Our scattering-length parametrisation of $S$-wave $I=0$ scattering similarly does not lead to a virtual bound state in the region where we are constraining the amplitudes -- we find that a form with at least two parameters is required to give a virtual bound state in this energy region.

Ref.~\cite{Albaladejo:2016lbb} found agreement between a unitarized chiral perturbation theory analysis of $D \pi$, $D \eta$, $D_s \bar{K}$ scattering and the lattice results in Ref.~\cite{Moir:2016srx}, and from this predicted the presence of a virtual bound state in $S$-wave $D\bar{K}$ $I=0$.  Ref.~\cite{Du:2017zvv} goes further and, amongst other things, discusses how this state should appear at the SU(3) flavour symmetric point for various quark masses.

If the charm and strange quark masses were increased, this exotic-flavour $I=0$ $D\bar{K}$ $J^P = 0^+$ channel would resemble an $I=0$ $bb\bar{u}\bar{d}$ system often considered as a possible tetraquark candidate. The related $J^P=1^+$ $I=0$ $bb\bar{u}\bar{d}$ systems have been observed to contain a bound state in lattice calculations~\cite{Francis:2016hui,Francis:2018jyb,Junnarkar:2018twb,Leskovec:2019ioa}, but there was no evidence seen for a bound state in $J^P=1^+$ $I=0$ $cc\bar{u}\bar{d}$ in Ref.~\cite{Cheung:2017tnt}. Future investigations mapping out the mass, spin, and flavour dependence of these exotic systems will be particularly interesting.

\subsection{$DK$, $D\bar{K}$ and $D\pi$ scattering and SU(3) flavour symmetry}

In the limit of SU(3) flavour symmetry, where the up, down and strange quarks have the same mass, various channels in $DK$, $D\bar{K}$, $D\pi$, etc scattering are related. In the calculations presented here, the up and down quarks are degenerate and, particularly for the $m_\pi = 391$~MeV ensembles, are not so much lighter than the strange quark. SU(3) may therefore not be too badly broken and provide a useful framework for interpreting the results.

The combination of a $D_{(s)}$ meson (SU(3) $\overline{\bm{3}}$) and a light meson ($\bm{8}$ or $\bm{1}$) gives,
\begingroup
\setlength\abovedisplayskip{3pt}
\setlength\belowdisplayskip{3pt}
$$\overline{\bm{3}}~\otimes~\bm{8}~\to~\overline{\bm{3}}~\oplus~\bm{6}~\oplus~\overline{\bm{15}} \, ,
\quad 
\overline{\bm{3}} \otimes \bm{1} \to \overline{\bm{3}} \, .$$
\endgroup
Using Ref.~\cite{Kaeding:1995vq}, the various scattering channels decompose into SU(3) multiplets as,
\begin{center}
\begin{tabular}{lll}
$(I=0)$ $DK$-$D_s\eta$: $\overline{\bm{3}} \oplus \overline{\bm{15}}$ & $\quad$ & $(I=\tfrac{1}{2})$ $D\pi$-$D\eta$-$D_s\bar{K}$: $\overline{\bm{3}} \oplus \bm{6} \oplus \overline{\bm{15}}$ \\[5pt]
$(I=1)$ $DK$-$D_s\pi$: $\bm{6} \oplus \overline{\bm{15}}$             & $\quad$ & $(I=0)$ $D\bar{K}$: ${\bm{6}}$ \\[5pt]
\multicolumn{3}{l}{
$(I=\tfrac{1}{2})$ $D_s K$,  $(I=1)$ $D\bar{K}$, $(I=\tfrac{3}{2})$ $D\pi$: $\overline{\bm{15}}$
}
\end{tabular}
\end{center}

We found $S$-wave $D\bar{K}$ $(I=1)$ to be weakly repulsive with no poles in the energy region considered. This is very similar to the results for $D\pi$ $(I=3/2)$ presented in Ref.~\cite{Moir:2016srx}, consistent with them being identical in the SU(3) limit. We conclude that, in this energy region, $S$-wave $\overline{\bm{15}}$ exhibits weak repulsion and there are no poles in the scattering amplitude. From our results for $DK$ $(I=0)$, we conclude that the $S$-wave $\overline{\bm{3}}$ contains a bound state for the light-quark masses used here. From $D\bar{K}$ $(I=0)$, the $S$-wave $\bm{6}$ shows weak attraction and may contain a virtual bound state.

We would expect $D\pi$ $(I=1/2)$ to show behaviour which is a superposition of $DK$ $(I=0)$ and $D\bar{K}$ $(I=0)$, and so it could contain two poles. In Ref.~\cite{Moir:2016srx}, with $m_\pi = 391$~MeV we found only one robustly-determined pole in $S$-wave $D\pi$ $(I=1/2)$ in the energy region considered, a bound state consistent in mass with the $D\pi$ threshold, but we did not exclude the possibility of another pole.
This picture appears to have some similarity with Refs.~\cite{Albaladejo:2016lbb,Du:2017zvv,Guo:2018tjx} which suggest a ``two-pole'' structure in $I=1/2$ $D\pi$,~$D\eta$,~$D_s \bar{K}$, a bound-state pole and a resonance (complex-conjugate pair of poles) at $m_\pi = 391$~MeV which evolve to two resonances for physical quark masses.\footnote{So that the number of poles remains constant when the light-quark masses are varied, the bound-state pole at unphysical quark masses should come with a virtual pole. $DK$ $(I=0)$ should contain a virtual pole in addition to the bound-state pole and $D\bar{K}$ $(I=0)$ should have two virtual poles.} Those references also discuss how states evolve as the quark masses are varied and related scattering channels; see Ref.~\cite{Meissner:2020khl} for a more general discussion of two-pole structures.
The effect of SU(3) flavour symmetry breaking is apparent here: because $m_K > m_\pi$, the bound state in $DK$ is more bound than in $D\pi$. For physical light-quark masses we would expect even less binding in $D\pi$, consistent with experiment where the $D_0^\ast(2400)$ is a resonance rather than a bound state.

The $P$ and $D$-wave channels appear to be weakly interacting in the energy region we are considering, with the exception of the $\overline{\bm{3}}$ which contains a deeply bound state in $P$-wave and a narrow resonance in $D$-wave.


\section{Summary and Outlook}
\label{sec:conclusion}

\begin{figure}[tb]
\centering
\includegraphics[width=0.8\textwidth]{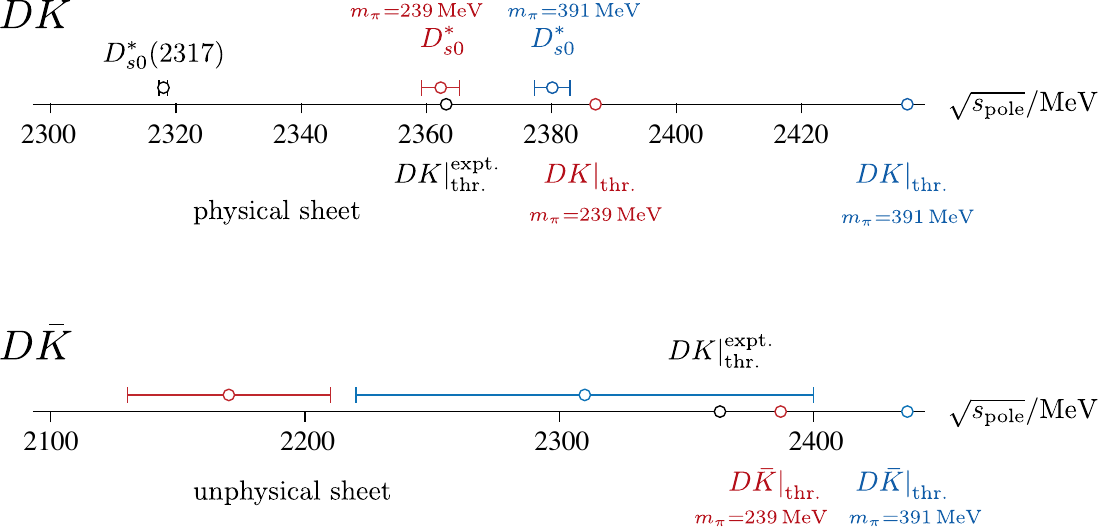} 
\caption{A summary of the $S$-wave poles found in this work. The top panel shows the bound-state poles found on the physical sheet in the $I=0$ $DK$ scattering amplitudes with the experimental $D_{s0}^\ast(2317)$ and threshold from Ref.~\cite{Tanabashi:2018oca}. The lower panel shows the virtual bound-state poles found on the unphysical sheet in the exotic $I=0$ $D\bar{K}$ scattering amplitudes; note that not all the amplitudes for $m_\pi=239$~MeV produced such a pole.}
\label{fig:summary_poles}
\end{figure}

Scattering amplitudes for $I=0$ $DK$ and $I=0,1$ $D\bar{K}$ have been determined using lattice QCD
with light-quark masses corresponding to $m_\pi = 239$ MeV and $m_\pi = 391$ MeV.
As described, the techniques we use enable us to reliably extract a large number of finite-volume energy levels and hence to map out precisely the energy dependence of the amplitudes in the elastic scattering region.
Analytically continuing these amplitudes in the complex plane
reveals the presence of a bound state in $S$-wave $I=0$ $DK$ and, for the first time in a lattice calculation,
evidence for a virtual bound state in $S$-wave $I=0$ $D\bar{K}$. We summarise the $S$-wave poles found in this work in Fig.~\ref{fig:summary_poles}.

The $J^P = 0^+$ $DK$ bound state, corresponding to the $D_{s0}^\ast(2317)$, is found close to $DK$ threshold: 57(3) MeV below threshold for the larger $m_\pi$, reducing to 25(3) MeV for the smaller $m_\pi$. It has a large coupling to the $DK$ channel and, as seen in Fig.~\ref{fig:DK:varying_amps}, it leaves a clear signature on the elastic $DK$ scattering amplitude which rises sharply from threshold for both $m_\pi$. In $I=0$ $DK$ we also find a deeply bound vector in $P$-wave, corresponding to the $D_s^\ast$, and in addition there is some evidence for $2^+$ (see Fig.~\ref{fig:DK:840:Dwave:phase}) and $1^+$ resonances at higher energies.

The exotic $I=0$ $D\bar{K}$ amplitudes are mapped out for the first time in a lattice calculation and we
find that these interactions are smaller in magnitude than in $DK$ with only the $S$-wave determined
to be significant, as shown in Fig.~\ref{fig:DKbar:varying_amps}.
As described above, we see hints of a $0^+$ $I=0$ virtual bound state pole below $D\bar{K}$ threshold -- this is present in all the acceptable $m_\pi = 391$~MeV amplitudes and the majority of those at
$m_\pi = 239$~MeV. The lighter pion mass produces an amplitude with larger uncertainties and more energies might constrain this better to produce a completely unambiguous result. The $I=1$ $D\bar{K}$ amplitudes are found to be weakly repulsive. It is interesting to note that lattice calculations have found signs of a bound state in
similar $1^+$ $I=0$ exotic-flavour channels when the valence $c,s$-quarks are replaced by $b$-quarks~\cite{Francis:2016hui,Francis:2018jyb,Junnarkar:2018twb,Leskovec:2019ioa}. Further calculations at a range of quark masses and in various flavour and spin channels would help map out the QCD dynamics at work in these systems.

The work described here opens the way to many possible avenues of exploration.
For example, $D^\ast K$ amplitudes can be computed in an analogous way
and a study of these would be particularly interesting since two $D_{s1}$ states are expected near $D^\ast K$ threshold -- experimentally one is observed just below threshold and one just above threshold.
More broadly, lattice calculations offer the possibility to map out the quark-mass dependence of the scattering
amplitudes, yielding insight into the dynamics of QCD.

\bigskip

\begin{acknowledgments}
We thank our colleagues within the Hadron Spectrum Collaboration (www.hadspec.org).
GKCC, CET, DJW and GM acknowledge support from the U.K. Science and Technology Facilities Council (STFC) [grant number ST/P000681/1].
GKCC acknowledges support from the Cambridge European Trust and St John's College, Cambridge.
DJW acknowledges support from a Royal Society University Research Fellowship.
GM acknowledges support from the Herchel Smith Fund at the University of Cambridge.

This work used the Cambridge Service for Data Driven Discovery (CSD3), part of which is operated by the University of Cambridge Research Computing Service (www.csd3.cam.ac.uk) on behalf of the STFC DiRAC HPC Facility (www.dirac.ac.uk). The DiRAC component of CSD3 was funded by BEIS capital funding via STFC capital grants ST/P002307/1 and ST/R002452/1 and STFC operations grant ST/R00689X/1. Other components were provided by Dell EMC and Intel using Tier-2 funding from the Engineering and Physical Sciences Research Council (capital grant EP/P020259/1).
This work also used the DiRAC Data Analytic system at the University of Cambridge, operated by the University of Cambridge High Performance Computing Service on behalf of the STFC DiRAC HPC Facility (www.dirac.ac.uk). This equipment was funded by BIS National E-infrastructure capital grant (ST/K001590/1), STFC capital grants ST/H008861/1 and ST/H00887X/1, and STFC DiRAC Operations grant ST/K00333X/1.
DiRAC is part of the National e-Infrastructure.
This work also used clusters at Jefferson Laboratory under the USQCD Initiative and the LQCD ARRA project, and the authors acknowledge support from the U.S. Department of Energy, Office of Science, Office of Advanced Scientific Computing Research and Office of Nuclear Physics, Scientific Discovery through Advanced Computing (SciDAC) program, and the U.S. Department of Energy Exascale Computing Project.

The software codes {\tt Chroma}~\cite{Edwards:2004sx}, {\tt QUDA}~\cite{Clark:2009wm,Babich:2010mu}, {\tt QPhiX}~\cite{Joo:2013lwm}, and {\tt QOPQDP}~\cite{Osborn:2010mb,Babich:2010qb} were used to compute the propagators required for this project.
This work used the Wilkes GPU cluster at the University of Cambridge High Performance Computing Service (www.hpc.cam.ac.uk), provided by Dell Inc., NVIDIA and Mellanox, and part funded by STFC with industrial sponsorship from Rolls Royce and Mitsubishi Heavy Industries.
Propagators were also computed on clusters at Jefferson Laboratory under the USQCD Initiative and the LQCD ARRA project.
This research was supported in part under an ALCC award, and used resources of the Oak Ridge Leadership Computing Facility at the Oak Ridge National Laboratory, which is supported by the Office of Science of the U.S. Department of Energy under Contract No. DE-AC05-00OR22725. This research is also part of the Blue Waters sustained-petascale computing project, which is supported by the National Science Foundation (awards OCI-0725070 and ACI-1238993) and the state of Illinois. Blue Waters is a joint effort of the University of Illinois at Urbana-Champaign and its National Center for Supercomputing Applications. This work is also part of the PRAC “Lattice QCD on Blue Waters”. This research used resources of the National Energy Research Scientific Computing Center (NERSC), a DOE Office of Science User Facility supported by the Office of Science of the U.S. Department of Energy under Contract No. DEAC02-05CH11231. The authors acknowledge the Texas Advanced Computing Center (TACC) at The University of Texas at Austin for providing HPC resources that have contributed to the research results reported within this paper.

Gauge configurations were generated using resources awarded from the U.S. Department of Energy INCITE program at Oak Ridge National Lab, NERSC, the NSF Teragrid at the Texas Advanced Computer Center and the Pittsburgh Supercomputer Center, as well as at Jefferson Lab.

\end{acknowledgments}

\clearpage

\appendix
\section*{Appendices}
\section{Operator Lists}
\label{app:operators}

In Tables~\ref{tab:ops:860:DK}, \ref{tab:ops:840:DK}, \ref{tab:ops:860:DKbar} and \ref{tab:ops:840:DKbar} we list the interpolating operators used to determine the finite-volume energy levels in the $DK$ isospin-$0$ channel shown in Figs.~\ref{fig:860spectrum:DK} and \ref{fig:840spectrum:DK}, and the $D\bar{K}$ isospin-$0$ and $1$ channels shown in Figs.~\ref{fig:860spectrum:DKbar}, \ref{fig:840spectrum:DKbarI0} and \ref{fig:840spectrum:DKbarI1}.

\begin{table}
\begin{center}
\begin{tabular}{c|c|c|c|c}
\multicolumn{1}{c|}{$[000]A_1^+$} & \multicolumn{1}{c|}{$[001]A_1$} & \multicolumn{1}{c|}{$[011]A_1$} & \multicolumn{1}{c|}{$[111]A_1$}  & \multicolumn{1}{c}{$[002]A_1$} \\
\hline \hline
$D_{[000]}K_{[000]}$        & $D_{[001]}K_{[000]}$         & $D_{[011]}K_{[000]}$       & $D_{[111]}K_{[000]}$      & $D_{[002]}K_{[000]}$  \\  
$D_{[001]}K_{[001]}$        & $D_{[000]}K_{[001]}$         & $D_{[000]}K_{[011]}$       & $D_{[000]}K_{[111]}$      & $D_{[000]}K_{[002]}$  \\
$D_{[011]}K_{[011]}$        & $D_{[011]}K_{[001]}$         & $D_{[001]}K_{[001]}$       & $D_{[011]}K_{[001]}$      & $D_{[001]}K_{[001]}$  \\
$D_{[111]}K_{[111]}$        & $D_{[001]}K_{[011]}$         & $D_{[001]}K_{[111]}$       & $D_{[001]}K_{[011]}$      & $D_{[011]}K_{[011]}$  \\
$D_{s[000]}\eta_{[000]}$    & $D_{[002]}K_{[001]}$         & $D_{[111]}K_{[001]}$       & $D_{[112]}K_{[001]}$      & $D_{[111]}K_{[111]}$  \\
$D_{s[001]}\eta_{[001]}$    & $D_{s[001]}\eta_{[000]}$     & $D_{[011]}K_{[011]}$       & $D_{s[111]}\eta_{[000]}$   & $D_{[012]}K_{[001]}$  \\
$D_{s[011]}\eta_{[011]}$    & $D_{s[000]}\eta_{[001]}$     & $D_{[012]}K_{[001]}$       & $D_{s[011]}\eta_{[001]}$   & $D_{s[002]}\eta_{[000]}$  \\
                         & $D^\ast_{s[001]}f_{0[000]}$   & $D_{s[011]}\eta_{[000]}$   & $D^\ast_{[011]}K_{[001]}$   & $D_{s[001]}\eta_{[001]}$  \\
                         &                          & $D_{s[001]}\eta_{[001]}$    & $D^\ast_{s[111]}f_{0[000]}$ & $D^\ast_{s[002]}f_{0[000]}$  \\
                         &                          & $D^\ast_{[001]}K_{[001]}$    &                        & $D^\ast_{s[001]}f_{0[001]}$  \\
                         &                          & $D^\ast_{s[011]}f_{0[000]}$  &                        &  \\ 
[1ex]
$(\singleop) \times 18$  & $(\singleop) \times 32$  & $(\singleop) \times 52$ & $(\singleop) \times 36$ & $(\singleop) \times 32$ \\
\hline
\end{tabular}

\vspace{0.5cm}

\begin{tabular}{c|c|c|c|c|c}
\multicolumn{1}{c|}{$[000]T_1^-$} & \multicolumn{1}{c|}{$[000]E^+$} & \multicolumn{1}{c|}{$[001]E_2$} & \multicolumn{1}{c|}{$[011]B_1$}  & \multicolumn{1}{c|}{$[011]B_2$} & \multicolumn{1}{c}{$[111]E_2$} \\
\hline \hline
$D_{[001]}K_{[001]}$        & $D_{[001]}K_{[001]}$        & $D_{[001]}K_{[011]}$       & $D_{[001]}K_{[001]}$       & $D_{[111]}K_{[001]}$  & $D_{[011]}K_{[001]}$  \\  
$D_{[011]}K_{[011]}$        & $D_{[011]}K_{[011]}$        & $D_{[011]}K_{[001]}$       & $D_{[012]}K_{[001]}$       & $D_{[011]}K_{[011]}$  & $D_{[001]}K_{[011]}$  \\
$D_{s[001]}\eta_{[001]}$    & $D_{[002]}K_{[002]}$        & $D_{s[011]}f_{0[001]}$      & $D_{[011]}K_{[011]}$       & $D_{s[001]}f_{0[001]}$  & $D_{s[011]}f_{0[001]}$  \\
$D^\ast_{[001]}K_{[001]}$    & $D_{s[001]}\eta_{[001]}$    & $D^\ast_{[001]}K_{[000]}$   & $D_{s[111]}f_{0[001]}$      & $D^\ast_{[011]}K_{[000]}$ & $D^\ast_{[111]}K_{[000]}$  \\
$D^\ast_{s[000]}f_{0[000]}$  & $D_{s[011]}\eta_{[011]}$     & $D^\ast_{[000]}K_{[001]}$   & $D^\ast_{[011]}K_{[000]}$   & $\{2\}D^\ast_{[001]}K_{[001]}$ &  \\
                         &                          & $D^\ast_{s[001]}f_{0[000]}$  & $D^\ast_{s[011]}f_{0[000]}$ & $D^\ast_{s[011]}f_{0[000]}$ &  \\
[1ex]
$(\singleop) \times 16$  & $(\singleop) \times 52$  & $(\singleop) \times 52$ & $(\singleop) \times 52$ & $(\singleop) \times 52$ & $(\singleop) \times 68$ \\
\hline
\end{tabular}
\end{center}
\caption{The interpolating operators used in each irrep, $[\vec{P}] \Lambda^{(P)}$, of the $DK$ $I=0$ channel on the $m_\pi = 239$~MeV ensemble.
The subscripts on the meson-meson operators refer to the momentum types.
The number in braces, $\{N_\text{mult}\}$, denotes the multiplicity of linearly independent two-meson operators if this is larger than one.
The number of $\bar{q}q$ operators used, $\mathfrak{n}$, is indicated by $(\singleop) \times \mathfrak{n}$, where $\mathbf{\Gamma}$ represents some combination of Dirac $\gamma$-matrices and up to three (two) spatial covariant derivatives at zero (non-zero) momentum (except only up to two spatial derivatives for $[000]T_1^-$).}
\label{tab:ops:860:DK}
\end{table}

\begin{table}
\begin{center}
\begin{tabular}{c|c|c|c}
\multicolumn{1}{c|}{$[000]A_1^+$} & \multicolumn{1}{c|}{$[001]A_1$} & \multicolumn{1}{c|}{$[011]A_1$} & \multicolumn{1}{c}{$[111]A_1$} \\
\hline \hline
$D_{[000]}K_{[000]}$           & $D_{[001]}K_{[000]}$           & $D_{[011]}K_{[000]}$           & $D_{[111]}K_{[000]}$  \\
$D_{[001]}K_{[001]}$           & $D_{[000]}K_{[001]}$           & $D_{[000]}K_{[011]}$           & $D_{[000]}K_{[111]}$  \\
$D_{[011]}K_{[011]}$           & $D_{[011]}K_{[001]}$           & $D_{[001]}K_{[001]}$           & $D_{[011]}K_{[001]}$  \\
$D_{[111]}K_{[111]}$           & $D_{[001]}K_{[011]}$           & $D_{[001]}K_{[111]}$           & $D_{[001]}K_{[011]}$  \\
                            & $D_{[111]}K_{[011]}$           & $D_{[111]}K_{[001]}$           & $D_{[002]}K_{[111]}$  \\
                            & $D_{[011]}K_{[111]}$           & $D_{[011]}K_{[011]}$           & $D_{[111]}K_{[002]}$  \\
                            & $D_{[002]}K_{[001]}$           & $D_{[002]}K_{[011]}$           &  \\
                            & $D_{[001]}K_{[002]}$           & $D_{[011]}K_{[002]}$           &  \\ [1ex]

$(\singleop) \times 8$      & $(\singleop) \times 32$     & $(\singleop) \times 52$     & $(\singleop) \times 36$ \\
\hline
\end{tabular}

\vspace{0.5cm}

\begin{tabular}{c|c|c|c|c}
\multicolumn{1}{c|}{$[000]T_1^-$} & \multicolumn{1}{c|}{$[001]E_2$} & \multicolumn{1}{c|}{$[011]B_1$}  & \multicolumn{1}{c|}{$[011]B_2$} & \multicolumn{1}{c}{$[111]E_2$} \\
\hline \hline
$D_{[001]}K_{[001]}$        & $D_{[011]}K_{[001]}$       & $D_{[001]}K_{[001]}$       & $D_{[111]}K_{[001]}$       & $D_{[011]}K_{[001]}$  \\
$D_{[011]}K_{[011]}$        & $D_{[001]}K_{[011]}$       & $D_{[011]}K_{[011]}$       & $D_{[001]}K_{[111]}$       & $D_{[001]}K_{[011]}$  \\
$D_{[111]}K_{[111]}$        & $D_{[111]}K_{[011]}$       & $D_{[002]}K_{[011]}$       & $D_{[011]}K_{[011]}$       & $D_{[002]}K_{[111]}$  \\
                         & $D_{[011]}K_{[111]}$       & $D_{[011]}K_{[002]}$       &                         & $D_{[111]}K_{[002]}$  \\
[1ex]
$(\singleop) \times 16$  & $(\singleop) \times 52$ & $(\singleop) \times 52$ & $(\singleop) \times 52$ & $(\singleop) \times 68$ \\
\hline
\end{tabular}

\vspace{0.5cm}

\begin{tabular}{c|c|c|c}
\multicolumn{1}{c|}{$[000]E^+$} &\multicolumn{1}{c|}{$[000]T_2^+$} & \multicolumn{1}{c|}{$[001]B_1$} & \multicolumn{1}{c}{$[001]B_2$}  \\
\hline \hline
$D_{[001]}K_{[001]}$        & $D_{[011]}K_{[011]}$        & $D_{[011]}K_{[001]}$       & $D_{[111]}K_{[011]}$  \\  
$D_{[011]}K_{[011]}$        & $D_{[111]}K_{[111]}$        & $D_{[001]}K_{[011]}$       & $D_{[011]}K_{[111]}$  \\
[1ex]
$(\singleop) \times 8$  & $(\singleop) \times 10$  & $(\singleop) \times 20$  & $(\singleop) \times 20$  \\
\hline
\end{tabular}
\end{center}
\caption{As Table~\ref{tab:ops:860:DK} but for $DK$ $I=0$ on the $m_\pi = 391$~MeV ensembles.
The $\bar{q}q$ operators include some combination of Dirac $\gamma$-matrices and up to two spatial covariant derivatives.
The irreps in the bottom row were not used on the $16^3$ volume.}
\label{tab:ops:840:DK}
\end{table}

\begin{table}
\begin{center}
\begin{tabular}{c|c|c|c|c|c|c}
\multicolumn{1}{c|}{$[000]A_1^+$} & \multicolumn{1}{c|}{$[000]T_1^-$} & \multicolumn{1}{c|}{$[000]E^+$} & \multicolumn{1}{c|}{$[001]A_1$} & \multicolumn{1}{c|}{$[011]A_1$} & \multicolumn{1}{c|}{$[111]A_1$}  & \multicolumn{1}{c}{$[002]A_1$} \\
\hline \hline
$D_{[000]}K_{[000]}$ & $D_{[001]}K_{[001]}$ & $D_{[001]}K_{[001]}$ & $D_{[001]}K_{[000]}$ & $D_{[011]}K_{[000]}$ & $D_{[111]}K_{[000]}$ & $D_{[001]}K_{[001]}$  \\
$D_{[001]}K_{[001]}$ & $D_{[011]}K_{[011]}$ & $D_{[011]}K_{[011]}$ & $D_{[000]}K_{[001]}$ & $D_{[000]}K_{[011]}$ & $D_{[000]}K_{[111]}$ & $D_{[011]}K_{[011]}$  \\
$D_{[011]}K_{[011]}$ & $D_{[111]}K_{[111]}$ & $D_{[002]}K_{[002]}$ & $D_{[011]}K_{[001]}$ & $D_{[001]}K_{[001]}$ & $D_{[011]}K_{[001]}$ & $D_{[111]}K_{[111]}$  \\
$D_{[111]}K_{[111]}$ &                   &                   & $D_{[001]}K_{[011]}$ & $D_{[111]}K_{[001]}$ & $D_{[001]}K_{[011]}$ & $D_{[002]}K_{[000]}$  \\
$D_{[002]}K_{[002]}$ &                   &                   & $D_{[002]}K_{[001]}$ & $D_{[001]}K_{[111]}$ & $D_{[002]}K_{[111]}$ & $D_{[000]}K_{[002]}$  \\
$^*D_{[012]}K_{[012]}$ &                 &                   & $D_{[001]}K_{[002]}$ & $D_{[011]}K_{[011]}$ & $D_{[012]}K_{[011]}$ & $D_{[012]}K_{[001]}$  \\
$^*D_{[112]}K_{[112]}$ &                 &                   & $D_{[111]}K_{[011]}$ & $D_{[002]}K_{[011]}$ & $D_{[112]}K_{[001]}$ & $D_{[112]}K_{[011]}$  \\
                  &                   &                   & $D_{[011]}K_{[111]}$ & $D_{[012]}K_{[001]}$ &                   &  \\
                  &                   &                   & $D_{[012]}K_{[011]}$ & $D_{[112]}K_{[011]}$ &                   &  \\
                  &                   &                   & $^*D_{[011]}K_{[012]}$ &                 &                   &  \\
                  &                   &                   & $^*D_{[012]}K_{[002]}$ &                 &                   &  \\
                  &                   &                   & $D_{[112]}K_{[111]}$ &                   &                   &  \\
\hline
\end{tabular}
\end{center}
\caption{As Table~\ref{tab:ops:860:DK} but for $D\bar{K}$ $I=0,1$ on the $m_\pi = 239$~MeV ensemble.
The $^*$ indicates operators that were only used in the $I=1$ channel.}
\label{tab:ops:860:DKbar}
\end{table}

\begin{table}
\begin{center}
\begin{tabular}{c|c|c|c}
\multicolumn{1}{c|}{$[000]A_1^+$} & \multicolumn{1}{c|}{$[001]A_1$} & \multicolumn{1}{c|}{$[011]A_1$} & \multicolumn{1}{c}{$[111]A_1$} \\
\hline \hline
$D_{[000]}K_{[000]}$           & $D_{[001]}K_{[000]}$           & $D_{[011]}K_{[000]}$           & $D_{[111]}K_{[000]}$  \\
$D_{[001]}K_{[001]}$           & $D_{[000]}K_{[001]}$           & $D_{[000]}K_{[011]}$           & $D_{[000]}K_{[111]}$  \\
$D_{[011]}K_{[011]}$           & $D_{[011]}K_{[001]}$           & $D_{[001]}K_{[001]}$           & $D_{[011]}K_{[001]}$  \\
$D_{[111]}K_{[111]}$           & $D_{[001]}K_{[011]}$           & $D_{[001]}K_{[111]}$           & $D_{[001]}K_{[011]}$  \\
$D_{[002]}K_{[002]}$           & $D_{[111]}K_{[011]}$           & $D_{[111]}K_{[001]}$           & $D_{[002]}K_{[111]}$  \\
                            & $D_{[011]}K_{[111]}$           & $D_{[011]}K_{[011]}$           & $D_{[111]}K_{[002]}$  \\
                            & $D_{[002]}K_{[001]}$           & $D_{[002]}K_{[011]}$           &  \\
                            & $D_{[001]}K_{[002]}$           & $D_{[011]}K_{[002]}$           &  \\
\hline
\end{tabular}

\vspace{0.5cm}

\begin{tabular}{c|c|c|c|c}
\multicolumn{1}{c|}{$[000]T_1^-$} & \multicolumn{1}{c|}{$[001]E_2$} & \multicolumn{1}{c|}{$[011]B_1$}  & \multicolumn{1}{c|}{$[011]B_2$} & \multicolumn{1}{c}{$[111]E_2$} \\
\hline \hline
$D_{[001]}K_{[001]}$        & $D_{[011]}K_{[001]}$       & $D_{[001]}K_{[001]}$       & $D_{[111]}K_{[001]}$       & $D_{[011]}K_{[001]}$  \\
$D_{[011]}K_{[011]}$        & $D_{[001]}K_{[011]}$       & $D_{[011]}K_{[011]}$       & $D_{[001]}K_{[111]}$       & $D_{[001]}K_{[011]}$  \\
$D_{[111]}K_{[111]}$        & $D_{[111]}K_{[011]}$       & $D_{[002]}K_{[011]}$       & $D_{[011]}K_{[011]}$       & $D_{[002]}K_{[111]}$  \\
$D_{[002]}K_{[002]}$        & $D_{[011]}K_{[111]}$       & $D_{[011]}K_{[002]}$       &                         & $D_{[111]}K_{[002]}$  \\
\hline
\end{tabular}

\vspace{0.5cm}

\begin{tabular}{c|c|c|c}
\multicolumn{1}{c|}{$[000]E^+$} &\multicolumn{1}{c|}{$[000]T_2^+$} & \multicolumn{1}{c|}{$[001]B_1$} & \multicolumn{1}{c}{$[001]B_2$}  \\
\hline \hline
$D_{[001]}K_{[001]}$        & $D_{[011]}K_{[011]}$        & $D_{[011]}K_{[001]}$       & $D_{[111]}K_{[011]}$  \\  
$D_{[011]}K_{[011]}$        & $D_{[111]}K_{[111]}$        & $D_{[001]}K_{[011]}$       & $D_{[011]}K_{[111]}$  \\
$D_{[002]}K_{[002]}$        &                          &                         &  \\
\hline
\end{tabular}
\end{center}
\caption{As Table~\ref{tab:ops:860:DK} but for $D\bar{K}$ $I=0,1$ on the $m_\pi = 391$~MeV ensembles.}
\label{tab:ops:840:DKbar}
\end{table}


\clearpage
\section{Principal correlators}
\label{app:prin_corrs}

\begin{figure}
\includegraphics[width=1.0\textwidth]{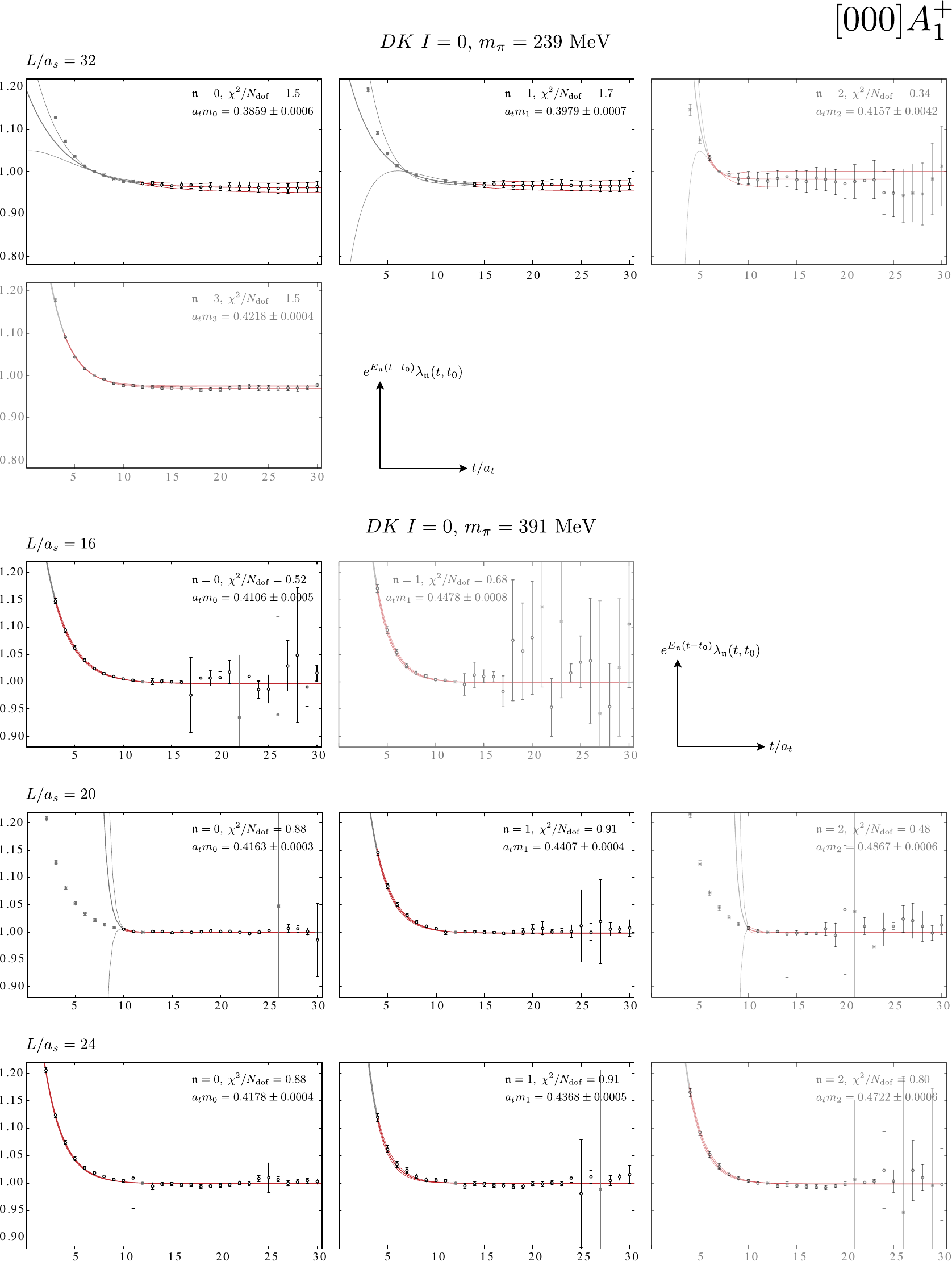}
\caption{The principal correlators, plotted as $e^{E_\mathfrak{n}(t-t_0) }\lambda_\mathfrak{n}(t,t_0)$, obtained for $DK$ $I=0$ in the $[000]A_1^+$ irrep with $m_\pi = 239$~MeV (top) and $m_\pi = 391$~MeV (bottom). Points shown in grey were not included in the fit. The red curves show the result of two-exponential fits as detailed in Section~\ref{sec:calculation} (this fit form enforces $\lambda_\mathfrak{n}(t_0,t_0)=1$). The mass extracted from the leading exponential and the $\chi^2/N_{\mathrm{dof}}$ are indicated for each fit.
These principal correlators correspond to the levels shown in Figs.~\ref{fig:860spectrum:DK} and \ref{fig:840spectrum:DK}; faded plots correspond to levels shown in grey in those figures and these are not used in the scattering analyses.}
\label{fig:prin_corrs:DK}
\end{figure}

\begin{figure}
\includegraphics[width=1.0\textwidth]{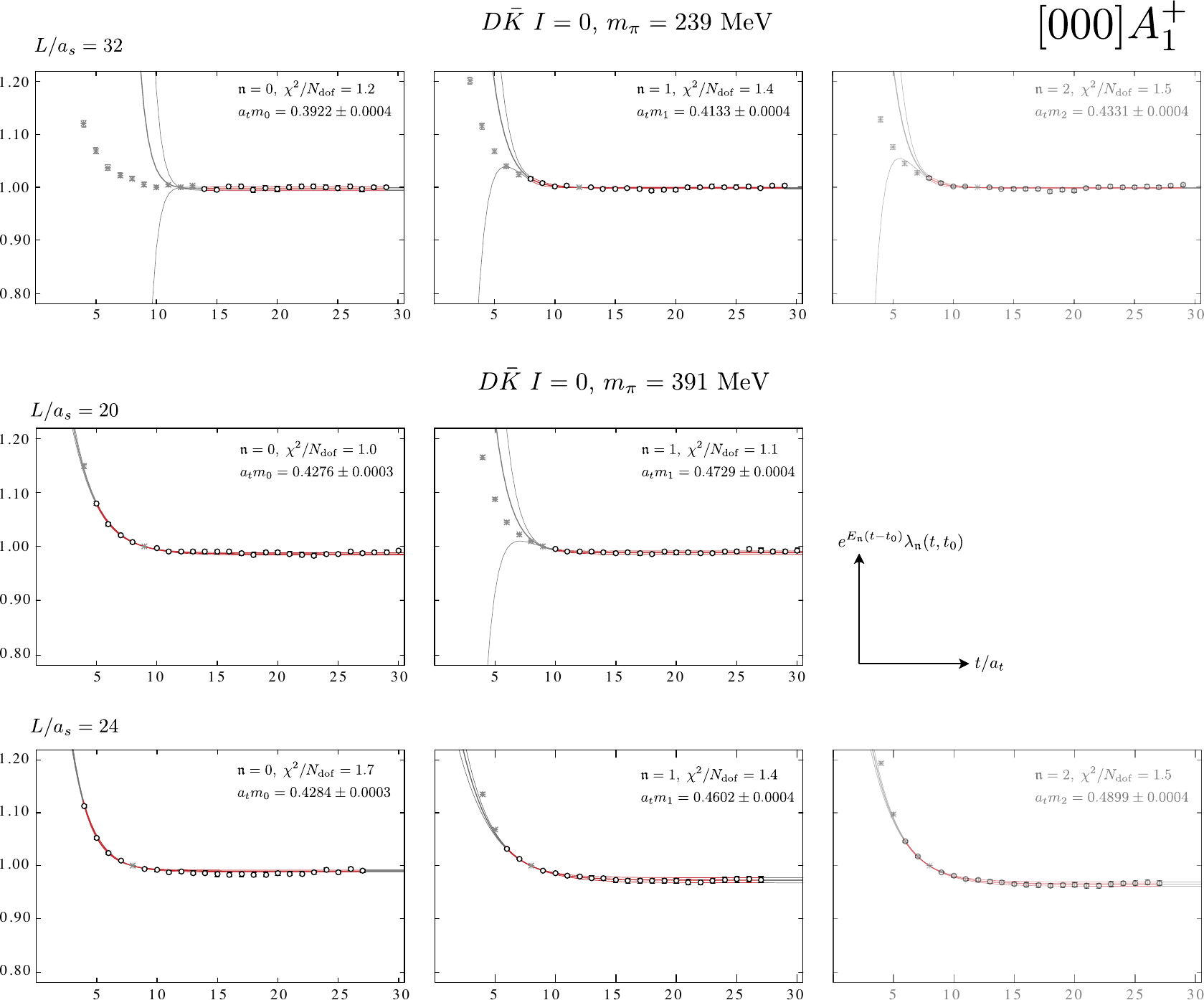}
\caption{As Fig.~\ref{fig:prin_corrs:DK} but for $D\bar{K}$ $I=0$. These correspond to levels shown in Figs.~\ref{fig:860spectrum:DKbar} and \ref{fig:840spectrum:DKbarI0}.}
\label{fig:prin_corrs:DKbarI0}
\end{figure}

\begin{figure}
\includegraphics[width=1.0\textwidth]{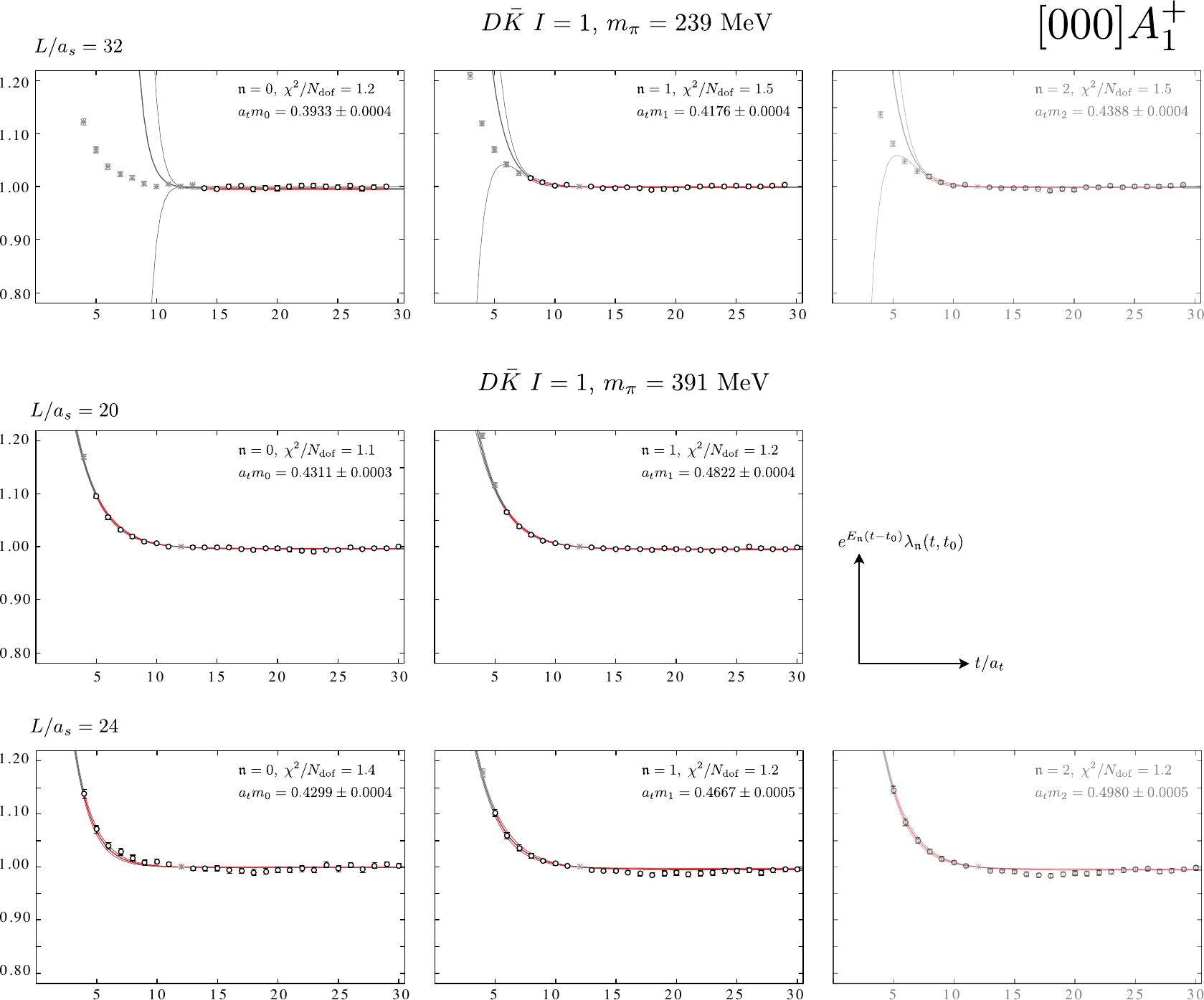}
\caption{As Fig.~\ref{fig:prin_corrs:DK} but for $D\bar{K}$ $I=1$. These correspond to levels shown in Figs.~\ref{fig:860spectrum:DKbar} and \ref{fig:840spectrum:DKbarI1}.}
\label{fig:prin_corrs:DKbarI1}
\end{figure}

In Figs.~\ref{fig:prin_corrs:DK}, \ref{fig:prin_corrs:DKbarI0} and \ref{fig:prin_corrs:DKbarI1}, as a representative example we show the principal correlators from the $[000]A_1^+$ irrep for $DK$ $I=0$, $D\bar{K}$ $I=0$, and $D\bar{K}$ $I=1$ respectively. All of the levels shown are included in the spectrum figures in Section~\ref{sec:spectrum}, but the faded principal correlator plots correspond to levels shown in grey in those figures and these levels were not used in the scattering analyses. Higher-energy levels obtained in the variational analysis are not shown. Table~\ref{tab:prin_corrs} summarises the fitting ranges and $t_0$ values used. Note that we add an additional uncertainty to the raw $m_\pi = 391$ MeV energies shown in Figs.~\ref{fig:prin_corrs:DK},~\ref{fig:prin_corrs:DKbarI0} and \ref{fig:prin_corrs:DKbarI1} before performing any scattering analyses, as detailed above.

\begin{table}[ht]
\begin{center}
\begin{tabular}{cccccc}
$[000]A_1^+$                      & $m_\pi$/MeV & $L/a_s$ & $t_0$ & $t_\mathrm{min}$ & $t_\mathrm{max}$\\
\hline
\multirow{4}{*}{$DK$ $I=0$}       & 239 & 32      & 7     & 4     & 30 \\ \cline{2-6}
                 & \multirow{3}{*}{391} & 16      & 12    & 3     & 31 \\
                                  &     & 20      & 12    & 4     & 30 \\
                                  &     & 24      & 12    & 2     & 30 \\
\hline
\multirow{3}{*}{$D\bar{K}$ $I=0$} & 239 & 32      & 12    & 8     & 29 \\ \cline{2-6}
                 & \multirow{2}{*}{391} & 20      & 9     & 5     & 30 \\
                                  &     & 24      & 8     & 4     & 27 \\
\hline
\multirow{3}{*}{$D\bar{K}$ $I=1$} & 239 & 32      & 12    & 8     & 29 \\ \cline{2-6}
                 & \multirow{2}{*}{391} & 20      & 12    & 5     & 30 \\
                                  &     & 24      & 12    & 4     & 30 \\
\end{tabular}
\caption{The maximum overall fitting ranges and $t_0$ values used for the fits presented in Figs.~\ref{fig:prin_corrs:DK}, \ref{fig:prin_corrs:DKbarI0} and \ref{fig:prin_corrs:DKbarI1}.}
\label{tab:prin_corrs}
\end{center}
\end{table}

\clearpage
\section{Covariance-adjusted and uncorrelated $D\bar{K}$ fits}
\label{app:DKbarfits}

Tables \ref{tab:860DKbarI0:limiteigs}, \ref{tab:840DKbarI0:limiteigs}, \ref{tab:860DKbarI1:limiteigs} and \ref{tab:840DKbarI1:limiteigs} present details of $D\bar{K}$ $I=0$ and $I=1$ scattering amplitude fits discussed in Sections \ref{sec:scattering:DKbarI0} and \ref{sec:scattering:DKbarI1}.

\begin{table}[tb]
\begin{center}
\small
\setlength\tabcolsep{2pt}
\begin{tabular}{c|cccc|c} 
  & $r=0.001$ & $r=0.002$ & $r=0.005$ & $r=0.01$ & uncorr. \\
\hline
$a_0/a_t$ & $9.4 \pm 2.0$     & $9.5 \pm 2.4$      & $8.9 \pm 2.9$     & $8.3 \pm 3.4$      & $5.3 \pm 2.4$      \\
$r_0/a_t$ & $32 \pm 12$       & $33 \pm 14$        & $31 \pm 18$       & $26 \pm 24$        & $-9 \pm 55$        \\
$a_1/a_t$ & $-65 \pm 95$      & $-62 \pm 108$      & $-100 \pm 140$     & $-114 \pm 240$     & $-240 \pm 220$     \\
$a_2/a_t$ & $-15100 \pm 8700$ & $-17200 \pm 10500$ & $-20500 \pm 14500$ & $-23000 \pm 18000$ & $-36000 \pm 32000$ \\
corr. &
$\begin{smallbmatrix}
    & 1   & 0.93  &  0.57 &  0.53  \\
    &     & 1     &  0.30 &  0.36  \\
    &     &       &  1    &  0.68  \\
    &     &       &       &  1
\end{smallbmatrix}$ &
$\begin{smallbmatrix}
    & 1   & 0.94  &  0.63 &  0.61  \\
    &     & 1     &  0.40 &  0.45  \\
    &     &       &  1    &  0.67  \\
    &     &       &       &  1
\end{smallbmatrix}$ &
$\begin{smallbmatrix}
    & 1   & 0.95  &  0.69 &  0.65  \\
    &     & 1     &  0.54 &  0.55  \\
    &     &       &  1    &  0.61  \\
    &     &       &       &  1
\end{smallbmatrix}$ &
$\begin{smallbmatrix}
    & 1   & 0.96  &  0.74 &  0.67  \\
    &     & 1     &  0.63 &  0.61  \\
    &     &       &  1    &  0.55  \\
    &     &       &       &  1
\end{smallbmatrix}$ &
$\begin{smallbmatrix}
    & 1   & 0.97  & -0.10 &  0.01  \\
    &     & 1     & -0.08 &  0.06  \\
    &     &       &  1    & -0.30  \\
    &     &       &       &  1
\end{smallbmatrix}$
\\
$N_\text{adj}$ & 3 & 7 & 10 & 12 & -- \\
$\chi^2/\Ndf$ & $\frac{11.2}{18 - 4} = 0.80$ & $\frac{10.3}{18 - 4} = 0.73$ & $\frac{7.46}{18 - 4} = 0.53$ &
  $\frac{5.75}{18 - 4} = 0.41$ & $\frac{1.51}{18 - 4} = 0.11$ \\
\end{tabular}
\caption{$D\bar{K}$ $I=0$ scattering amplitude fits on the $m_\pi = 239$~MeV ensemble, as in Eq.~\ref{eqn:860DKbarI0:ref} but with small eigenvalues adjusted as described in the text (four columns labelled by $r$) or correlations between finite-volume energies set to zero (final column). $N_\text{adj}$ is the number of small eigenvalues adjusted. While adjusting the eigenvalues of the covariance matrix could be considered to have reduced the number of degrees of freedom, here the $\chi^2/\Ndf$ are presented with the original $\Ndf$.}
\label{tab:860DKbarI0:limiteigs}
\end{center}
\end{table}

\begin{table}[tb]
\begin{center}
\small
\setlength\tabcolsep{2pt}
\begin{tabular}{c|ccc|c} 
  & $r=0.002$ & $r=0.005$ & $r=0.01$ & uncorr. \\
\hline
$a_0/a_t$ & $16.1 \pm 2.3$  & $15.9 \pm 2.4$  & $16.0 \pm 2.5$  & $18.3 \pm 2.2$  \\
$r_0/a_t$ & $36.4 \pm 1.8$  & $36.6 \pm 1.8$  & $36.7 \pm 1.8$  & $37.3 \pm 3.4$  \\
$a_1/a_t$ & $133 \pm 27$    & $129 \pm 29$    & $128 \pm 30$    & $160 \pm 39$    \\
$a_2/a_t$ & $4600 \pm 1300$ & $4400 \pm 1400$ & $4400 \pm 1400$ & $4600 \pm 1400$ \\
corr. &
$\begin{smallbmatrix}
    & 1   &  0.36  &  0.75 &  0.75  \\
    &     &  1     & -0.14 & -0.14  \\
    &     &        &  1    &  0.91  \\
    &     &        &       &  1 
\end{smallbmatrix}$ &
$\begin{smallbmatrix}
    & 1   &  0.39  &  0.74 &  0.75  \\
    &     &  1     & -0.12 & -0.11  \\
    &     &        &  1    &  0.97  \\
    &     &        &       &  1
\end{smallbmatrix}$ &
$\begin{smallbmatrix}
    & 1   &  0.37  &  0.75 &  0.76  \\
    &     &  1     & -0.13 & -0.13  \\
    &     &        &  1    &  0.94  \\
    &     &        &       &  1
\end{smallbmatrix}$ &
$\begin{smallbmatrix}
    & 1   &  0.64  & -0.06 & -0.01  \\
    &     &  1     &  0.05 &  0.04  \\
    &     &        &  1    & -0.04  \\
    &     &        &       &  1
\end{smallbmatrix}$
\\
$N_\text{adj}$ & 2 & 5 & 7 & -- \\
$\chi^2/\Ndf$ & $\frac{18.6}{29 - 4} = 0.74$ & $\frac{16.6}{29 - 4} = 0.66$ &
  $\frac{15.6}{29 - 4} = 0.62$ & $\frac{29.7}{29 - 4} = 1.19$ \\
\end{tabular}
\caption{As Table~\ref{tab:860DKbarI0:limiteigs} but for $D\bar{K}$ $I=0$ on the $m_\pi = 391$~MeV ensembles, Eq.~\ref{eqn:840DKbarI0:ref}.}
\label{tab:840DKbarI0:limiteigs}
\end{center}
\end{table}

\begin{table}[tb]
\begin{center}
\small
\setlength\tabcolsep{2pt}
\begin{tabular}{c|ccc|c} 
  & $r=0.001$ & $r=0.002$ & $r=0.005$ & uncorr. \\
\hline
$a_0/a_t$ & $-4.61 \pm 0.24$  & $-4.68 \pm 0.27$  & $-4.73 \pm 0.34$   & $-6.21 \pm 0.65$   \\
$a_1/a_t$ & $-227 \pm 70$     & $234 \pm 75$      & $-226 \pm 88$      & $-530 \pm 190$     \\
$a_2/a_t$ & $-18700 \pm 9300$ & $19400 \pm 10100$ & $-21000 \pm 12000$ & $-54000 \pm 29000$ \\
corr. &
$\begin{smallbmatrix}
    & 1   & 0.80  &  0.63 \\
    &     & 1     &  0.68 \\
    &     &       &  1
\end{smallbmatrix}$ &
$\begin{smallbmatrix}
    & 1   & 0.73  &  0.58 \\
    &     & 1     &  0.62 \\
    &     &       &  1
\end{smallbmatrix}$ &
$\begin{smallbmatrix}
    & 1   & 0.58  &  0.49 \\
    &     & 1     &  0.45 \\
    &     &       &  1
\end{smallbmatrix}$ &
$\begin{smallbmatrix}
    & 1   & -0.15 & -0.20 \\
    &     &  1    & -0.20 \\
    &     &       &  1
\end{smallbmatrix}$
\\
$N_\text{adj}$ & 4 & 7 & 11 & -- \\
$\chi^2/\Ndf$ & $\frac{14.7}{18-3} = 0.98$ & $\frac{13.5}{18-3} = 0.90$ & $\frac{10.3}{18-3} = 0.68$ 
 & $\frac{7.01}{18-3} = 0.47$ \\
\end{tabular}
\caption{As Table~\ref{tab:860DKbarI0:limiteigs} but for $D\bar{K}$ $I=1$ on the $m_\pi = 239$~MeV ensembles, Eq.~\ref{eqn:860DKbarI1:ref}.}
\label{tab:860DKbarI1:limiteigs}
\end{center}
\end{table}

\begin{table}[tb]
\begin{center}
\small
\setlength\tabcolsep{2pt}
\begin{tabular}{c|ccc|c} 
  & $r=0.002$ & $r=0.005$ & $r=0.01$ & uncorr. \\
\hline
$a_0/a_t$ & $-4.39 \pm 0.38$ & $-4.24 \pm 0.48$ & $-4.02 \pm 0.57$ & $-2.78 \pm 0.56$ \\
$r_0/a_t$ & $16.2 \pm 3.7$   & $-15.0 \pm 4.7$  & $-13.1 \pm 6.2$  & $9 \pm 19$       \\
$a_1/a_t$ & $-46 \pm 16$     & $-44 \pm 19$     & $-41 \pm 22$     & $-18 \pm 34$     \\
$a_2/a_t$ & $-2710 \pm 750$  & $-2500 \pm 890$  & $2330 \pm 1020$  & $2100 \pm 1300$  \\
corr. &
$\begin{smallbmatrix}
    & 1   &  0.80  &  0.66 &  0.66  \\
    &     &  1     &  0.21 &  0.21  \\
    &     &        &  1    &  0.91  \\
    &     &        &       &  1 
\end{smallbmatrix}$ &
$\begin{smallbmatrix}
    & 1   &  0.82  &  0.69 &  0.70  \\
    &     &  1     &  0.31 &  0.32  \\
    &     &        &  1    &  0.85  \\
    &     &        &       &  1 
\end{smallbmatrix}$ &
$\begin{smallbmatrix}
    & 1   &  0.85  &  0.68 &  0.71  \\
    &     &  1     &  0.38 &  0.39  \\
    &     &        &  1    &  0.79  \\
    &     &        &       &  1 
\end{smallbmatrix}$ &
$\begin{smallbmatrix}
    & 1   &  0.93  & -0.14 & -0.05  \\
    &     &  1     & -0.11 & -0.04  \\
    &     &        &  1    & -0.03  \\
    &     &        &       &  1 
\end{smallbmatrix}$
\\
$N_\text{adj}$ & 7 & 13 & 17 & -- \\
$\chi^2/\Ndf$ & $\frac{20.2}{28-4} = 0.84$ & $\frac{18.5}{28-4} = 0.77$ &
  $\frac{15.2}{28-4} = 0.63$ & $\frac{29.6}{28-4} = 1.23$ \\
\end{tabular}
\caption{As Table~\ref{tab:860DKbarI0:limiteigs} but for $D\bar{K}$ $I=1$ on the $m_\pi = 391$~MeV ensembles, Eq.~\ref{eqn:840DKbarI1:ref}.}
\label{tab:840DKbarI1:limiteigs}
\end{center}
\end{table}


\clearpage
\newpage

\bibliography{DK_paper}
\bibliographystyle{JHEP}

\end{document}